\documentclass{aa}

\usepackage{txfonts}
\usepackage{textgreek}
\usepackage{graphicx, color}
\usepackage[draft]{hyperref}
\usepackage{pdflscape}
\usepackage{multirow,bigdelim}
\usepackage{calc}
\usepackage{threeparttable}
\usepackage{amsmath}
\usepackage{booktabs}
\usepackage[utf8]{inputenc}
\usepackage{array}
\usepackage{tabularx}
\usepackage{caption}
\usepackage[font=small]{caption}
\usepackage[labelformat = empty,position=top]{subcaption}
\usepackage[export]{adjustbox}
\usepackage{mathptmx}
\usepackage{grffile}
\usepackage{pdflscape}

\definecolor{dgreen}{rgb}{0., 0.7, 0.}

\definecolor{tiff}{rgb}{0.04, 0.73, 0.71}

\definecolor{darkgreen}{RGB}{0, 113, 0}
\definecolor{darkyellow}{RGB}{204, 204, 0}
\definecolor{orange}{RGB}{255,165,0}

\newcommand{\feh}{\mbox{\rm [{\rm Fe}/{\rm H}]}}
\newcommand{\ofe}{\mbox{\rm [{\rm O}/{\rm Fe}]}}
\newcommand{\mgfe}{\mbox{\rm [{\rm Mg}/{\rm Fe}]}}
\newcommand{\cafe}{\mbox{\rm [{\rm Ca}/{\rm Fe}]}}
\newcommand{\sife}{\mbox{\rm [{\rm Si}/{\rm Fe}]}}

\newcommand{\afe}{\mbox{\rm [{\textalpha}/{\rm Fe}]}}

\newcommand{\Msun}{\mbox{${M}_{\odot}$}}
\newcommand{\Zsun}{\mbox{${Z}_{\odot}$}}

\newcommand{\Mi}{\mbox{$M_{\rm i}$}}
\newcommand{\Zi}{\mbox{$Z_{\rm i}$}}
\newcommand{\MUP}{\mbox{${M_{\rm UP}}$}}
\newcommand{\M}{\mbox{$M$}}
\newcommand{\Mfin}{\mbox{${M}_{\rm fin}$}}
\newcommand{\Mrem}{\mbox{${M}_{\rm rem}$}}
\newcommand{\Mcut}{\mbox{${M}_{\rm cut}$}}
\newcommand{\Mhe}{\mbox{${M}_{\rm He}$}}
\newcommand{\Mco}{\mbox{${M}_{\rm CO}$}}
\newcommand{\Mvmo}{\mbox{${M}_{\rm VMO}$}}

\usepackage[T1]{fontenc}
\usepackage{ae,aecompl}
{}

\usepackage{fixltx2e}
{\left\lbrace\begin{array}{@{}l@{}}}%
{\end{array}\right.}

\DeclareUnicodeCharacter{2212}{-}

\begin{document}

\title{On the effects of the Initial Mass Function on Galactic chemical enrichment}  

\subtitle{}




\author{S. Goswami
          \inst{1}
          \and
          A. Slemer
         \inst{2,3,4}
          \and
          P. Marigo 
         \inst{3}
          \and
          A. Bressan
          \inst{1}
          \and
          L. Silva
          \inst{5,6}
          \and
           M. Spera
          \inst{1, 3, 7, 8}
          \and
		   L. Boco
          \inst{1,5,6}
          \and
           V. Grisoni
          \inst{1,5}
          \and
           L. Pantoni
          \inst{1}
          \and
           A. Lapi
          \inst{1,5,6,9}
          }


       \institute{
        SISSA, Via Bonomea 265, I-34136 Trieste, Italy\\
        \email{ sgoswami@sissa.it; alessandro.bressan@sissa.it;
        mario.spera@sissa.it; vgrisoni@sissa.it; lboco@sissa.it; lapi@sissa.it; lpantoni@sissa.it} 
        \and 
        CNR-IFN Padova, Via Trasea 7, I-35131 Padova, Italy \\
        \email{alessandra.slemer@pd.ifn.cnr.it}
        \and
        Dipartimento di Fisica e Astronomia, Universit\`a degli studi di Padova,
        Vicolo Osservatorio 3, Padova, Italy\\
        \email{paola.marigo@unipd.it}
        \and
        INAF-OAPD, Vicolo dell'osservatorio 5, Padova, Italy
        \and
        INAF-OATs, Via G. B. Tiepolo 11, I-34143 Trieste, Italy\\
        \email{laura.silva@inaf.it}
        \and
        IFPU - Institute for fundamental physics of the Universe, Via Beirut 2, 34014 Trieste, Italy
        \and
        INFN, Sezione di Padova, Via Marzolo 8, I--35131, Padova, Italy
        \and
        Center for Interdisciplinary Exploration \& Research in Astrophysics (CIERA) and Department of Physics \& Astronomy, Northwestern University, Evanston, IL 60208, USA
        \and 
        INFN-Sezione di Trieste, via Valerio 2, 34127 Trieste,  Italy
        }

 \date{Accepted by A\&A}
    
  \abstract{
 There is mounting evidence that the stellar initial mass function (IMF) could extend much beyond the canonical $\Mi \sim 100\, \Msun$ limit, but the impact of such hypothesis on the chemical enrichment of galaxies still remains to be clarified.}
   {We aim to address this question by analysing the observed abundances of thin- and thick-disc stars in the Milky Way with 
   chemical evolution models that account for the contribution of very massive stars dying as pair instability supernovae.}
   {We built new sets of chemical yields from massive and very massive stars up to  $\Mi \sim 350\, \Msun$, by combining the wind ejecta extracted from our hydrostatic stellar evolution models with explosion ejecta from the literature. 
   Using a simple chemical evolution code we  analyse the effects of adopting different yield tables by comparing predictions against observations of stars in the solar vicinity.}
   {
   After several tests, we focus on the [O/Fe] ratio which best separates the chemical patterns of the two Milky Way components. We find that with a {\sl standard} IMF, truncated at $\Mi \sim 100\, \Msun$, we can reproduce various observational constraints for thin-disc stars,  but the same IMF  fails to account for the [O/Fe] ratios of thick-disc stars. 
   The best results are obtained by extending the IMF up to  $\Mi = 350\, \Msun$ and including the chemical ejecta of very massive stars, in the form of winds and  pair instability supernova explosions.}
   {Our study indicates that PISN could have played a significant role in shaping the chemical evolution of the Milky Way thick disc. By including their chemical yields it is easier to reproduce not only the level of the \textalpha-enhancement but also the observed slope of thick-disc stars in the [O/Fe] vs. [Fe/H] diagram.
   The bottom line is that the contribution of very massive stars to the chemical enrichment of galaxies is potentially quite important and should not be neglected in chemical evolution models.}
   \maketitle
%
%
%
%
%
%
%
%
%
%
%
%
%
%
%
%
%
%
%
%
\section{Introduction}
The chemical evolution of the Milky Way is  one of the most
important astrophysical topics because it provides direct hints on the more general question of how galaxies formed and evolved. It is also an anchor for galaxy models because of the possibility to study their properties through the analysis of individual stars.
Indeed this field of research is continuously growing providing on one side an increasing
amount of observational abundance data that contribute to enlarge and sharpen
the whole picture \citep{Gilmore2012,Bensbyetal14,Majewski2017,Laverny2013} and, on the other, a growing number of interpretative tools that go from simple chemical evolutionary models to more complex chemo-hydrodynamical models \citep[e.g.][]{Valentini2019}. A key ingredient to interface model predictions with observations are the stellar chemical yields, describing the contribution of stars of different types to the metal enrichment of the galaxies. Other physical processes play of course an important role in the chemical evolution of the galaxies, like the functional forms adopted to describe the stellar birthrate function,  the gas inflows and outflows, the mixing of newly ejected elements with the surrounding medium, the relative displacement of individual stars from their original  positions, etc.
However stellar yields keep a key role because, being the result of the evolution of stars,  they may provide a tight link between stellar and galactic time-scales.
The contribution of individual stars to the metal enrichment has been the subject of many studies in the past \citep[e.g.][]{Matteucci2014}. The chemical enrichment from stars takes place when elements newly produced by nuclear reactions in the deep stellar interiors are ejected into the interstellar medium, via stellar winds or supernova explosions. 

Low- and intermediate-mass  stars, $\Mi \simeq 0.8\,\Msun - 6\,\Msun$, never reach the carbon ignition temperature and enrich the interstellar medium (ISM) mainly during the Red Giant Branch (RGB) and  Asymptotic Giant Branch (AGB) phases \citep[][and references therein]{Marigo01,Cristallo_etal09,Cristallo_etal_11,Cristallo_etal_15,Karakas_10,Karakas_etal_16,Karakas_etal_18,Ventura_etal13,Ventura_etal_17, Ventura_etal_18,Pignatari2016, Slemer_etal_17, Ritter_etal18}. 

Stars in a narrow mass interval, $\Mi \simeq 6\,\Msun-10\,\Msun$, are able to burn carbon in their non- or mildly degenerate cores and experience the so-called Super-AGB phase \citep{Ritossa_etal_96,GarciaBerro_etal_97,Iben_etal_97}.
Depending on the efficiency of stellar winds and the growth of the core mass, the final fate of Super-AGB stars splits in two channels that lead either to the formation of an O-Ne-Mg white dwarf ($7 \lesssim \Mi/\Msun \lesssim$ 9) or to an electron capture supernova   \citep[$9 \lesssim \Mi/\Msun \lesssim$ 10; ][]{Hurley2000,Siess2007,Poelarends2008}. 

Massive stars ($\Mi \simeq 10 - 120\,\Msun$) experience more advanced nuclear burnings (Ne, O, Si)  up to the formation of an iron core, that eventually implodes producing either a successful core-collapse supernova (CCSN) explosion,
or the direct collapse into a black hole as a \textit{failed}~SN    \citep{Woosley_Weaver1995,Fryer1999,Limongi&Chieffi2004, Limongi&Chieffi2006,Nomoto2006,Fryer2006, Fryer2001,Heger_Woosley2002y,Heger2003,Fryer2012,Janka2012, Ugliano2012,Ertl2015,Pignatari2016, Ritter_etal18, Limongi_etal18}.  
On rare occasions that depend on the characteristics of the progenitor and the details of the explosion, stars in this mass range may give rise to powerful hypernovae \citep{Izzoetal2019}.

Very massive objects (VMO), $100 \lesssim \Mi/\Msun \lesssim 300\, \Msun$, enter the pair instability regime during central oxygen burning, which may trigger their final thermonuclear explosion \citep{Heger_Woosley2002y, Umeda2002, Kozyreva2014a, Woosley_Heger2015, Woosley2017ApJ}. 

The chemical ejecta contributed by all stars of different initial masses and evolutionary stages are key ingredients of galactic chemical evolution studies.
In most cases, the adopted stellar IMF is truncated at around $\Mi \simeq 100\, \Msun$, which implies that no chemical contribution from VMO is taken into account. This common assumption is due to the fact that very few evolutionary models exist for stars with $\Mi > 100 \,\Msun$.  With the exception of zero-metallicity stars
\citep[e.g., ][]{Haemmerle_etal_18, Yoon_etal_12, Ohkubo_etal_09,Ohkubo_etal_2006, Lawlor_etal_08, Marigo03_zeta0,Marigo01_zeta0},  the lack of systematic evolutionary studies of VMO has effectively limited the exploration area of chemical evolution models, which focussed on the role of population III stars only \citep[e.g., ][]{CherchneffDwek_10,Cherchneff_etal_09,Rollinde_etal_09, Ballero_etal_06, MatteucciPipino_05, Ricotti_04}.
In fact, the occurrence of VMO and their final fates were thought to apply only to primordial population III stellar populations \citep{Bond1982,Bond1984,Heger_Woosley2002y,Nomoto2013}.

In recent years our understanding of the evolution of massive and very massive stars has dramatically changed as a result of important discoveries.
We can now rely on studies focussed on young super star clusters \citep{Evans2010, Walborn2014, Schneider2018, Crowther2019, Crowther2016} and on  the identification of massive stellar black holes hosted in binary systems \citep{Abbottetal2016,Abbottetal2020,Spera_etal2017,Spera_etal2015}. All this evidence points to an IMF that extends up to VMO, either as a genuine top-heavy IMF, or in virtue of an early 
efficient merging in binary systems \citep{Senchynaetal20}.
Further support for a top-heavy IMF in  certain environmental conditions comes also from recent finding that the low observed $^{16}$O/$^{18}$O isotopic ratios in starburst galaxies can be reproduced only by models that assume an excess of massive stars \citep{Romano2017}.  
However, while there is strong evidence supporting the existence of stars with initial mass up to $\Mi \simeq 350\, \Msun$, other evidence exists in favour of the opposite scenario, i.e. that the IMF is bottom-heavy, like in the case of
observations of the gravity-sensitive narrow-band integrated indices in local ellipticals 
\citep{van_Dokkum_2012,van_Dokkum_2017}. These latter observations are quite challenging from the chemical evolution point of view because massive elliptical galaxies are  among the most metal-rich stellar systems known and it is difficult to explain their chemical pattern with such a bottom-heavy IMF \citep{Bressan94,Matteucci94,Thomas2005}.
In parallel, it has also been suggested  that the IMF may be a self regulating process 
leading to a variation of its form in different environmental conditions. Early models based on an IMF varying with time and with galactic environment \citep{Padoan1997}  could better explain many of the observed properties of elliptical and low surface brightness galaxies, such as $\alpha$-enhancement, downsizing, fundamental plane etc. \citep{Chiosi+1998,Jimenez98}. Since then, more evidence of a dependence of the IMF on the environment has been accumulated, leading to the definition of  galaxy wide IMF (gwIMF) and time integrated gwIMF (IGIMF)
that may deviate from a canonical IMF, depending on metallicity and star formation activity
\citep{Kroupa2003ApJ,Weidner2005,Kroupa2008,Recchi2009,marks_imf_2012MNRAS, jerabkova2018, hoseketal_IMF2019}.

Motivated by these studies our group has recently carried out the first systematic analysis of stellar models for massive stars and VMO, extending up to $\Mi = 350\, \Msun$, for a large grid of initial metallicity, from $\Zi = 0.0001$ to $\Zi = 0.06$ \citep{Tang2014,Chen2015,costa2020}. 
These models, computed with the \texttt{PARSEC} code, are based on updated input physics and, above all, rely on the most recent advances in the theory of stellar winds, which critically affect the evolution of massive stars and VMO.

In this work we use the same \texttt{PARSEC} models of massive stars and VMO to derive their stellar wind ejecta, which are then incorporated  in a chemical evolution model to analyse the impact of varying the slope of the IMF and its upper mass limit.
Additionally, we test different sets of chemical ejecta and check their ability in reproducing the chemical characteristics of the thin and thick-disc stars.

For this purpose  we first collected from the literature various sets of ejecta, that cover the contributions from low-, intermediate-mass and  massive stars, the latter being usually provided up to $\Mi \simeq 100\, \Msun$.
Then, we constructed new tables of chemical ejecta, extending the initial mass range to include the contribution of VMO,  up to $\Mi = 350\, \Msun$. 
The new set of chemical yields of massive stars and VMO  contains the stellar wind ejecta, based on the \texttt{PARSEC} models, suitably tailored to explosive yields for CCSN, pulsational pair instability SN (PPISN), and of pair instability SN (PISN) taken from the recent literature. 

The structure of the paper is as follows. Basic stellar classification and relevant mass limits are recalled in Sect.~\ref{sec_masslim_final_fates}.
In Sect.~\ref{sec_parsec} we introduce our new set of chemical ejecta for massive and very massive stars.  After an outline of our \texttt{PARSEC} code,  we describe the method adopted to combine \texttt{PARSEC} stellar evolution models of massive stars with explosive models available in the literature. Then we discuss the resulting ejecta due to both stellar winds and explosions (CCSN, PPISN, and PISN). The full content of the ejecta tables, made available to the user, is detailed in Appendix~\ref{tables_ejecta}. Section~\ref{agb_yields}  summarizes the main characteristics of the AGB yields obtained with our \texttt{COLIBRI} code.  Section \ref{sec_comp_ej} introduces and compares various sets of chemical ejecta of AGB  and massive stars taken from the recent literature.
Section~\ref{sec_chem_mod} briefly describes 
the observational sample of thin- and thick-disc stars in the solar vicinity and the adopted chemical evolution code used for the interpretation of their abundances. 
In Sect.~\ref{sec_chem_prop} we analyse the predictions of chemical evolution models calculated adopting  different sets of chemical ejecta.  Their performance is tested through the use of various diagnostics, with particular focus on the observed [O/Fe] vs. [Fe/H] diagram populated by thin- and thick-disc stars.
Finally, Sect.~\ref{conclusions} recaps our study and its main conclusions.
\section{Stellar classes and mass limits}
\label{sec_masslim_final_fates}
The final fate of stars depends primarily on their initial mass, \Mi, and metallicity, \Zi.
To characterise the chemical contributions of stars it is convenient to group them in classes as a function of \Mi, according to evolutionary paths and final fates.
Let us introduce a few relevant limiting masses that define each stellar family.  Mass limits and  other relevant quantities used throughout the paper are also defined  in Table \ref{tab_quantities}. 

It should be noticed that the mass ranges specified below should not be considered as strict, but rather approximate limits, since they significantly depend on the efficiency of processes like convective mixing and stellar winds and, especially for massive stars, also on the initial chemical composition.

We define with $M_{\rm AGB} \simeq 6\, \Msun$ the maximum initial mass  for a star to build a highly electron-degenerate C-O core after the end of the He-burning phase. This class comprises low- and intermediate-mass stars, which then proceed through the AGB phase leaving a C-O WD as compact remnant.

Stars with $\Mi > M_{\rm AGB}$
are able to burn carbon in mildly or non-degenerate conditions.
Those stars that build an electron-degenerate O-Ne-Mg core are predicted to enter the Super-AGB phase, undergoing recurrent He-shell flashes and  powerful mass loss, similarly to the canonical AGB phase. If stellar winds are able to strip off the entire H-rich envelope while the core mass is still lower than $\simeq 1.38\, \Msun $, then the evolution will end as an O-Ne-Mg WD \citep{Nomoto1984,Iben_etal_97}.
We denote with $M_{\rm SAGB} \simeq 7\, \Msun$ the upper  mass limit of this class of stars \citep{herwig2005,Siess2006,Siess2007,Doherty_etal14}.

Stars with $\Mi > M_{\rm SAGB}$ and having an electron-degenerate O-Ne-Mg core that is able to grow in mass up to the critical value of $\simeq 1.38\Msun$, are expected to explode as electron capture supernovae \citep[ECSN;][]{Nomoto1984, Poelarends2008, Leung2020}.
 
Let us denote with $M_{\rm mas}$ the minimum initial mass for a star to avoid electron degeneracy in the core after carbon burning.
We note that following this definition the progenitors of electron capture SN cover the range $M_{\rm SAGB} < \Mi < M_{\rm mas}$.

Stars with $M_{\rm mas}\leq \Mi < M_{\rm VMO} \simeq 100\, \Msun$
are able to proceed through all hydrostatic nuclear stages up to Si-burning, with the formation of a Fe core that eventually undergoes a dynamical collapse triggered by electron-captures and photodisintegrations \citep{Woosley_Weaver1995,Thielemann2011}.

Very massive stars  with  $\Mi \ge M_{\rm VMO}$
may experience electron-positron pair creation instabilities before and during oxygen burning, with a final fate that is mainly controlled by the mass of the helium core, $\Mhe$ \citep{Heger_Woosley2002y, Heger2003,Nomoto2013,Kozyreva2014a,Woosley2016,Woosley2017ApJ,Leung2019}, resulting in a successful/failed CCSN or thermonuclear explosion. 
For further details, we refer to Sect.~\ref{ssec_pisn}.


\section{Chemical ejecta of massive and very massive stars using PARSEC models}
\label{sec_parsec}

In this work our reference set of evolutionary tracks for massive and very massive stars is taken from the large database of Padova and TRieste Stellar Evolution Code (\texttt{PARSEC})\footnote{\url{https://people.sissa.it/~sbressan/parsec.html}}$^{,}$\footnote{\url{http://stev.oapd.inaf.it/cgi-bin/cmd}}. The \texttt{PARSEC} code is extensively described elsewhere \citep{Bressan2012,costaetal2019AA,costaetal2019MNRAS} and here we will provide only a synthetic description of the relevant input physics.   

\subsection{Evolutionary models}
\label{ssec_parsecmod}
The \texttt{PARSEC} database includes stellar models with initial masses from $8\, \Msun$  to $350\, \Msun$ and metallicity values $\Zi = 0.0001, 0.0005, 0.001, 0.004, 0.006, 0.008, 0.017, 0.02, 0.03$ \citep{Chen2015,Tang2014}.
 The adopted reference solar abundances are taken from \cite{Caffau_etal11}, with a present-day solar metallicity $Z_{\odot}=0.01524$. Note that the latter value does not correspond to the initial metallicity of the Sun, which instead is predicted to be $Z_{\rm protoSUN}=0.017$ (see Sect.~\ref{modelcostraints}).
 For all metallicities the initial chemical composition of the models is assumed to be scaled-solar. The isotopes included in the code are: $\mathrm{H}$, $\mathrm{D}$, $\mathrm{^{3}He}$, $\mathrm{^{4}He}$, $\mathrm{^{7}Li}$, $\mathrm{^{7}Be}$, $\mathrm{^{12,13}C}$,$\mathrm{^{14,15}N}$, $\mathrm{^{16,17,18}O}$, $\mathrm{^{19}F}$, $\mathrm{^{20,21,22}Ne}$,$\mathrm{^{23}Na}$ $\mathrm{^{24,25,26}Mg}$, $\mathrm{^{26,27}Al}$, $\mathrm{^{28}Si}$.
Opacity tables are from Opacity Project At Livermore (OPAL)\footnote{\url{http://opalopacity.llnl.gov/}} team \citep[][and references  therein]{Iglesias1996} for $4.2 \leq \log(T/\mathrm{K}) \leq 8.7$, and from 
\textsc{\AE SOPUS} tool\footnote{\url{http://stev.oapd.inaf.it/aesopus}} \citep{Marigo2009}, for  $3.2 \leq \log(T/\mathrm{K}) \leq 4.1$. Conductive opacities are included following \citet{Itoh2008}.
Neutrino energy losses by electron neutrinos are taken from \citet{Munakata1985}, \citet{Itoh1983} and  \citet{Haft1994}.
The equation of state is from \textsc{freeeos}\footnote{\url{http://freeeos.sourceforge.net/}}
code version 2.2.1 by Alan W. Irwin.

The mass loss prescriptions employed in \texttt{PARSEC} are the law of \cite{DeJager1988} for red super giants (RSG; $T_{\rm eff} \le 12\,000$ K), the \cite{Vink2000} relations for blue super giants (BSG; $T_{\rm eff} > 12\,000$ K), and \cite{Grafener2008} and \cite{Vink2011} during the transition phase from O-type to Luminous Blue Variables (LBV) and RSG, and finally to Wolf Rayet (WR) stars.


\begin{figure}
\centering
\resizebox{1.02\hsize}{!}{
\includegraphics[angle=0]{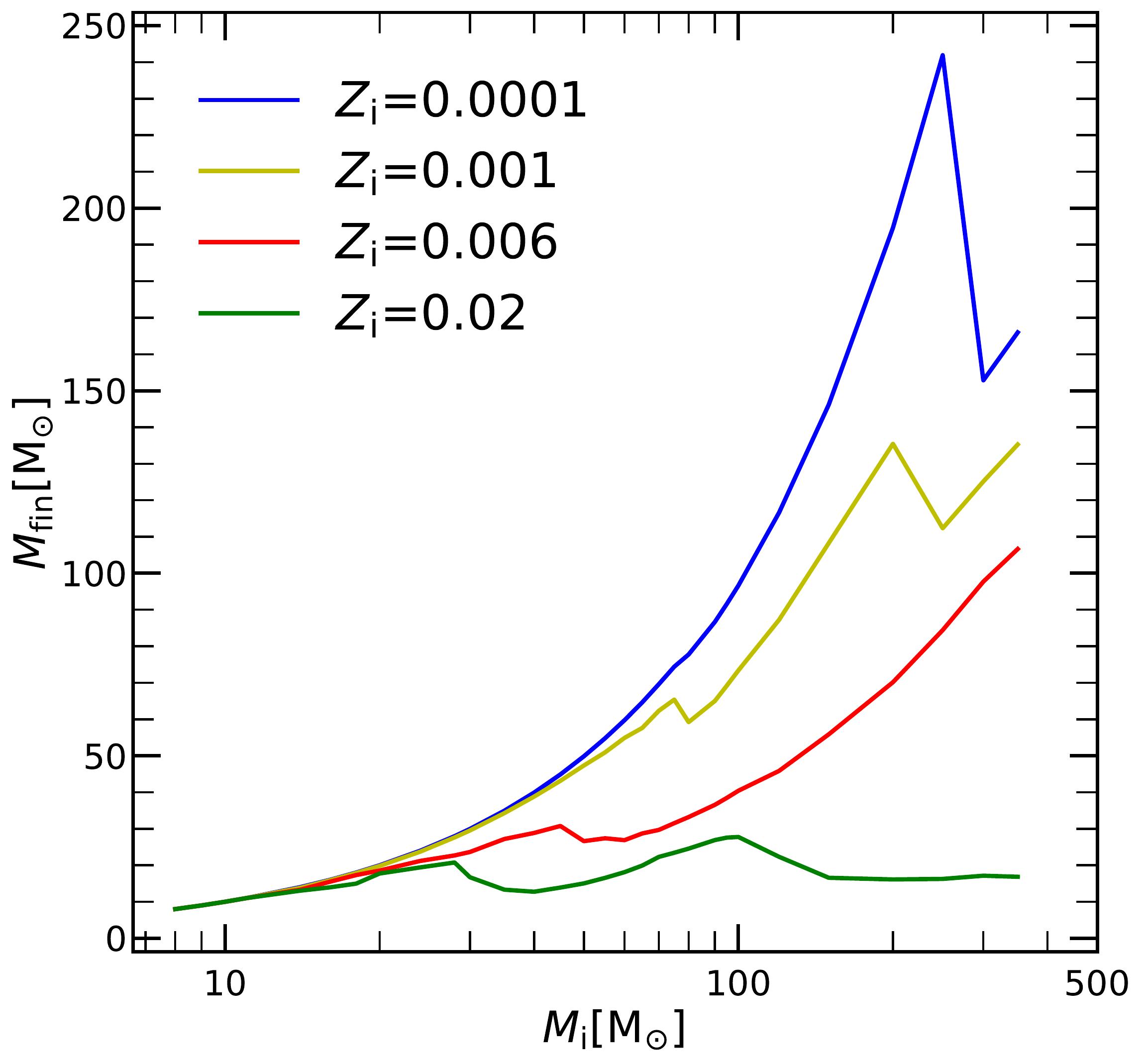}}
 \caption{Pre-SN mass $M_{\rm fin}$ as function of initial $\Mi$ for different values of the initial metallicity \Zi, as indicated. Stellar tracks are taken from \texttt{PARSEC} $\mathrm{VI.1}$ models.}
 \label{Mfin-Miniz}
\end{figure}

\begin{table*}
\begin{center}
\caption{Description of the main quantities used in this work.\label{tab_quantities}}
\begin{tabularx}{\linewidth}{lX}
\hline\hline
\multicolumn{1}{c}{\textbf{name}} &
\multicolumn{1}{c}{\textbf{definition}} \\
\hline
WD   & White dwarf \\
ECSN & Electron capture supernova \\
CCSN  & Core collapse supernova \\
PPISN & Pulsation pair instability supernova \\
PISN & Pair instability supernova \\
DBH & Direct collapse to black hole \\
$\Zi$ & Initial metallicity. \\
$\Mi$ & Mass of the star at the zero-age main sequence \\
$M_{\rm fin}$ & Mass of the star at the beginning of central carbon burning (almost equivalent to the pre-SN mass) \\
$M_{\rm rem}$ & Mass of the remnant \\
$M_{\rm cut}$ & Mass-cut, in a pre-supernova model, enclosing the entire mass that will collapse and form the compact remnant \\
$M_{\rm He}$ & Mass of the He-core at the beginning of central carbon burning \\
$M_{\rm CO}$ & Mass of CO-core at the beginning of central carbon burning \\
$M_{\rm AGB}$ & Maximum mass for a star to experience the AGB phase and leave a C-O WD \\
$M_{\rm SAGB}$ & Maximum mass for a star to evolve through the Super-AGB and leave an O-Ne-Mg WD \\
$M_{\rm mas}$ & Minimum mass for a star to experience all hydrostatic nuclear burnings up to the Si-burning stage, with the formation of an iron core which eventually collapses, leading to either a successful CCSN or a failed SN\\
$M_{\rm VMO}$ & Mass boundary between massive stars and VMO \\
%
%
 \hline\hline
\end{tabularx}
\end{center}
\end{table*}

\subsection{Calculation of the chemical ejecta} 
\label{EJECTA}
Since with the adopted version of the \texttt{PARSEC} code, models of massive and very massive stars are evolved until central carbon exhaustion, we combine our evolutionary tracks with extant explosive models
covering a range of initial masses that corresponds to different final fates (CCSN, failed SN, PPISN and PISN).
Following the work by \citet{Slemer2016},
for each stellar model of given initial mass $M_{\rm i}$ we first compute
the amount of ejected mass of the element \textit{j} due to the stellar winds, $E_j^{\rm w}(M_{\rm i})$, and then
the contributions of the  associated supernova channels, $E_j^{\rm sn}(M_{\rm i})$, as detailed in Sect.~\ref{sec_explo}.
The total ejecta $E_j(M_{\rm i})$ are given by
\begin{equation}
E_j(M_{\rm i})=E_j^{\rm w}(M_{\rm i}) +E_j^{\rm sn}(M_{\rm i})
\label{eijtot}
\end{equation}
The complete tables of wind and explosion ejecta for massive and very massive stars  ($8 \le \Mi/\Msun \le 350$) and four values of the initial metallicity ($0.0001 \le \Zi \le 0.02$) are described in Appendix~\ref{tables_ejecta}. They are available online \footnote{\url{http://gofile.me/55RoU/qnITvlSQU}}.

We consider the most important chemical species and their isotopes from H to Zn.
The species explicitly included in \texttt{PARSEC} nuclear networks are all the isotopes from $\mathrm{^{1}H}$ to $\mathrm{^{28}Si}$. Heavier elements are present in the initial chemical composition, according to the adopted scaled-solar mixture (see Sect.~\ref{sec_parsec}), and are not affected by the nuclear reactions and mixing events during the hydrostatic H- and He-burning phases.

In the following we detail how we calculate the wind and explosion ejecta,
$E_j^{\rm w}(M_{\rm i})$ and  $E_j^{\rm sn}(M_{\rm i})$,  that appear in  Eq.(\ref{eijtot}).
\subsection{Wind ejecta}\label{wind}
%
%
The wind ejecta of a species $j$ contributed by a star of initial mass \Mi\ is computed with the following equation:
\begin{equation}
E^{\rm w}_j(M_{\rm i}) = \int_{0}^{\tau_{\rm C}} \dot{M}(M_{\rm i},t) X_j^{\rm s}(t) dt \,
\label{ejecta_wind}
\end{equation}
where the integral is performed over the stellar lifetime, from the zero age main sequence (ZAMS) up to the stage of carbon ignition, $\tau_{\rm C}$. For a given \Mi\ the quantities $\dot{M}(M_{\rm i},t)$ and $X_j^{\rm s}(t)$
denote, respectively,  the mass-loss rate and surface abundance (in mass fraction) of the species \textit{j}, at the current time \textit{t}.

The total amount of mass lost by a star during its hydrostatic evolution, $\Mi-\Mfin$,  can be appreciated from 
Fig.~\ref{Mfin-Miniz},  which shows the pre-SN mass ($M_{\rm fin}$) as a  function of \Mi, for a few selected values of the initial metallicity.

Figure~\ref{ej-wind1}, in the appendix, illustrates the fractional wind ejecta, $E^{\rm w}_j(M_{\rm i})/M_{\rm i}$, of the main chemical species considered in the \texttt{PARSEC} models,  as a function of \Mi\ and \Zi. 
 For $M_{\rm i} \le 100\, \Msun$ the  wind ejecta generally increase with initial mass and metallicity, which is explained by the strengthening of stellar winds at higher luminosities and larger abundances of metals. This applies to H, He, N, Ne, Na, Mg, Al, and Si.

The metallicity trends for C and O reverse in the case of VMO. Compared to the predictions for $\Zi = 0.02$, at low metallicities and high initial masses, $\Zi \leq 0.006$ and $\Mi > 100-200\, \Msun$, the wind  ejecta may be larger by even one order of magnitude for C and up to two orders of magnitude for O. This result is explained when considering the stage at which stars of different \Mi\ and \Zi\ enter the WC and WO phases, which are characterized by powerful winds enriched in C and O.

%
At $\Zi = 0.02$ all VMO  experience high mass loss before entering in the WC regime, which is attained close to the end of the He-burning phase.
We note that these models are not expected to go through the WO regime. As a consequence, their ejecta are characterized by low amounts of primary C and O.
Conversely, at lower metallicities, $\Zi = 0.0001$ and $0.004$,  due to the relatively weak stellar winds during the early evolutionary stages, VMO reach the WC and WO regimes with a much larger mass, hence producing higher ejecta of C and O.
\subsection{Explosion ejecta of electron capture supernovae}
\label{ssec_ecsn}
To account for the ECSN channel we take advantage of the recent revision of the \texttt{PARSEC} code \citep{costa2020}, who extended the sequence of hydrostatic nuclear burnings up to oxygen. In this way we can check which models develop a degenerate O-Ne-Mg core after the carbon burning phase. As mentioned in Sect.~\ref{agb_yields}, we did not follow the Super-AGB phase and the corresponding yields are taken from \citet{Ritter_etal18} using the models with $\Mi=6 \,\Msun$ and $ 7 \,\Msun$. 

As to the ECSN channel we proceed as follows.
Given the severe uncertainties that affect the definition of the mass range for the occurrence of ECSN \citep[e.g.,][]{Doherty_etal_17, Poelarends2008} and the modest chemical contribution expected from the explosive nucleosynthesis \citep[e.g.,][]{Wanajo_etal_09}, we adopt a simple approach.
For each \Zi\, we look over the  mass range $8 \le \Mi/\Msun \le 10$, and  assign the ECSN channel to the  \texttt{PARSEC} models that, after the carbon-burning phase, develops a degenerate core with mass close to the critical value of $1.38\,\Msun$. In the metallicity range under consideration ($0.0001 \le \Zi \le 0.02$), this condition is met by  \texttt{PARSEC} models with $8 \le \Mi/\Msun \le 9$.

The ECSN explosion ejecta are taken from the work of \citet[][their table 2]{Wanajo_etal_09}
using the FP3 model as suggested by the authors. The nucleosynthesis results derive from a neutrino-driven explosion of a collapsing O–Ne–Mg core of mass $=1.38 \,\Msun$, with a stellar progenitor of $\Mi=8.8\,\Msun$ \citep{Nomoto1984}. According to this model, the total mass ejected by the explosion is quite low, $\simeq 1.39\times 10^{-2}\,\Msun$. This fact, together with the neutron-richness of the ejecta,  lead to a modest production of radioactive $^{56}$Ni and hence of stable $^{56}$Fe ($\approx 0.002-0.004\, \Msun$). In addition, the ECSN yields are characterized by a minor production of \textalpha-elements (e.g., O and Mg), and an appreciable synthesis of heavier species like $^{64}$Zn and some light p-nuclei (e.g., $^{74}$Se, $^{78}$Kr, $^{84}$Sr, and $^{92}$Mo).
Assuming no fall back during the explosion, the ECSN event is expected to produce a neutron star with mass $\Mrem=1.366\,\Msun$.
Finally, we add the ejecta of the layers above \Mrem\, with the chemical composition predicted by the corresponding \texttt{PARSEC} model.

The specific ECSN model adopted for each \Zi\ serves as a bridge between AGB  and massive stars, to avoid a coarse mass-interpolation of the ejecta in the transition region $\M_{\rm AGB} < \Mi/\Msun < M_{\rm mas}$ (see Table~\ref{tab_quantities}). 
We note that, with a canonical IMF extending up to 350 $ \Msun$, the weight of the $\M_{\rm AGB}-M_{\rm mas}$ range should be not that large and only a few elements would be affected, that are not the focus of the present paper.
A more careful consideration of this mass interval is postponed to a future study.

\subsection{Explosion ejecta of core collapse supernovae}
\label{sec_explo}
\begin{figure*}
\centering
\includegraphics[angle=0,width=84mm]{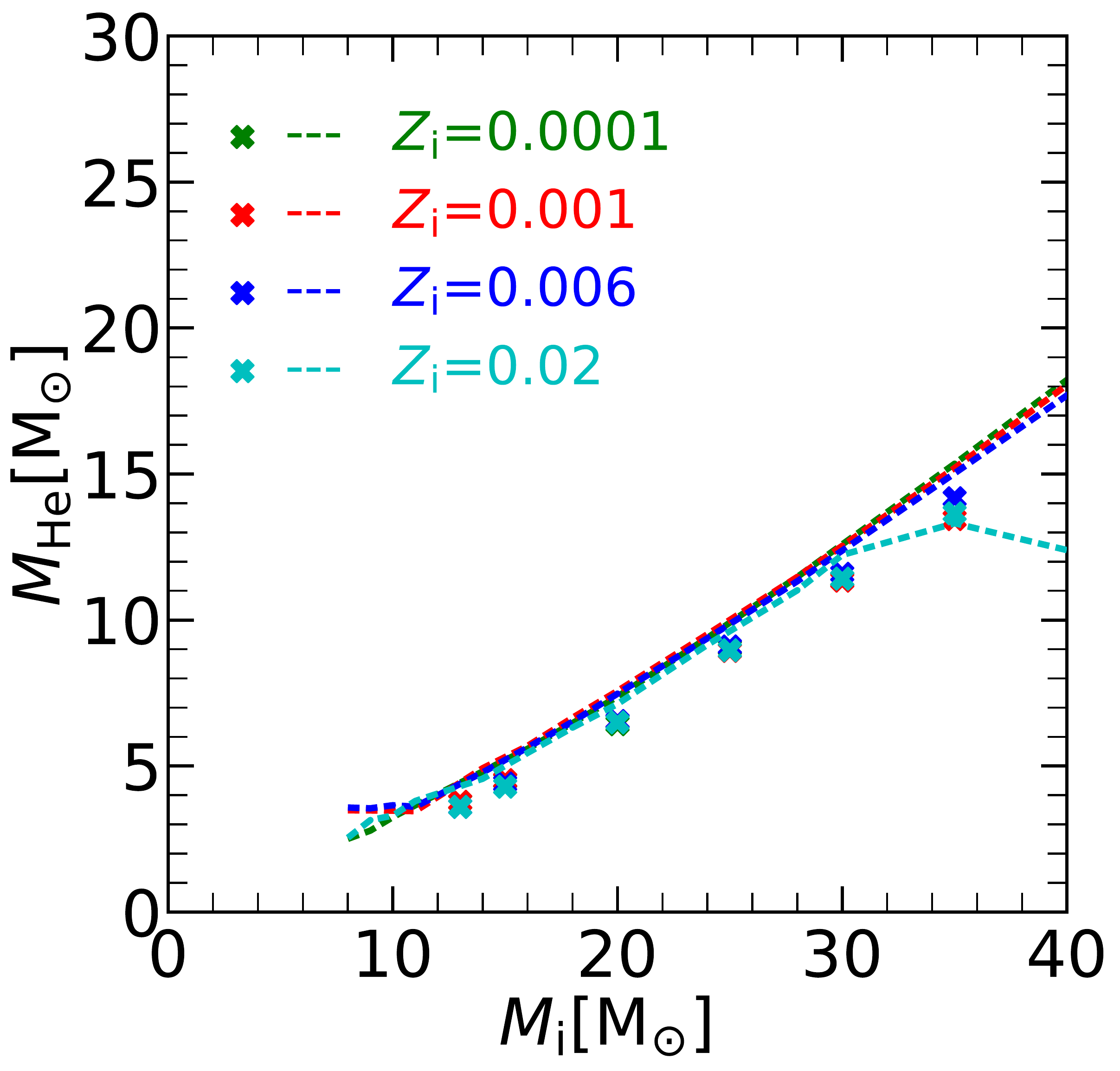}
\includegraphics[angle=0,width=84mm]{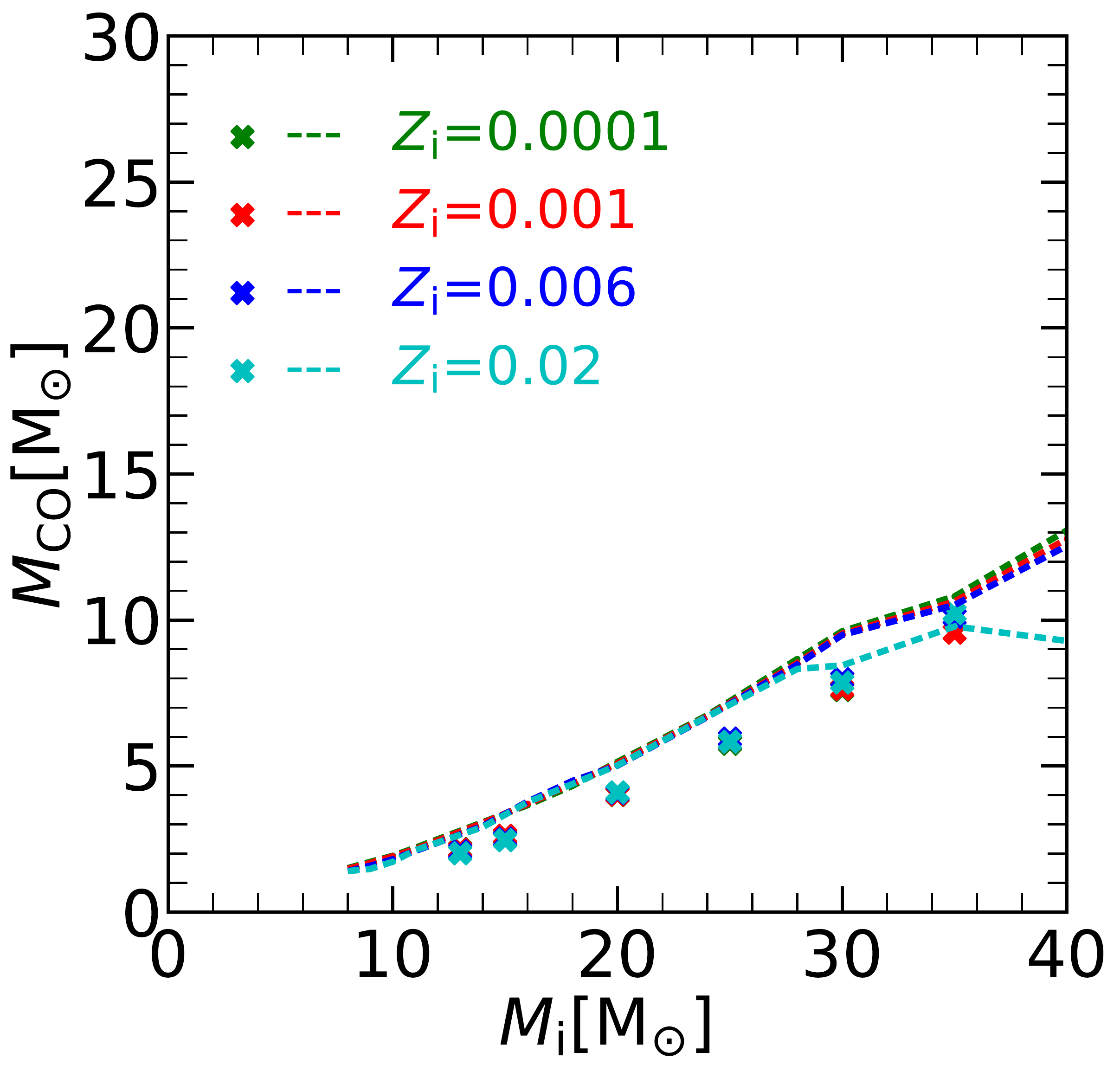}
\caption{\Mhe (left panel) and \Mco (right panel) as a function of \Mi, for different values of \Zi. Lines show  the data extracted from \texttt{PARSEC} stellar evolution models, while crosses represent the models of \protect\cite{Limongi&Chieffi2004}.
}\label{Mco-Miniz}
\end{figure*}

Stars in this class have $M_{\rm mas} < \Mi < \Mvmo$.
The upper limit, \Mvmo, corresponds to a star that reaches $M_{\rm He}\sim 32\Msun$ after central He burning and enters the pair-instability regime during O-burning \citep{Woosley2017ApJ}, thus avoiding the standard evolutionary path to the silicon burning stage (see Sect.~\ref{ssec_pisn}).
At solar composition, this mass limit is typically $\Mvmo \sim 100\,\Msun$ but it is expected to vary with metallicity,  as it is affected by mass loss during the early evolutionary phases.

Of particular relevance for this mass range is the determination of the explodability of a model, i.e. the conditions that lead to a successful SN or to a failed SN.
In recent years there have been many attempts to explore the dependence of the  outcome of the supernova collapse on the input physics, with the final goal to possibly determine a relation between the explodability and the main stellar parameters, in particular the initial or pre-supernova mass of the star \citep{Fryer1999,Fryer2006,Fryer2001,Heger_Woosley2002y,Heger2003,oconnor2011,
Fryer2012,Janka2012,Ugliano2012,Ertl2015}.

We briefly recall that \citet{Fryer2012} provided simple relations of the explodability with  the final C-O core mass ($\Mco$), \citet{oconnor2011} introduced the compactness criterion as a threshold for the explodability and \citet{Ertl2015} introduced  a two-parameter explodability criterion.  Fryer et al's models depend only on the C-O core mass after carbon burning and on the pre-supernova mass of the star. Instead, the other models
predict a non-monotonic  behaviour of the explodability with the core mass,
with the existence of islands of explodability intermixed with islands of failures.

A related issue concerns the material that may fall back onto the surface of the proto-neutron star after the explosion, leading eventually to the formation of a black hole and possibly also to a failed SN \citep{Fryer2012}.

It is clear that the present theoretical scenario is heterogeneous and there is no unanimous consensus of different authors on the explodability of a massive star following the collapse of its Fe core.
This is mainly due to the fact that when the models are near the critical conditions for explosion they become critically sensitive to slight variations in the input micro-physics and numerical treatments \citep[][]{Burrows_etal_18}.

All these facts make it hard to unambiguously set a threshold mass between successful and failed explosions. However, since indications exist that a reasonable limit could be in the range $25\, \Msun \lesssim \Mi \lesssim 30\, \Msun$ \citep[e.g.][]{oconnor2011, Sukhbold2016}, we assume that massive stars with  $\Mi \geq 30\, \Msun$ will fail to explode.

The other important parameter needed to obtain the ejecta is the remnant mass, $\Mrem$.
To derive this quantity one could use the observed relation between the ejecta of $\mathrm{^{56}Ni}$ and the pre-supernova mass of CCSN \citep{Umeda2008,Utrobin&Chugai2009,Utrobin2010}.
We note that also this relation is affected by some uncertainty, in particular on the determination of the pre-supernova mass \Mfin. For this reason, we prefer to use, as the calibrating value for the models, the estimated value of the $\mathrm{^{56}Ni}$ mass ejected
by SN1987A, $\mathrm{^{56}Ni}\sim0.07\Msun$  \citep{Nomoto2013,Prantzos_18}.

Given the explodability criterion and the ejected mass of $\mathrm{^{56}Ni}$, we adopt suitable explosion models to derive the corresponding ejecta.
For this purpose we use the CCSN models by \cite{Limongi_Chieffi2003} and \cite{Limongi&Chieffi2004} (hereafter CL04) because they  tabulate the explosion isotopes as a function of the internal mass coordinate.

Each stellar model of the \texttt{PARSEC} grid is characterized by four known parameters, namely:  \Mi, \Zi, \Mfin    and \Mco.  We use the mass of the C-O core, \Mco, to match the \texttt{PARSEC} models to CL04 ones for $\Zi=0.0001, 0.001, 0.006, 0.02$. These are the only values of \Zi\ in common between CL04 and  \texttt{PARSEC}. Once identified the CL04 explosion model that corresponds to a given \Mco, it is straightforward to integrate  from the external layers inward until the desired ejecta of $\mathrm{^{56}Ni}$ is reached.
The corresponding mass coordinate of the inner layer provides the mass cut, \Mcut, and hence the explosion ejecta.

This scheme needs to be made a bit more articulated to take into account that for the same \Mi\ the \texttt{PARSEC} and CL04 models do not predict  exactly the same \Mco.
Simple  interpolations are therefore applied.

We proceed as follows.
For each \texttt{PARSEC} model of metallicity \Zi\ we identify in the corresponding CL04 grid the two explosive models that bracket  the mass of the core, $M_{\rm CO1} <\Mco<M_{\rm CO2}$, with  pre-explosive masses $M_{\rm fin1}$ and $M_{\rm fin2}$, respectively.
 Using the $\mathrm{^{56}Ni}$ criterion we derive the corresponding mass cuts, $M_{\rm cut}(M_{\rm CO1})$ and  $M_{\rm cut}(M_{\rm CO2})$, and the explosive ejecta, integrating from
$M_{\rm cut}(M_{\rm CO1,2})$ to $M_{\rm fin1,2}$.
Finally, we use \Mco\ of the \texttt{PARSEC} model  as interpolating variable to obtain $M_{\rm cut}(\Mi, \Zi, \Mco)$ and the explosion ejecta  $E^{\rm sn}_j(\Mi, \Zi, \Mco)$ for all chemical species under consideration.

To estimate the mass of the remnant, \Mrem, we assume that in successful CCSN the efficiency of fall-back is negligible, as shown by recent hydrodynamical simulations \citep{Ertl2015}.
It follows that $\Mrem = \Mcut$ for successful CCSN and $\Mrem = \Mfin$ for failed SN.
As to the nature of the compact remnant, we assign a neutron star for  $\Mrem < 2.9\,\Msun$, or a black hole otherwise \citep{Tewsetal20,kalogera_baym96}.

Before closing this section, a few remarks are worth.
The first applies to the matching parameter \Mco.
Figure~\ref{Mco-Miniz} compares  the \texttt{PARSEC} values of $\Mhe$ and $\Mco$ with  those derived from CL04 models, as a function of \Mi.
We note that the values of $\Mhe$  and $\Mco$ of our \texttt{PARSEC} models are slightly larger than predicted by CL04. This is due to the fact that in \texttt{PARSEC}  we adopt a slightly more efficient core overshooting.
An implication of this difference will be discussed later (Sect.~\ref{sec_comp_ej}).

The second is that we assume that 
differences in the internal configurations between \texttt{PARSEC} models and pre-explosive models with the same $M_{\rm CO}$, do not impact the final nucleosynthesis outcome. Differences are in fact expected because the two evolutionary models adopt different mass-loss rates, mixing schemes and distribution of heavy elements. 

The latter difference is considered when computing the yields of newly produced elements by properly accounting for the initial composition. As to the differences in input physics, we note the following.
The stellar models of the CL04 grid were computed at constant mass, while our \texttt{PARSEC} tracks include mass loss by stellar winds for $\Mi \geq 14\, \Msun$. However, this difference should not affect our results because,  besides the fact that we match the models using \Mco (which somewhat alleviates the problem of the different mixing scheme),
mass loss is not so important for the progenitors of  successful CCSN with $\Mi \leq 25\, \Msun$, and especially for $\Zi \leq 0.006$. 
Powerful stellar winds affect the pre-supernova evolution of more massive stars, but in this case  the matching with CCSN explosive models is not required, as these stars fail to explode and only their wind ejecta are considered.
We conclude this section noting that, for example, differences between the new yields of massive stars obtained in this work and those of \citet[][LC18]{Limongi_etal18} for zero rotational velocity are generally less than those between LC18 models with different rotational velocities, as discussed in Sect.~\ref{sec_comp_ej}.

\label{sec_PPISN}

\subsection{Explosion ejecta of pulsational pair instability and pair instability supernovae}
\label{ssec_pisn}
Very massive stars that develop a final helium core mass in the range between $\sim 32\,\Msun$ and $\sim 64\,\Msun$ are expected to enter the domain of pulsational pair-instability supernovae (PPISN), before ending their life with a successful or failed core-collapse supernova \citep{Woosley_Heger2002y, Chen2014, yoshida2016, Woosley2017ApJ}. During the pair-instability phase, several strong pulses may eject a significant fraction of the star's residual envelope and, possibly, a small fraction of the core mass. In contrast, the thermonuclear ignition of oxygen in stars with helium core masses between $\mathrm{\sim 64\Msun}$ and $\sim 135\,\Msun$ leads to a pair-instability supernova (PISN), assimilated to a single strong pulse that disrupts the entire star, leaving no remnant behind \citep{Heger_Woosley2002y, hegerfryer2003}. 

 PISN have been usually associated to the first, extremely metal-poor stellar generations \citep[e.g.][]{Karlsson_2008}. However, recent stellar evolution models suggest that PISN could occur also for stars with initial metallicity $\Zi \approx \Zsun/3$, which implies that they are potentially observable even in the local universe \citep{Yusof2013, Kozyreva2014a}. For these reasons, PPISN and PISN may play a key role to understand the chemical evolution of the Galaxy. 

While the physical mechanisms behind PPISN and PISN are quite well understood, severe uncertainties affect the range of helium (or, equivalently, carbon-oxygen) core masses that drive stars to enter the pair-instability regime \citep[e.g.][]{Leung2019,farmer2019,marchant2020,renzo2020,costa2020}.
In this work we adopt the indications from \citet{Woosley2017ApJ}, who suggests $32 \lesssim \Mhe/\Msun \lesssim 64$ for PPISN and $64 \lesssim \Mhe/\Msun \lesssim 135$ for PISN.

We model PPISN as a super-wind phase that ejects the surface layers,  without any appreciable synthesis of new elements, until the star collapses to a BH. For each \texttt{PARSEC} model with a given helium core mass,  the corresponding $\Mrem$ is obtained by interpolation in \Mhe, between the values tabulated by \citet{Woosley2017ApJ}. Then, the total PPISN ejecta are  estimated by integrating in mass the \texttt{PARSEC} structures from  \Mrem\ to \Mfin. 
Finally, we add the \texttt{PARSEC} wind ejecta.

For PISN, we calculate the explosion ejecta from the zero-metallicity pure-helium stellar models provided by \cite{Heger_Woosley2002y}, as a function of the mass $M_\mathrm{He-star}$. Similarly to the case of PPISN, the helium core mass, \Mhe, of our \texttt{PARSEC} tracks is taken as interpolating variable to perform the match with the explosion models and derive the ejecta. Finally, we add the  \texttt{PARSEC} 
contributions of all layers from $\Mhe$ to $\Mfin$ and the wind ejecta.

We use the  models of \cite{Heger_Woosley2002y}, computed at $\Zi=0$, to derive the PISN ejecta for VMO with $\Zi>0$. This assumption is reasonable since 
the total ejected mass of metals\footnote{According to standard terminology, metals refer to chemical species heavier than helium.}, $M_\mathrm{metals}$, comprises most of the PISN ejecta, with a fractional contribution, $M_{\mathrm{metals}}/M_{\mathrm{He-star}}$, that is generally larger than $97\%$ for all tabulated models. 

\begin{figure*}
\centering
\resizebox{\hsize}{!}{
\includegraphics[angle=0,width=\hsize]{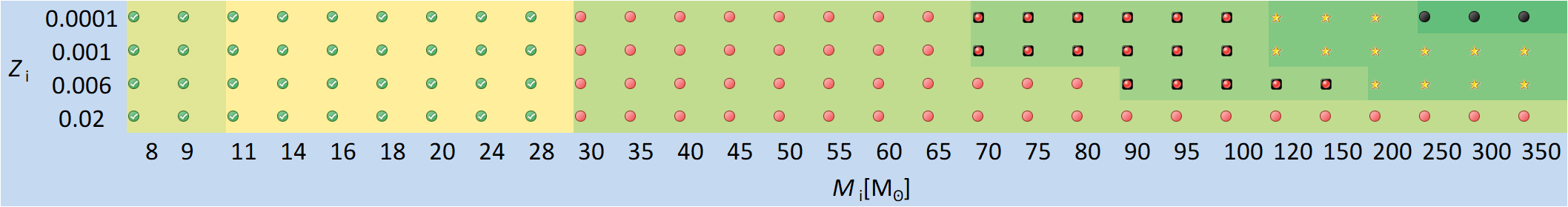}
}
\caption{Final fate of massive and very massive stars as a function of \Mi\ and \Zi. Green dot is a successful SN, from ECSN if the background is light green or CCSN if it is yellow; red dot is a BH from a failed CCSN; red dot in a black box is a BH from PPISN; yellow star is a thermonuclear explosion from PISN and black dot is a DBH.} \label{fateSN}
\end{figure*}

It is worth noticing that stars with $\Mi \geq \ 90\, \Msun$ enter the PPISN and PISN regimes already at $\Zi = 0.006$ (see Fig.~\ref{fateSN}). Our models agree with earlier theoretical findings \citep[e.g.,][]{Kozyreva2014a} and support the hypothesis that some superluminous supernovae recently observed at metallicity $\approx \Zsun/3$, may be explained through the pair-instability mechanism, provided the IMF extends to VMOs \citep{Woosley07,Kozyreva2014b}.
In this respect, we note that, according to IGIMF analysis by \citet{jerabkova2018},  galaxies with metallicity $\feh  < 0$ dex and $\mathrm{SFR > 1}$ $\Msun/{\rm yr}$ are characterized by a top-heavy and  bottom-light IGIMF, as compared to the canonical one.  Using the IGIMF grid provided by \citet{jerabkova2018} and assuming, for example, a  metallicity \feh \ =~-1 dex and a SFR $\sim$2~Msun/yr, the slope in the high mass tail would be  \rm{x}~$\sim$~-2.1. Extrapolating this slope up to 200 \Msun \ instead of 150 \Msun , the value adopted in the grid, this will allow about 66700 stars with 10~\Msun  $\le$ \Mi \ $\leq$120~\Msun and 2000 stars with 120~\Msun  $\le$ \Mi \ $\leq$200~\Msun , to be born every 3 Myr (the lifetime of the most massive stars).

\subsection{Ejecta of very massive stars that directly collapse to  black holes}
\label{sec_dcl}

If a star is massive enough to build a helium  core with $\Mhe > 135\, \Msun$, no material will be able to avoid the direct  collapse into a black hole (DBH), induced by the pair creation instability. Under these conditions no explosive ejecta  
are produced \citep{Fryer2001,Heger_Woosley2002y,Nomoto2013}, and the only chemical contribution comes from wind ejecta.
With the adopted mass-loss rates in \texttt{PARSEC}, these objects appear only at a low  metallicity, $\Zi = 0.0001$, and initial masses $\Mi > 200\, \Msun$.
Conversely, at larger metallicities stars with $\Mi > 200\, \Msun$ avoid the DBH channel, since mass loss is efficient  enough to drive their He-core masses into the regimes of PISN or failed CCSN.

\section{Ejecta of AGB stars}
\label{agb_yields}
We have complemented the ejecta of massive stars with those of AGB stars computed with the \texttt{COLIBRI} code \citep{Marigo2013}, in the mass range $0.7\lesssim \Mi/\Msun \lesssim 6$  and for 
the same values of the  \texttt{PARSEC} metallicities, already
mentioned in Sect.~\ref{ssec_parsecmod}. These models follow the whole thermally pulsing phase, TP-AGB, up to the ejection of the entire envelope by stellar winds. The initial conditions are taken from the \texttt{PARSEC} grid of stellar models at the first thermal pulse  or at an earlier stage on the Early-AGB. \texttt{COLIBRI} and \texttt{PARSEC} share the same input physics (e.g., opacity, equation of state, nuclear reaction rates, mixing-length parameter) and the numerical treatment to solve the structures of the atmosphere and the convective envelope.  For these reasons, the \texttt{PARSEC}+\texttt{COLIBRI}  combination provides a dense, homogeneous and complete grid of models for low- and intermediate-mass stars (roughly $\simeq 70$ values of \Mi\ for each metallicity value).

In \texttt{COLIBRI} models, the parameters describing the main processes that affect the TP-AGB phase, such as the mass-loss rates and the efficiency of the third dredge-up, have been thoroughly calibrated with observations of AGB stars in the Galaxy, Magellanic Clouds, and low-metallicity nearby galaxies
\citep{Girardi2010,Rosenfield_etal_14, Rosenfield_etal_16, marigoetal2017,Lebzelter2018,pastorellietal2019MNRAS,Pastorelli_etal_20, Marigo_etal_20}. The \texttt{COLIBRI} yields account for the chemical changes due to the first, second, third dredge-up episodes and hot-bottom burning in the most massive AGB stars ($\Mi \gtrsim 3-4 \, \Msun$), and include the same chemical species as in \texttt{PARSEC}, from $^{1}$H to $^{28}$Si (see Sect.~\ref{ssec_parsecmod}).

Finally, super-AGB stars are not treated explicitly here and their ejecta are taken from \cite{Ritter_etal18}, for stars with $\Mi=6,7\,\Msun$  and $\Zi=0.0001, 0.001, 0.006, 0.02$ \footnote{\url{https://github.com/NuGrid/NuPyCEE}}. The chemical composition of the ejecta is the result of third dredge-up episodes and hot-bottom burning. An overshoot scheme is applied to the borders of convective regions, including the bottom of the pulse-driven convection zone. As a consequence, the intershell composition is enriched with primary $^{16}$O ($\approx 15 \%$) in \cite{Ritter_etal18} computations, much more than in standard models without overshoot ($^{16}\mathrm{O} \approx 1-2 \%$), like in \citet{Karakas_10}.

\begin{table*}
\caption{Sets of chemical ejecta adopted in the chemical evolution models}
\label{yields_comp}
\centering
\begin{tabular}{|c|c|cc|c|c|}
\hline
\hline
\multicolumn{1}{|c|}{\textbf{label}} &
\multicolumn{1}{|c|}{\textbf{AGB stars}} &
\multicolumn{1}{|c}{\textbf{massive stars}} &
\multicolumn{1}{c|}{\textbf{rotation}} &
\multicolumn{1}{|c|}{\textbf{PPISN/PISN/DBH}} &
\multicolumn{1}{|c|}{\textbf{colour/line}} \\
\hline
MTW& M20 & TW & No & Yes & \color{darkyellow}{\textbf{yellow}}\\
\hline
KTW&  K10 & TW & No  & No & \color{blue}{\textbf{blue (dashed)}}\\
\hline
R$_{\rm r}$&  R18  & R18$_{\rm r}$  & No & No&\color{darkgreen}{\textbf{green (continous)}} \\
\hline
R$_{\rm d}$&  R18  & R18$_{\rm d}$  & No & No&\color{darkgreen}{\textbf{green (dashed)} } \\
\hline
MLr& M20 & LC18 & Yes  & No&\color{cyan}{\textbf{cyan}}\\
\hline
\end{tabular}
\end{table*}

\section{Chemical ejecta from other authors}
\label{sec_comp_ej}
Here we present various combinations of chemical ejecta taken from the literature and compare the main trends as a function of \Mi\ and \Zi.  They are summarized in Table~\ref{yields_comp}.
The different sets of ejecta are then incorporated in our chemical evolution model of the Milky Way (see Sect.~\ref{ssec_chemevol}).

As to the yields from AGB stars, we consider three sets, namely: M20 (from the \texttt{COLIBRI} code, Sect.~\ref{agb_yields}), K10 \citep{Karakas_10} and R18 \citep{Ritter_etal18}.
K10 provides the ejecta of AGB stars in the mass range $1 \lesssim \Mi/\Msun \lesssim 6$ for four metallicities ($\Zi= 0.0001, 0.004, 0.008, 0.02$). To obtain the yields at $\Zi=0.001$ and $\Zi=0.006$, we interpolate in metallicity  between their original tables.
R18  provide the ejecta of AGB and Super-AGB stars in the mass range $1 \leq \Mi/\Msun \leq 7$ for five metallicities ($\Zi= 0.0001, 0.001, 0.006, 0.01, 0.02$).

As to the yields of massive stars, including both wind and explosion contributions, we consider three sets, namely: R18 \citep{Ritter_etal18}, L18 \citep{Limongi_etal18} with and without rotation, and TW that refers to the new ejecta from this work (see Sect.~\ref{sec_parsec}).
\cite{Nomoto2013} also published yields for massive stars, which however do not include the wind contributions, and therefore we did not consider them in our analysis.

R18 computed the ejecta of massive stars in the mass range $12 \leq \Mi/\Msun \leq 25$, for the same initial metallicities as their AGB models, i.e. $\Zi = 0.0001, 0.001, 0.006, 0.01, 0.02$, and for two models of explosion conditions, rapid (R$_ {\rm r}$) or delayed (R$_{\rm d}$), respectively \citep{Fryer2006}. 

LC18 calculated the ejecta in the mass range $13 \leq \Mi/\Msun \leq 120$ for three different rotational velocities ($V_{\rm rot} = 0, 150, 300$~km/s),  and four metallicities ($\feh = 0, -1, -2, -3$ dex). Here we use the version of their ejecta for $\Zi = 0.0001, 0.004, 0.008, 0.02$ publicly available on Github  NuPyCEE repository\footnote{\url{https://github.com/NuGrid/NuPyCEE}}. 
Both R18 and LC18 sets have the noticeable property that wind and explosion ejecta derive from homogeneous stellar evolution models.

The distinguishing feature of our TW ejecta is that they range in mass beyond the classical limit of $\Mi \simeq 100\,\Msun$, extending up to $\Mi = 350\, \Msun$, hence opening the possibility to investigate the chemical role of  VMO in terms of stellar winds, PPISN and PISN explosions, and DBH channel.


To make a meaningful comparison among the different sets of  ejecta we opt to use a dimensionless quantity, defined as the ratio between the newly produced yield of a given species \textit{j}  and the stellar initial mass, $P_j(\Mi)$:
\begin{equation}
P_j(\Mi) = \left[E_j(\Mi) - (\Mi - \Mrem)\,X_{j,0}\right]/\Mi
\end{equation}
where the total ejecta $E_{j}(\Mi)$ is defined by Eq.(\ref{eijtot}), and $X_{j,0}$ is the initial stellar abundance (in mass fraction) of the element \textit{j}.

The different sets are compared in Figs.~\ref{ejecta-comp0001} - \ref{ejecta-comp02}  as a function of \Mi\ and a few values of the initial metallicity, $\Zi = 0.0001, 0.001, 0.006, 0.02$ respectively. We note that, for comparison purposes only, the LC18 yields for $\Zi=0.006$ are obtained through a metallicity interpolation.
The various panels show the ejecta of  $^4$He, $^{12}$C, $^{14}$N, $^{16}$O, $^{20}$Ne, $^{24}$Mg, $^{28}$Si, S, Ar, Ca, Ti and Fe.

The AGB ejecta of $^4$He, $^{12}$C, $^{14}$N exhibit  significant differences among different sets.
At $\Zi=0.0001$ the K10 ejecta are much larger than those of R18 and M20. In general, the production of $^{12}$C and $^{14}$N predicted by K10 is much higher than M20, reaching a factor of ten for $^{14}$N at  $\Zi=0.006$.
At increasing metallicity the differences become less pronounced. At  $\Zi=0.02$ the trend reverses with K10 predicting the lowest yields, but for the most massive AGB stars with hot-bottom burning.
These discrepancies are mainly the result of different input physics (e.g., molecular opacities, mixing length parameter),  mass-loss prescriptions, as well as differences in the efficiency of the third dredge-up.

In the domain of massive stars, our TW ejecta  agree fairly well with non-rotating LC18 at both $\Zi = 0.0001$ and $\Zi = 0.006$, while  the comparison slightly worsens at $\Zi = 0.02$, likely because  the effect of mass loss becomes important at higher metallicity. We note that our TW set produces slightly larger fractions of $^{16}$O, $^{20}$Ne and $^{24}$Mg than non-rotating LC18 at any \Zi.

At $\Zi = 0.0001$, the rotating LC18 models yield a much larger fraction of $^{14}$N and, to a much less extent,  $^{16}$O and  $^{12}$C, compared to the non-rotating set. This trend remains at increasing metallicity ($\Zi = 0.006$ and $\Zi = 0.02$) but the over-production of $^{14}$N appears less pronounced. 
Conversely, species such as $^{20}$Ne and $^{24}$Mg are  produced less by stars with rotation. 

The comparison between TW and R18 shows that, at $\Zi = 0.0001$, there is a fairly good agreement for  $^{14}$N, $^{16}$O, $^{28}$Si, S, Ar and Ca. At the same time, our TW ejecta  produce less $^{12}$C,  more $^{20}$Ne and $^{24}$Mg than R18.
At $\Zi = 0.006$ the TW predictions for $^{12}$C, which agree well with non-rotating LC18, are about twice the R18 ejecta. 
We note that at this metallicity  R18 presents a notable Fe production, higher by more than a factor of three compared to TW and LC18.  A less pronounced, but still large Fe yield is predicted by R18 also at lower metallicity ($\Zi = 0.0001$). 
Possible consequences of such Fe over-production will be discussed later in the sections devoted to chemical evolution models.

We note that none of the  stellar models in the LC18 grid reaches \Mhe\ high enough to enter the pair-instability regime.
Recalling that stars with  $\Mi > 30\, \Msun$ fail to explode and collapse to a BH, it follows that in the mass range, $30 \leq \Mi/\Msun \leq 120$,  LC18 ejecta are only due to stellar winds and become null for $\Mi > 120\,\Msun$.  
Since the maximum mass in the R18 grid is \Mi \= 25 \Msun , beyond this limit all R18 ejecta are zero. It follows that we can analyse the total ejecta of VMO, $\Mi \gtrsim 100\, \Msun$, by only referring to our MTW set. 

At $\Zi = 0.02$,  the most important contributions from VMO correspond to $^4$He, $^{12}$C and $^{14}$N. For all the other elements there is no significant production because, due to the large mass-loss rates at this metallicity, our models do not enter the PISN channel.
The large production of $^{14}$N is due to the large convective cores of the high mass stars, that mix the CNO products into the external regions, subsequently exposed to the ISM by mass-loss. We note that, for \Mi $\le 30 \ \Msun$,  the $^{14}$N production in our models is only slightly larger than that of  LC18 models with zero rotational velocity and so significantly less than that of their models with high rotational velocities. However at \Mi $\gtrsim 30\Msun$, our models have a $^{14}$N wind production that is almost identical to that of LC18 models with high  rotational velocities. Since the latter is about 60 \% larger than that of their zero rotation models at the same \Mi \ , we identify the reason for this difference in the larger mass-loss rates and consequent more rapid pealing and faster core mass decrease in the LC18 models.
At $\Zi = 0.006$ VMO are expected to eject appreciable amounts of newly produced $^4$He and $^{12}$C, while the yield of $^{14}$N decreases considerably. At the same time, other species provide notable contributions, such as
$^{16}$O, $^{24}$Mg, $^{28}$Si, S, Ar, Ca, Ti and Fe. The yields of these nuclides increase at higher \Mi. 
This is a clear effect of the occurrence of PPISN and PISN  which is favoured  at lower metallicities. At $\Zi = 0.001$ and $\Zi = 0.0001$ the ejecta of all models with $\Mi > 100\, \Msun$  have typical signatures of these explosive events. We note that at these metallicities the yields of $^{56}$Fe  may reach extremely high values, up to $20\, \Msun$ and $40\, \Msun$, respectively.
\begin{figure*} 
\centering
\resizebox{0.85\hsize}{!}{\includegraphics[angle=0]{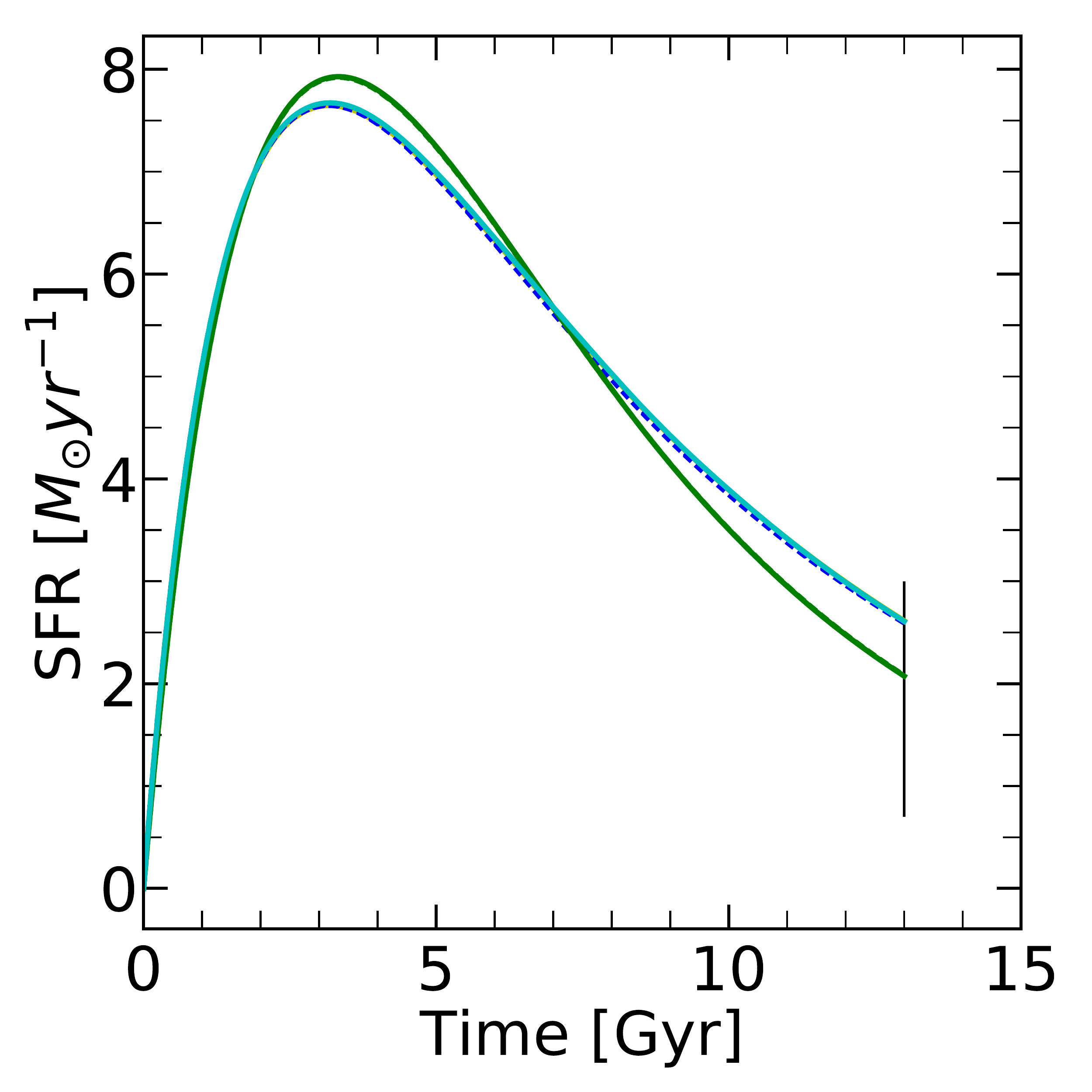},\includegraphics[angle=0]{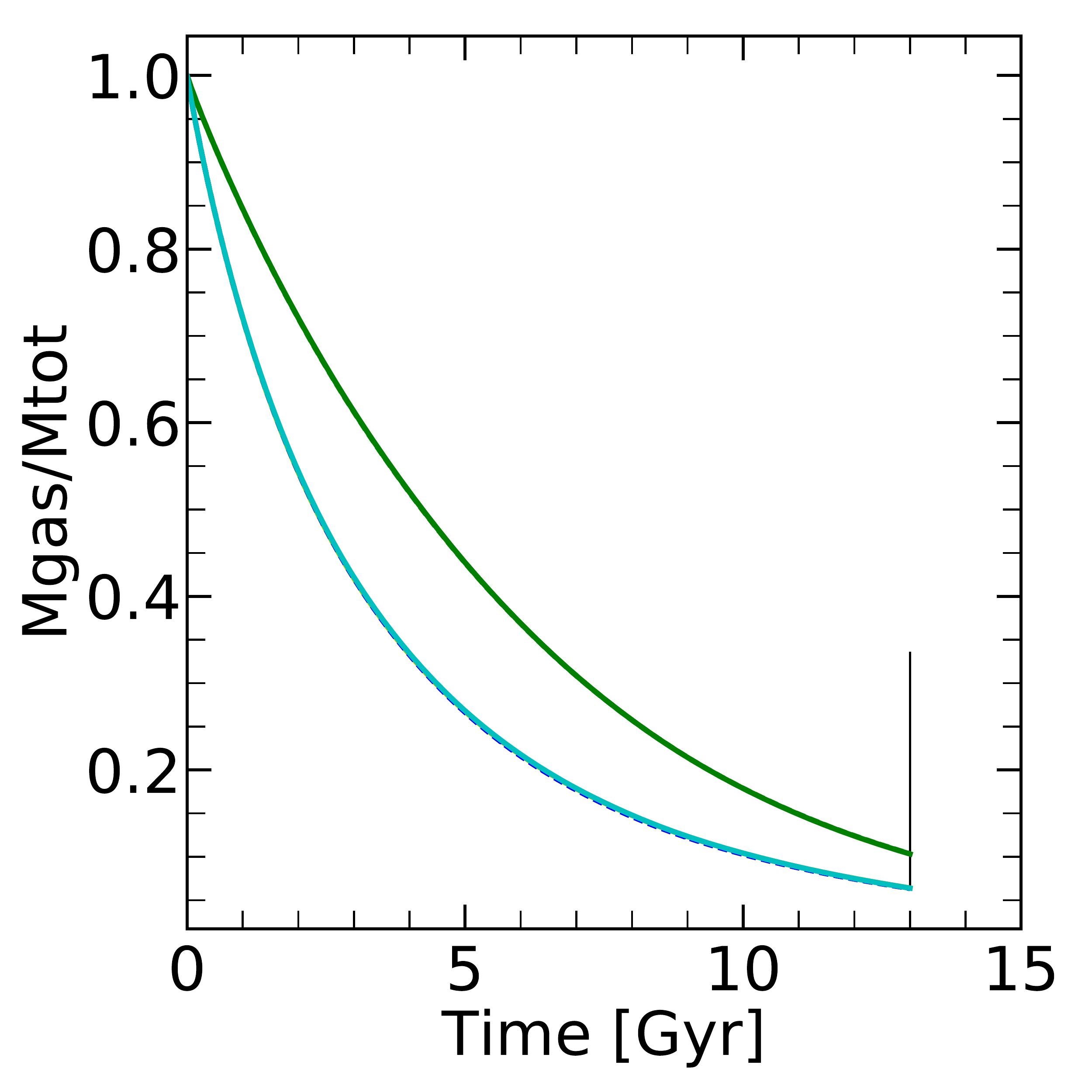}}
\resizebox{0.85\hsize}{!}{\includegraphics[angle=0]{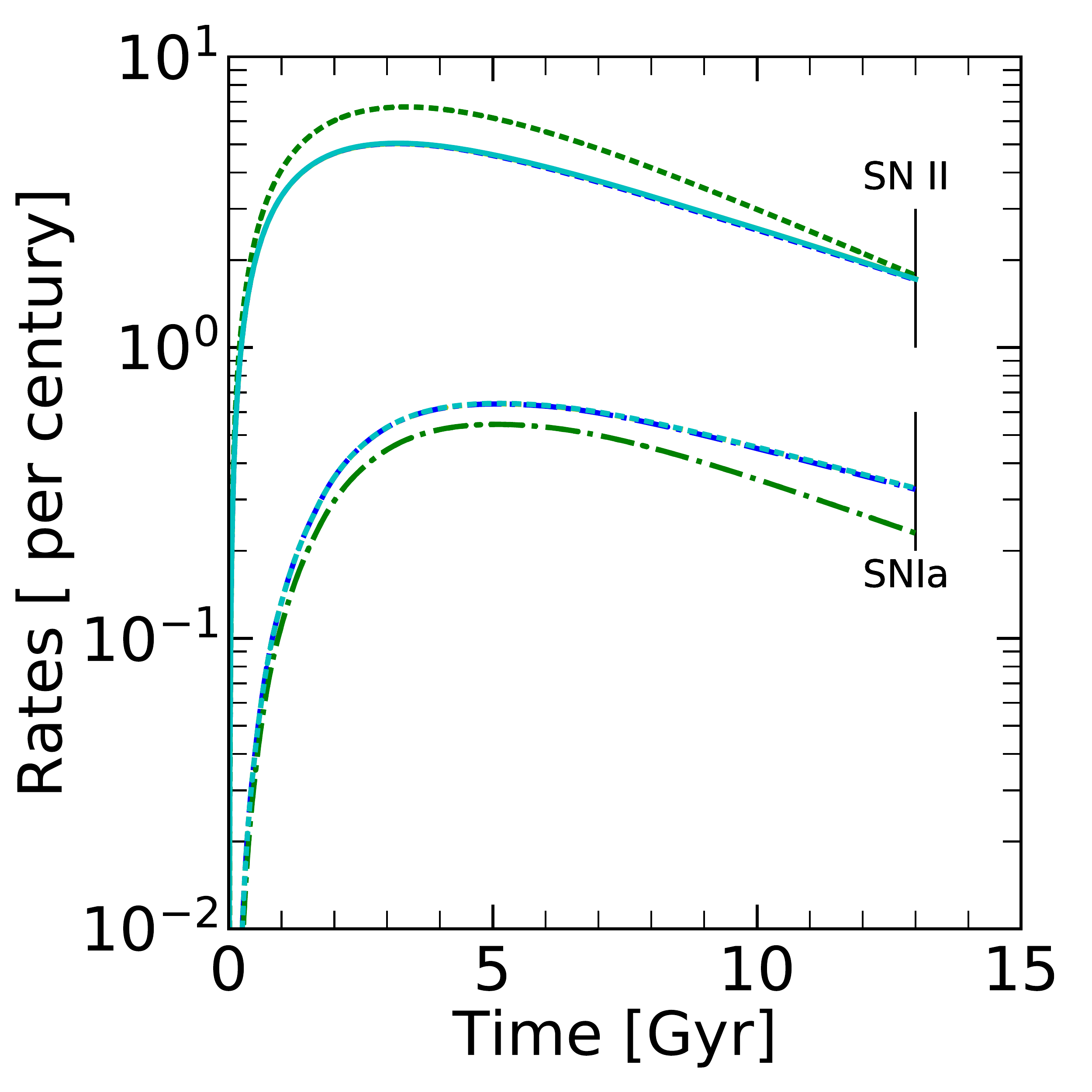},
\includegraphics[angle=0]{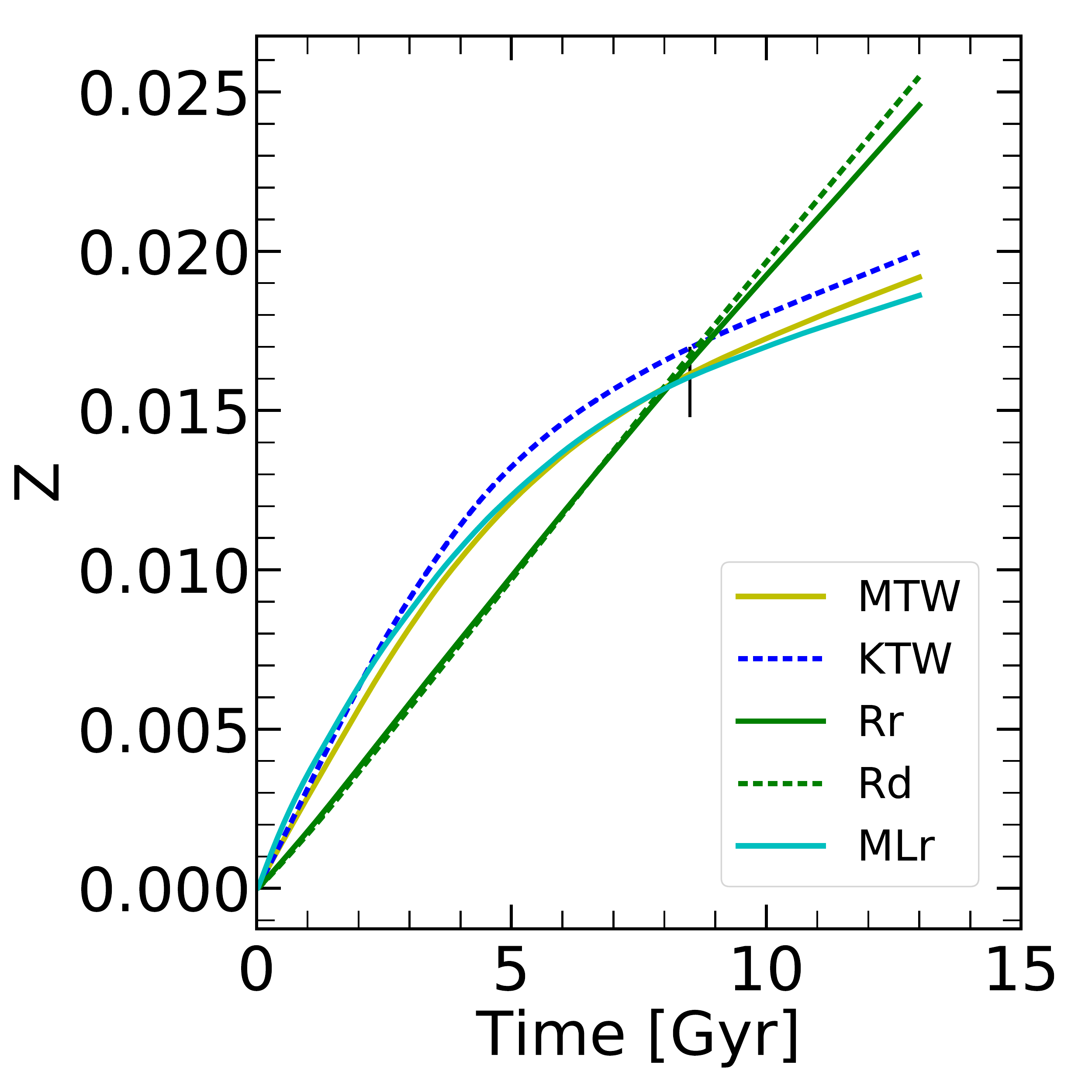}}
\caption{From top-left to bottom-right: evolution of SFR, gas fraction, SNII and SNIa rates, and total gas metallicity.
Different lines correspond to the best chemical evolution models obtained with different sets of yields, as indicated. Vertical bars at an age of 13 Gyr show their present-day estimated values. In the bottom-right panel the solid triangle at the age of 8.4 Gyr marks the protosolar metallicity, $Z_{\rm protoSUN} = 0.017$, resulting from the \texttt{PARSEC} calibration. The  parameters of the chemical evolution models are listed in Table~\ref{parameter}.}
\label{obs-TW}
\end{figure*}

\begin{figure} 
\centering
\resizebox{0.9\hsize}{!}{\includegraphics[angle=0]{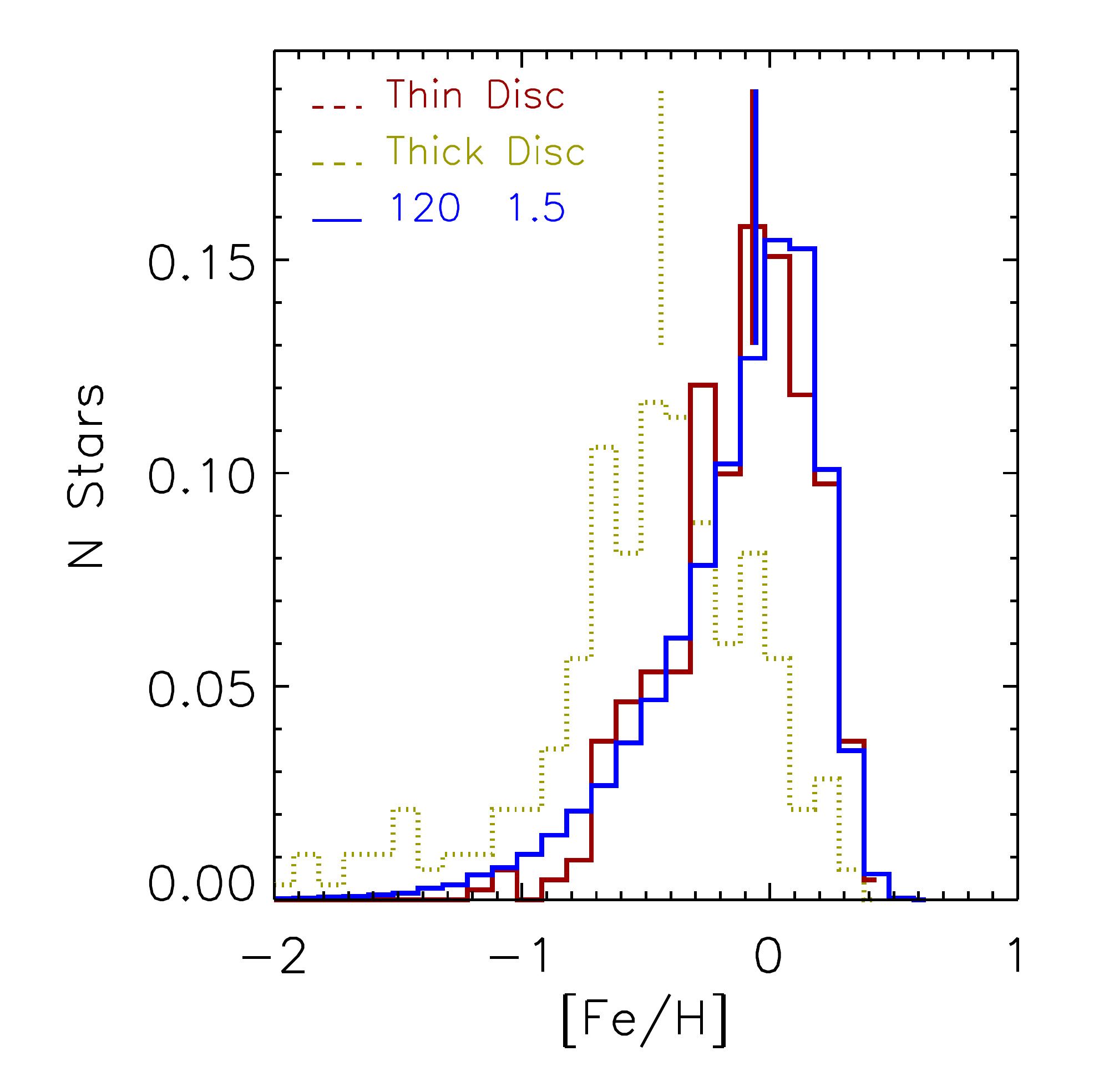}}
\caption{Observed \feh\ distributions of thin-disc stars (solid brown) and thick-disc stars (dotted yellow). Predicted thin-dick MDF, obtained with the MTW set (solid blue), are superimposed. All histograms are self-normalized. Vertical lines mark the median values of the distributions.}
\label{hist_thin_MTW}
\end{figure}

\section{Chemical evolution of the Milky Way: thin and thick discs}
\label{sec_chem_mod}
We aim to investigate how the adoption of different chemical ejecta affect our interpretation of the observed abundances in  Milky Way (MW) thin- and thick-disc stars.
\subsection{The observed abundance data}
\label{data}
The large amount of data collected over the years for stars in the solar vicinity led to the definition of different Galactic components, namely: the thin and thick discs, the halo and the  \textalpha-enhanced metal-rich population \citep{AllendePrieto2008,Gilmore2012,Zucker2012,Laverny2013}.
The populations of the MW disc can be distinguished in various ways. For example, by adopting kinematic parameters, \cite{Juric_etal08} pointed out that the stellar number density distribution of the MW could be well reproduced with two components with different scale heights above the Galactic plane: the thick disc with a scale height of $\mathrm{\simeq 900\,pc}$, and the thin disc with a scale height $\mathrm{\simeq 300\,pc}$.

At the same time, chemical abundances  reveal the existence of clearly separate sequences of \textalpha-elements as a function of \feh,  with
 thick-disc stars generally belonging to a high \afe\ (\textalpha-enhanced) sequence, while thin-disc stars exhibiting a lower \afe\ ratio at the same \feh\  \citep[e.g.][]{plevneetal20,Grisoni_2017, Chiappini_etal16,Bekki_etal10,Feltzing_2003,Prochaska_2000}.

The classification based on kinematical properties
and the one based on abundance measurements provide somewhat different results so that it is not clear which is the best way to group these stars  \citep{Navaro_etal13}. In this respect,
it has often been pointed out that chemical evolution leaves a persistent imprint that hardly changes,  while kinematic properties are more likely to vary as they may be affected by dynamical interactions \citep{Schonrich_2017,Ciro_2016}. 

In recent years, ages of individual stars have been measured with sufficient accuracy to be used as robust population indicators, like in the case of star clusters \citep{Fuhrman10}. 
We emphasize that age cannot be taken as a proxy for  metallicity, rather it is a complementary independent  parameter that concurs to define the full population box, i.e. the distribution of stars in age and abundances which, together with spatial and kinematic parameters, gives the information necessary to reconstruct the star formation history in a galaxy.

In this work, we  use the homogeneous set of data of disc stars provided by \cite{Bensbyetal14}, who conducted a high-resolution and high signal-to-noise spectroscopic analysis of 714 F and G dwarf and subgiant stars in the solar neighbourhood.
This study is particularly suited for our purpose because, based on the analysis of the their kinematical properties by \citet{Casa_etal11}, each star in the sample is 
assigned a relative membership probability, $TD/D$  defined as the ratio between the thick-disc  and thin-disc probability, and $TD/H$ defined as the ratio between the thick-disc and halo probability.

\cite{Bensbyetal14} classified stars with $TD/D > 2$ -- having the probability of belonging to the thick disc  of least twice that of belonging to the thin disc --, as potential thick-disc stars, while those with  $TD/D < 0.5$ as potential thin-disc stars. Some of the thick disc stars
where then assigned to the halo population following the  $TD/H < 0.5$  kinematical criterion.

Adopting the same kinematical criteria, we count 387 thin disc stars, 203 thick disc stars and 36 halo stars. We discard  88 stars with $0.5 < TD/D < 2$. 
The existence of at least two distinct disc sequences is clearly visible in the abundance patterns that define the so-called \textalpha-enhancement, as illustrated  in the \ofe,\ \mgfe,\ \sife\ and \cafe\ vs. \feh\ diagrams of Fig.~ \ref{met-TW}.
Interestingly, the \ofe\ vs. \feh\ diagram clearly shows not only that the two disc populations draw  separate sequences, but also that the slopes of two branches are different at increasing \feh.
Interpreting the different slopes with chemical evolution models is quite challenging,  as already noted by \cite{Kubryk_etal15}.


\subsection{Chemical evolution models}
\label{ssec_chemevol}
To analyze the evolution of the thin and thick disc populations we use the chemical evolution model \texttt{CHE-EVO} \citep{Silva1998}, a one-zone open chemical evolution code, that follows the time evolution of the gas abundances of the elements, including infall of primordial gas. It has been used in several contexts to provide the input star formation and metallicity histories to interpret the spectro-photometric evolution of both normal and starburst galaxies \citep[e.g.][]{Vega2008,Silva2011,fontanot2009,lofaro2013,lofaro2015,hunt2019}.
The equation describing the evolution of the gas masses reads:
\begin{equation}
\dot{M}_{{\rm g},j}= \dot{M}_{{\rm g},j}^{\rm SF}  +  \dot{M}_{{\rm g},j}^{\rm FB} + \dot{M}_{{\rm g},j}^{\rm Inf}
\label{eq_chem_ev}
\end{equation}
where, for the  specie \textit{j},    $\dot{M}_{{\rm g},j}^{\rm SF}$ represents the rate of gas consumption by star formation, $\dot{M}_{{\rm g},j}^{\rm FB}$ is the rate of gas return to ISM by dying stars and $\dot{M}_{{\rm g},j}^{\rm Inf}$ refers to the infall rate of pristine material.
For the star formation rate (SFR) we adopt a modified Kennicutt-Schmidt law \citep{Schmidt1959,Kenni1998}:
\begin{equation}
\psi(t)= \nu \,  M_{\rm g}(t)^{k} 
\label{eq_SF_law}
\end{equation}
where $\nu$  is the efficiency of star formation, $M_{\rm g}$ is the mass of the gas and $k$ is the exponent of the star formation law. 
The quantity $\dot{M}_{{\rm g},j}^{\rm FB}$ is calculated from the yields tables described in Sect.~\ref{sec_comp_ej}, integrating the contributions of dying stars at any time-step.
The gas infall is assumed exponential, with an e-folding time scale $\tau_{\rm inf}$ and a chemical composition equal to the primordial one \citep[e.g.][]{Grisoni_2017}.

Type Ia supernovae are also taken into account according to the single degenerate scenario and computed following the  standard formalism first introduced by \cite{Matteucci1986}.
The contribution of these sources to the chemical enrichment is regulated by the parameter $A_{\rm SNIa}$ which sets the fraction of the number of binary systems with total mass in the 3\,\Msun-16\,\Msun range, effectively contributing to the SNIa rate.
Our adopted ejecta for SNIa are taken from \cite{1999ApJS..125..439I}. As to AGB, massive, and very massive stars, we adopt the ejecta described in Sects.~\ref{sec_parsec}-\ref{sec_comp_ej}, and reported in Table \ref{yields_comp}.
In addition to the evolution of elemental gas masses, metallicity and gas fraction, the code provides also  the evolution of  SNII and SNIa rates, and total mass in stars ($M_{\rm s}$).

A fundamental assumption of chemical evolution models is the IMF:
\begin{equation}
\phi(\Mi) =  \frac{dn}{d\log(\Mi)} \propto \Mi^{-x}.
\end{equation}

We use a Kroupa-like three-slope power law IMF with $x=0.3$ for $0.1 \le \Mi/\Msun \le\, 0.5$, $x=1.3$ for $0.5 \le \Mi/\Msun \le\, 1$,
while we vary the slope for $\Mi > 1\, \Msun$,  as well as the upper mass limit of the IMF, $\MUP$, to search for best fitting models with different ejecta combinations.
We consider IMF slopes between $x = 1.7$ \citep{1993MNRAS.262..545K} and  $x = 1.3$ \citep{Kroupa01,Chabrier2003}.

\subsection{Previous Analyses of the MW thin and thick discs}
There have been in the past and in the more recent literature many attempts to explain the different chemical evolutionary paths of  different MW components, in particular of the thin and thick discs. The outcome of these studies is that the observed different chemical evolutionary paths are related to  differences in the main physical processes that drive galaxy evolution, among which the most significant are the gas accretion  time-scale and the star formation efficiency and, possibly, radial migration \citep{Larson1972, Lynden-Bell1975,Pagel1981ARA&A,Matteucci1986, Matte1990,ferrini1994,Prantzos1995A&A,Chiappini1997,Portinari1999, Chiappini_01,Bekki_etal10, Micali_etal2013,Sahijpal2014,Snaith2014,Grisoni_2017,Grand2018MNRAS,Spitoni2021}. 
Good agreement between observations and theoretical predictions for the Galaxy is obtained by models assuming
that the disc formed by infall of gas \citep{Chiosi1980,Matteucci1989,Chiappini1997}. 
The formation of the different components is associated to distinct sequential main episodes of gas accretion (infall phases) that first rapidly accumulates in the central regions and then, more slowly, in the more external ones, according to the so-called  {\sl inside-out scenario} \citep{Chiappini_01}. 
In particular, the three-infall model, devised by \citet{Micali_etal2013}, is able to reproduce the abundance patterns of the MW halo, thick and thin disc at once.
In this model, 
the halo forms in a first gas infall episode of short timescale (0.2 Gyr) and mild star formation efficiency, $\nu$=2 Gyr$^{-1}$, lasting for about 0.4 Gyr. It is immediately followed by the thick disc formation, characterized by a somewhat longer infall timescale (1.2 Gyr), a larger duration of about 2 Gyr and a higher star formation efficiency, $\nu$=10 Gyr$^{-1}$. Finally, the star formation continues in the thin disc with a longer infall timescale (6 Gyr in the solar vicinity) and is still continuing nowadays, with star formation efficiency $\nu$=1 Gyr$^{-1}$. The [O/Fe] vs. [Fe/H]
path is thus continuous across the regions populated by halo, thick and thin disc stars.
While in  \citet{Micali_etal2013} the chemical enrichment is continuous across the three different infall stages,  \cite{Grisoni_2017} used also an alternative scheme where the thin and thick disc components evolve separately, in a parallel approach (see also \cite{Chiappini2009}).
In the parallel approach, the disc populations are assumed to form in parallel but to proceed at different rates. The gas infall exponentially decreases with a timescale 
that is 0.1~Gyr and 7~Gyr, for the thick and thin disc, respectively. This alternative approach better reproduces the presence of the metal-rich $\alpha$-enhanced stars in the \mgfe\ vs. \feh\ diagram obtained with the recent AMBRE data \citep{2017A&A...600A..22M}.

In our analysis, we will assume that the thin and thick disc populations evolve separately as in  the parallel model approach adopted by  \cite{Grisoni_2017}. This is clearly an oversimplification, because these stellar components occupy the same volume in the solar neighbourhood,  but nevertheless it will allow at least to check our models against individually well-separated populations. 
Alternatively, we could have considered the two populations together and tried to obtain a model that recovers a sort of average path, as done several times in the past. However, the evidence that the two populations are different is so strong that reproducing their {\sl averaged} properties is even less meaningful.

\section{Analysis of stellar abundances in the solar vicinity}
\label{sec_chem_prop}

\subsection{Model constraints for the thin disc}
\label{modelcostraints}
For each combination of ejecta in Table \ref{yields_comp}, we build a large library of chemical evolution models by varying the following parameters:
the star formation rate efficiency ($0.2 \leq \nu \leq 2.0$) 
the normalization factor for the SNIa ($0.02\leq A_{\rm Ia} \leq 0.1$), and  the infall timescale ($0.1 \leq \tau_{\rm inf} \leq 10$). For simplicity we set $k=1$ for all models.
With this choice for parameter ranges, the chemical evolution models are able to bracket a few basic observational constraints for the thin disc, namely:
\begin{itemize}
\item \textit{The current SFR of the MW.}  To a large degree it corresponds to that the thin disc, and is estimated as $\rm{SFR} = 0.65-3.0 \Msun\,{\rm yr}^{-1}$ \citep{Robi_etal10}.
\item \textit{The current gas fraction.} It is assumed to be $M_{\rm g}/(M_{\rm g} + M_{\rm s}) \sim 0.2$ \citep{Kubryk_etal15}.
\item \textit{The current SNII rate.} We set $R_{\rm SNII} = 2 \pm 1$ SNII events  per century \citep{Prantzos_etal11}.
\item \textit{The current SNIa rate.} We set $R_{\rm SNIa} = 0.4 \pm 0.2$ SNIa events per century \citep{Prantzos_18}.
\item \textit{The protosolar metallicity} $Z_{\rm protoSUN}$. 
Though there are in literature contrasting opinions concerning the significance of the differences between Sun's abundances and those of solar twin stars
\citep[e.g.][]{Bensbyetal14,Botelho2020}, we take the protosolar metallicity as representative of that of other disc stars with similar [Fe/H], and thus as a constraint for disc chemical evolution models. The protosolar metallicity represents the bulk metallicity of the molecular cloud out of which the Sun was born. It does not coincide with the current photospheric solar metallicity, $Z_{\odot} \simeq 0.0134-0.0152$, from \cite{Asplund2009} and \cite{Caffau_etal11} respectively, as a result of chemical sedimentation effects over a time of about 4.6 Gyr (the present Sun's age). Here we adopt $Z_{\rm protoSUN}=0.017$, which is the initial metallicity of the \texttt{PARSEC}   $1\, \Msun$ model that best reproduces the currently observed Sun's properties when using the \cite{Caffau_etal11} solar mixture \citep{Bressan2012}.
Assuming an age of $13\,\mathrm{Gyr}$ for the formation of the Galaxy \citep[e.g.][]{Savino2020}, it follows that the Galactic age at the birth of the proto-Sun is $t_{\rm protoSUN}=8.4\,\mathrm{Gyr}$. At this epoch, the  metallicity in the solar vicinity is $Z_{\rm protoSUN}$.
\end{itemize}
In addition to the constraints just mentioned above, we also require the models to reproduce the observed  (\feh) \textit{metallicity distribution function} (MDF) of thin-disc stars derived from the \cite{Bensbyetal14} data.

In total, we build a database of  about 1300 models for each combination
of yields. Then, after identifying the group of models that satisfy all the constraints within the uncertainty bars, for each set of yields we find the best model that fits the observed path in the \ofe\ vs. \feh\ plane \citep{Bensbyetal14}, with the aid of a $\chi$-square minimization technique.
Finally, we perform a fine tuning and group the best models so as to minimize the change of the parameters  obtained with different yield sets, without worsening the fits. This allows us to better isolate and analyze the effects produced by changing either the yields or the IMF.

The evolution of the SFR, the gas mass fraction, the SNII and SNIa  rates  and the gas metallicity ($Z$) of the selected models are shown in Fig.~\ref{obs-TW}. The predicted thin-disc MDF of our reference model, MTW, is compared with the observed one in Fig.~\ref{hist_thin_MTW}. In the same figure we also plot the observed MDF of the thick-disc MDF, for sake of comparison.
The vertical bars in the  figure mark the location of the median values of the distributions.
Let us now analyse the results obtained with the various yield combinations.
The adopted parameters of the selected models are summarized in Table \ref{parameter}.
 \begin{itemize}
\item {\textbf{MTW yields.}}
The MTW model, that adopt the yields determined in this work, is our reference model. It is shown in dark yellow in Fig.~\ref{obs-TW}.
This model reproduces all the constraints fairly well, but for the gas fraction  that approaches the observed lower limit.
The predicted MDF of the MTW model is very similar to the observed one, with a median value in very close agreement to that of the observed one. The thin disc MDF is significantly different from that of the thick disc.
The IMF used to reproduce the observed constraints has a slope of $x=1.5$ in the high-mass range \citep{1993MNRAS.262..545K},  and $\MUP = 120\, \Msun$. 
\item {\textbf{KTW yields.}}
The KTW model is shown in blue dashed in Fig.~\ref{obs-TW}.
The IMF used in this case is the same as in the MTW model.
Likewise, this model reproduces all the constraints satisfactorily, including the the observed MDF, which is not shown here for sake of clarity.
We note that the metallicity of the KTW model increases with time a bit faster than the MTW model. At ages of 2~Gyr and 8~Gyr the metallicity $Z$ in the KTW model is about 4\% and 6\% higher, respectively, than predicted with MTW. This small difference is due mainly to the larger AGB yields of  $^{12}$C and $^{14}$N in K10 than in M20.
\item {\textbf{Rr and Rd yields.}}
The Rr and Rd models are shown in green in Fig.~\ref{obs-TW}. Both provide similarly good fits as the other sets, but the model parameters are quite different from the other cases, as shown in Table~\ref{parameter}. First, the mass limit is set to $\MUP = 25\, \Msun$, in agreement with their yield tables. Second, since the Fe production by CCSN in some of these models is significantly higher than predicted by other authors,  
in order to reproduce the observed [Fe/H] distribution and $Z_{\rm protoSUN}$ at the Galaxy age, we need to decrease the SNIa efficiency factor to A$_{\rm SNIa}$=0.025 and to adopt a flatter IMF in the high-mass regime. 
\item {\textbf{MLr yields.}}
The MLr model is shown in cyan in Fig.~\ref{obs-TW}.
Also this model reproduces well the above observational constraints without the need to change the chemical evolution parameters.
\end{itemize}
In general, we conclude that with all yield combinations it is possible to find chemical evolution models able to nicely reproduce the observed constraints for thin-disc component in the solar vicinity. As to the metallicity distribution traced by \feh, models generally tend to slightly under-populate the central bins at lower metallicities, at the same time extending  a bit beyond the maximum observed \feh.
Being of opposite sign, the two deviations roughly compensate each other, allowing  a good reproduction of the observed median value. 
In principle, the agreement could be improved even more through a fine tuning of the chemical evolution parameters.
However, we note the discrepancies between the predicted and observed thin-disc MDFs are  much lower than the intrinsic differences between the thin- and thick-disc observed distributions.
Therefore, we may reasonably consider our results as fair models for the thin disc and proceed with the analysis of the predicted abundance ratios.
\begin{table}
\caption{Parameters of the best chemical evolution models for each ejecta combination, described in Table \ref{yields_comp}.}
\label{parameter}
\centering
\begin{tabular}{|c|c|c|c|c|c|c|} 
\hline\hline
\multicolumn{1}{|c|}{{Ejecta}} &
\multicolumn{4}{|c|}{{Chemical parameter}} &
\multicolumn{2}{|c|}{{IMF}}\\
\hline
\multicolumn{1}{|c|}{{SET}} &
\multicolumn{1}{|c|}{${\nu}$} &
\multicolumn{1}{|c|}{${k}$} &
\multicolumn{1}{|c|}{$\tau_{\rm inf}$} &
\multicolumn{1}{|c|}{$A_{\rm SNIa}$} &
\multicolumn{1}{|c|}{${\MUP}$}&
\multicolumn{1}{|c|}{${x}$} \\
\hline
MTW &  0.8 & 1.0 & 6.0& 0.04 &120&1.5\\  
\hline
KTW &  0.8 & 1.0 & 6.0& 0.04 &120&1.5\\   
\hline
R$_{\rm r}$&  0.4& 1.0 & 3.0& 0.025 &25&1.3\\  
R$_{\rm d}$&  0.4& 1.0 & 3.0& 0.025 &25&1.3\\  
\hline
MLr & 0.8 & 1.0 & 6.0& 0.04  &120&1.5\\   
\hline
\end{tabular}
    \begin{tablenotes}
        \footnotesize
        \item[*] The efficiency of star formation, \textnu, is in units of Gyr$^{-1}$, \\ ${\tau_{\rm inf}}$ is expressed in units of Gyr, and \MUP\ is given in \Msun.
    \end{tablenotes}
\end{table}
\begin{figure*} 
\centering
\resizebox{0.9\hsize}{!}{\includegraphics[angle=0]{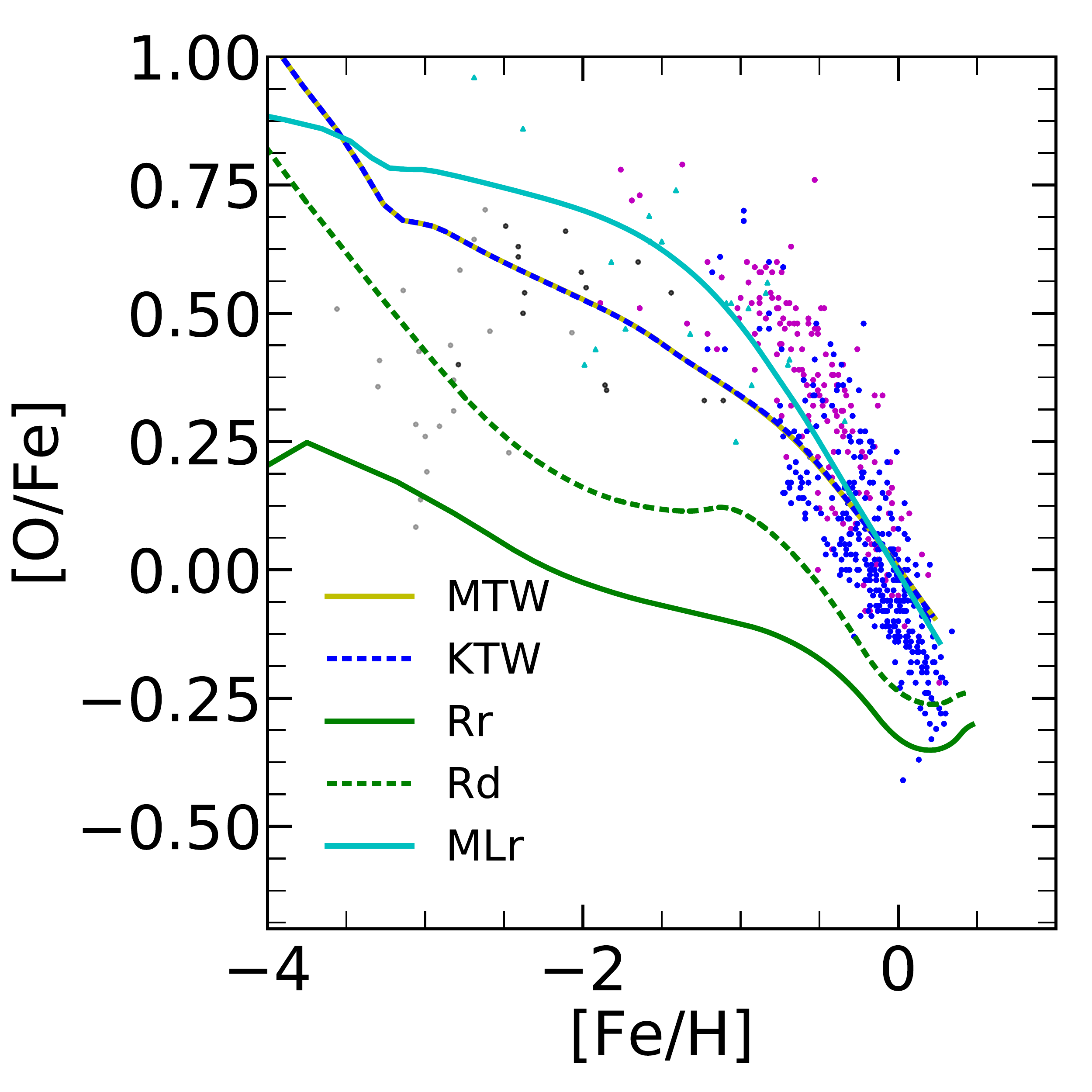},\includegraphics[angle=0]{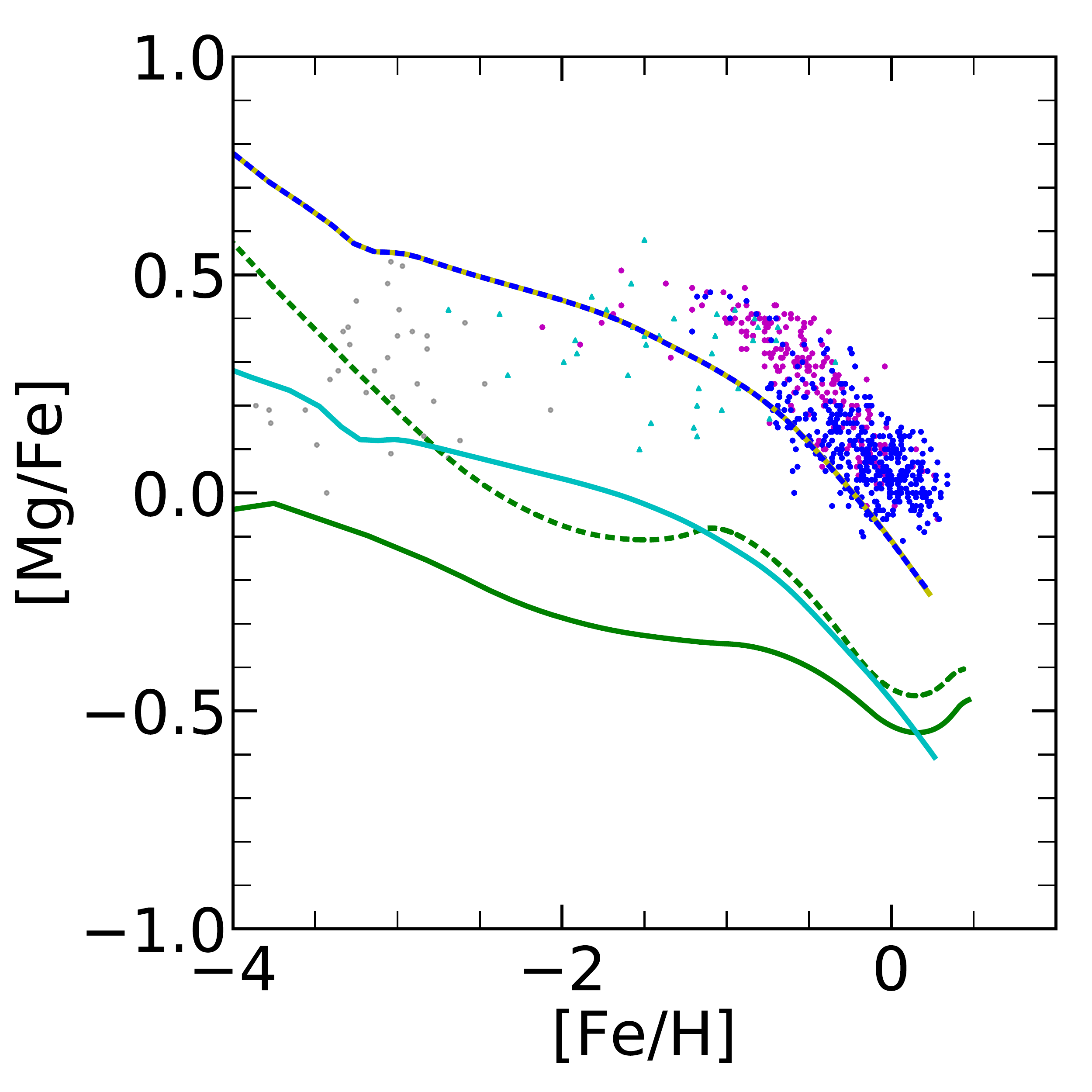}}
\resizebox{0.9\hsize}{!}{\includegraphics[angle=0]{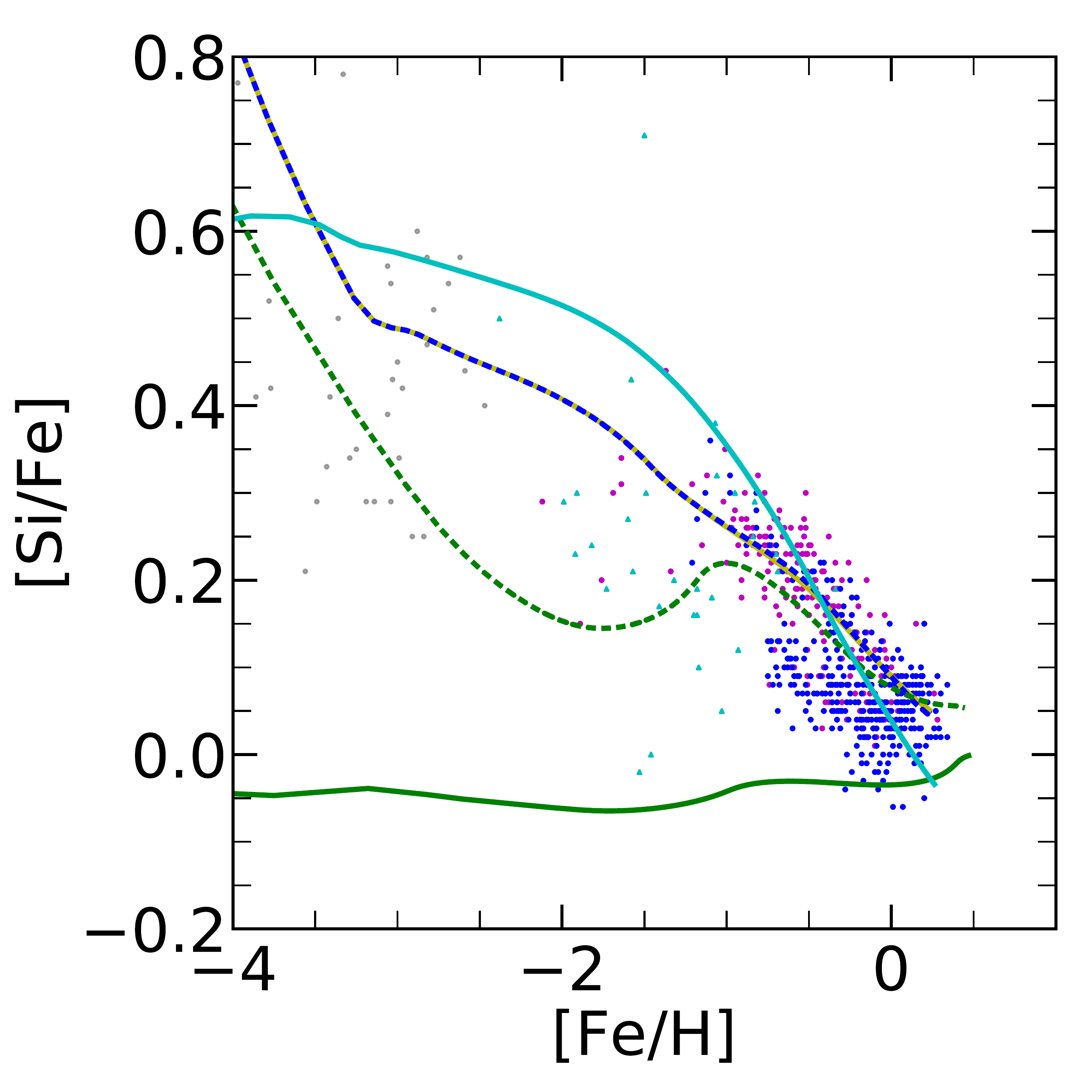},\includegraphics[angle=0]{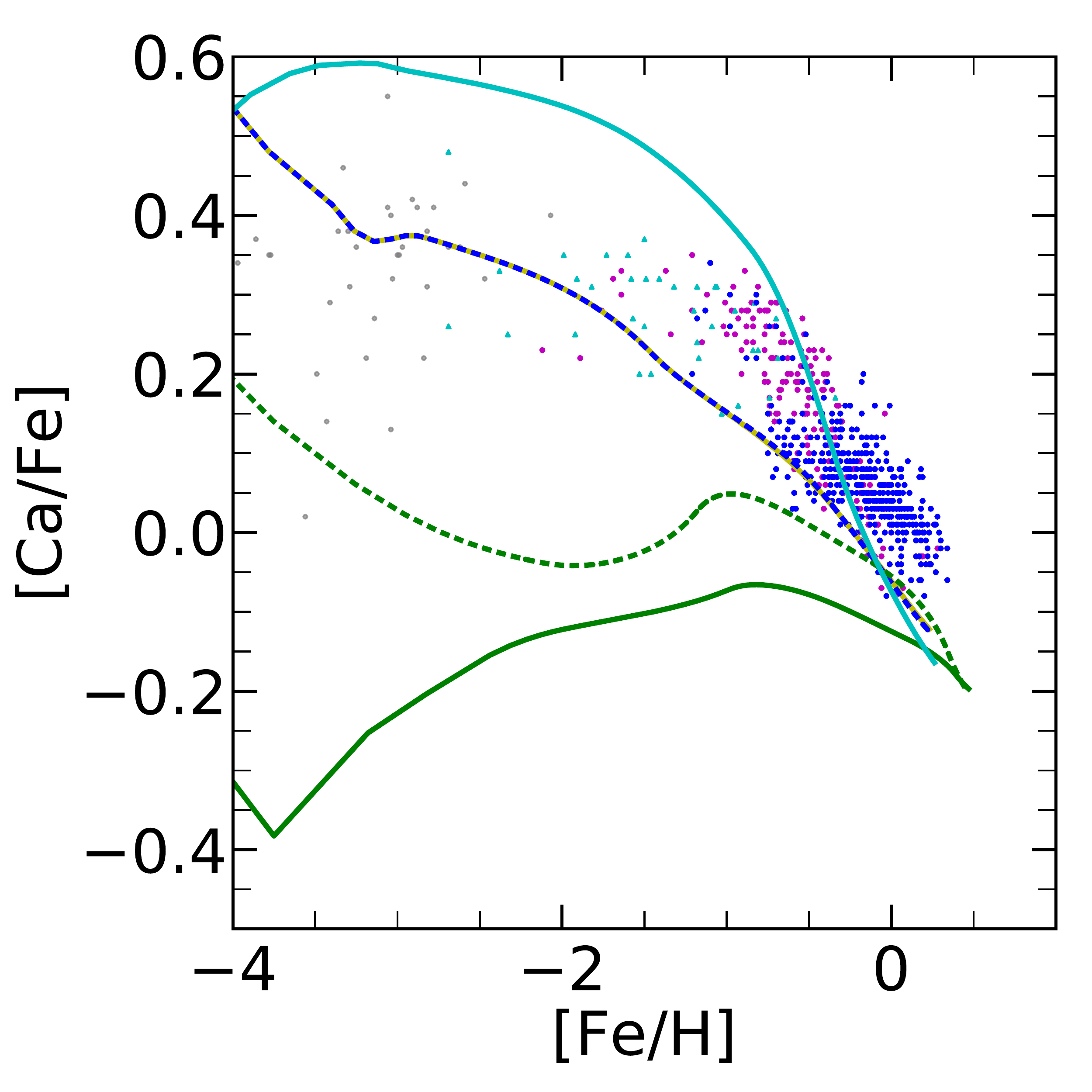}}
\caption{Comparison of predicted  abundances, derived with our  best chemical evolution models (see Fig.~\ref{obs-TW})  with observations of thin-disc stars (blue points) and thick-disc stars (magenta points) and halo stars (cyan triangles) in the solar neighbourhood from \cite{Bensbyetal14}. Plotted in the figure are also
the sample of metal-poor (black dots) and very metal-poor (gray dots) stars from \cite{Nissen2002} and \citep{Cayrel2004}, respectively. Note that the MTW and KTW models overlap almost completely.}
\label{met-TW} 
\end{figure*}
\subsection{Chemical evolution of the thin disc}
\label{predictedabundances}
For each set of chemical yields, we compare the elemental abundances predicted by the corresponding best model with observations of thin-disc, thick-disc and halo low-metallicity stars (Fig.~\ref{met-TW}).
The four diagrams show the trends of a few selected species, O, Mg, Si and Ca, as a function of \feh. 
 We recall that all the data are normalized according to the scaled-solar abundances given by \cite{Caffau_etal11}, which sets the reference solar mixture in \texttt{PARSEC}.
\begin{itemize}
\item {\textbf{MTW yields.}}
This model is able to reproduce fairly well the observed trend in the [O/Fe] diagram, with a modest over-production  at the highest \feh.
Such small discrepancy should be attributed to the chemical yields from massive stars. In fact, in the \texttt{COLIBRI} models used here (M20 yields),  TP-AGB stars  produce a negligible amount of primary oxygen, as the chemical composition of the intershell is the standard one and it contains no more than $1\%-2\%$ of $^{16}$O \citep[see, e.g. ][models without overshooting]{Herwig2000}.

Taking the MTW model as representative of the main thin-disc branch, we see that at decreasing \feh\ the difference in \ofe\ between thin and thick-disc stars increases.  At $\mathrm{\feh\leq-1}$ the thin-disc branch disappears, and the MTW model approaches the  lower boundary of the halo population. 

Following the probability $TD/D$-criterion \citep{Bensbyetal14}, we note that a second, less populated, thin-disc branch overlaps with the sequence of thick-disc stars in the same abundance diagram. Clearly, The MTW  model is not able to reproduce this secondary branch.

Concerning the \mgfe\ ratio, the MTW model runs through the lower border of the data, showing a well-known difficulty likely related with the $^{24}$Mg yields \citep{Romano2010}. The \sife\ ratio is fairly well recovered by this model, while the \cafe\ is  under-produced.
\item {\textbf{KTW yields.}}
This model behaves similarly to the MTW model for all the four abundance ratios. We recall that KTW and MTW models only differ for the AGB yields in use.
Small differences appear at the lowest \feh\ values where the increase of KTW metallicity with time  takes place somewhat faster than predicted by the MTW model (see Fig.~\ref{obs-TW} and the discussion in Sect~\ref{modelcostraints}). Since MTW and KTW models share the same chemical evolution parameters and  massive star yields, differences in the trends of chemical species should be likely ascribed to reaching somewhat different metallicities at the same evolutionary time.
\item{\textbf{Rr and Rd yields.}}
While successfully reproducing basic constraints of the MW thin disc, these models fail to recover the evolution of the selected abundance ratios (see Fig.~\ref{met-TW}). Even adopting a low SNIa efficiency parameter, models exhibit a substantial deficit in $^{16}$O, $^{24}$Mg, $^{28}$Si and Ca relative to Fe. In the attempt to solve the discrepancy we explored a wide range of chemical evolution parameters,  but we were unable to find better models than those shown in Figs.~\ref{obs-TW} and \ref{met-TW}. 
The results worsen for the rapid case and, since the ratios [\textalpha/Fe] run much flatter than observed, we argue that the issue may be linked to the iron yields of CCSN. In fact, we find that at $\Zi < 0.02$,  iron production by CCSNs predicted by R18 is significantly higher than in L18 explosive models (used in MTW, KTW and MLr), by a factor from two to four.
\item{\textbf{MLr yields.}}
The MLr model reproduces fairly well the [O/Fe] data of the thin disc.  The abundance of Mg is clearly under-predicted
as already found by \cite{Prantzos_18}, while predictions for Si and Ca are able to populate the regions of both thin- and thick-disc components. We note the MLr model shows a general tendency to produce abundance ratios running with slightly steeper slopes than observed.
\end{itemize}

From all the tests carried out with different combinations of chemical yields  we can draw a few conclusions.
First, the chemical species discussed here are marginally dependent on the chemical yields of AGB stars, and therefore cannot be considered useful diagnostics for testing the goodness of low- and intermediate-mass evolutionary models. This is not surprising since we do not expect that AGB stars synthesize Fe and Ca, while they may be contributors of Mg isotopes,  which are present both in the dredged material and involved in the Mg-Al cycle when hot-bottom burning is active in AGB stars with \Mi > 3-4 \Msun \citep{Slemer_etal_17,Marigo2013,VenturaDantona_09}.
As discussed above, the production of some primary oxygen by AGB stars depends on the inclusion of convective overshoot at the boundaries of the pulse-driven convective zone \citep{Herwig2000},
which applies to R18, but not to M20 and K10 yields. In the context of this work,  the role of AGB stars as oxygen producers is not critical irrespective of the selected yield set.
This reinforces the conclusion that we need to consider other more suitable elements, such as carbon and nitrogen, to  compare and check different sets of AGB yields. An in-depth analysis of AGB yields is postponed to a dedicated future work.

It follows that the abundance trends investigated in this work are critically dependent on the chemical yields from massive stars and VMO. Therefore, the reader should keep in mind that, even when not explicitly stated, the discussion that follows mainly deals with the effects produced by chemical yields of stars with $\mathrm{M_i>8\Msun}$. 

Once the chemical evolution models are calibrated on a few  basic observables of thin-disc stars, it is possible to reproduce fairly well the enrichment paths of \ofe\ and \sife\ with most of the yield sets. As to the \mgfe\ ratio, we meet the long-lasting problem of underproduction \citep[e.g.][]{Prantzos_18}, which appears somewhat less pronounced with the MTW yields.
As to the \cafe\,  the observations are better reproduced by massive star models with rotation. 
Using the R18  yields, none of the elemental ratios is well reproduced.The discrepancy is likely due to the high Fe yields in some of the  \cite{Ritter_etal18}  explosion models, which makes it hard to recover the observed [\textalpha/Fe] ratios at a given [Fe/H].



\begin{figure} 
\centering
\vskip-0.3truecm
\resizebox{0.99\hsize}{!}{
\includegraphics[angle=0]{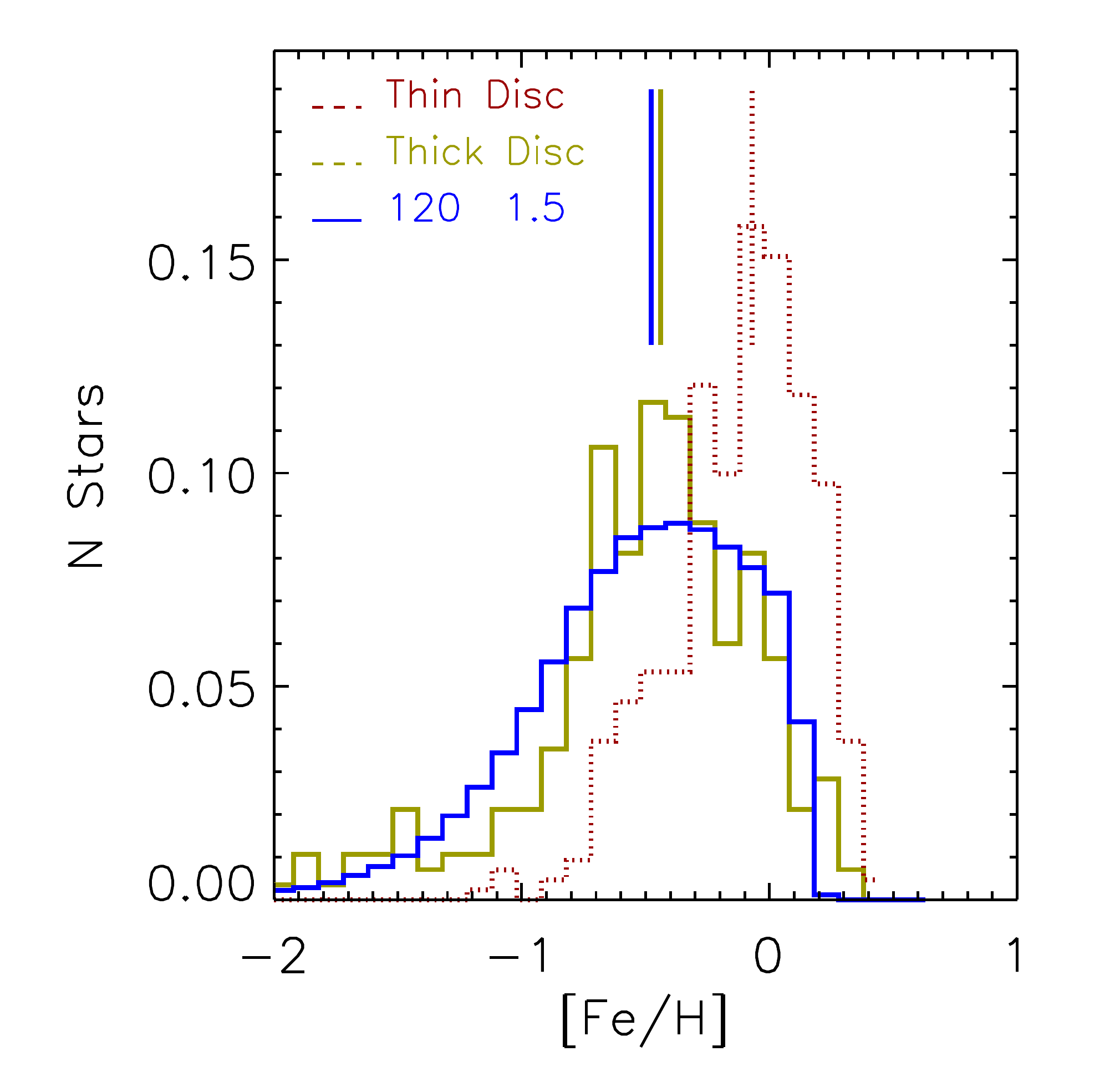}}
\vskip-0.3truecm
\caption{Observed thick-disc MDF  \citep[yellow thick solid line]{Bensbyetal14} compared with the predictions of model TD1  (Table \ref{model_comp}). The thin-disc MDF is also plotted for comparison.  Vertical lines are the median values of the corresponding distributions.}
\label{hist_thick_MTW}
\end{figure}
\subsection{Chemical evolution of the thick disc}
\label{chemthick}
We turn now our attention to the chemical evolution of thick-disc stars.
The diversity between thin (blue) and thick (magenta) disc stars is clearly visible in the \ofe\ vs. \feh\ diagram of Fig.~\ref{met-TW}.
The models adopted for the thin-disc stars are clearly not able to account for the [O/Fe] enrichment history of thick-disc stars.
Furthermore, stellar age determinations indicate that thick-disc stars are, on average, older than thin-disc stars. 

The origin of this marked difference may be linked to different star formation histories,  similarly to other components of the MW, such as the Galactic Bulge, whose stars are even more metal-rich than thick-disc stars \citep{Bensby_2017}.
Some chemo-dynamical models suggest that  thick-disc stars may have originated in the inner regions of the MW, characterized by an earlier and faster enrichment so that they had the time to migrate into the solar vicinity and become kinematically hotter. Under this hypothesis, these stars would be naturally characterized by chemical compositions and, in particular, levels of \textalpha-enhancement different from the native stellar populations of the solar neighbourhood \cite[e.g.][]{Aumer_2017,Schonrich_2017}. 

More recently the analysis of 
{\sl Gaia} data \citep{GaiaCollaboration2018}, provided  evidence that the peculiar  chemical composition of the thick disc could have originated in an early merger of the MW with a satellite galaxy, the Gaia Enceladus Sausage galaxy \citep{Belokurov_etal18,Helmi_etal18,Haywood_etal18}.



\begin{figure} 
\centering
\resizebox{0.9\hsize}{!}{\includegraphics[angle=0]{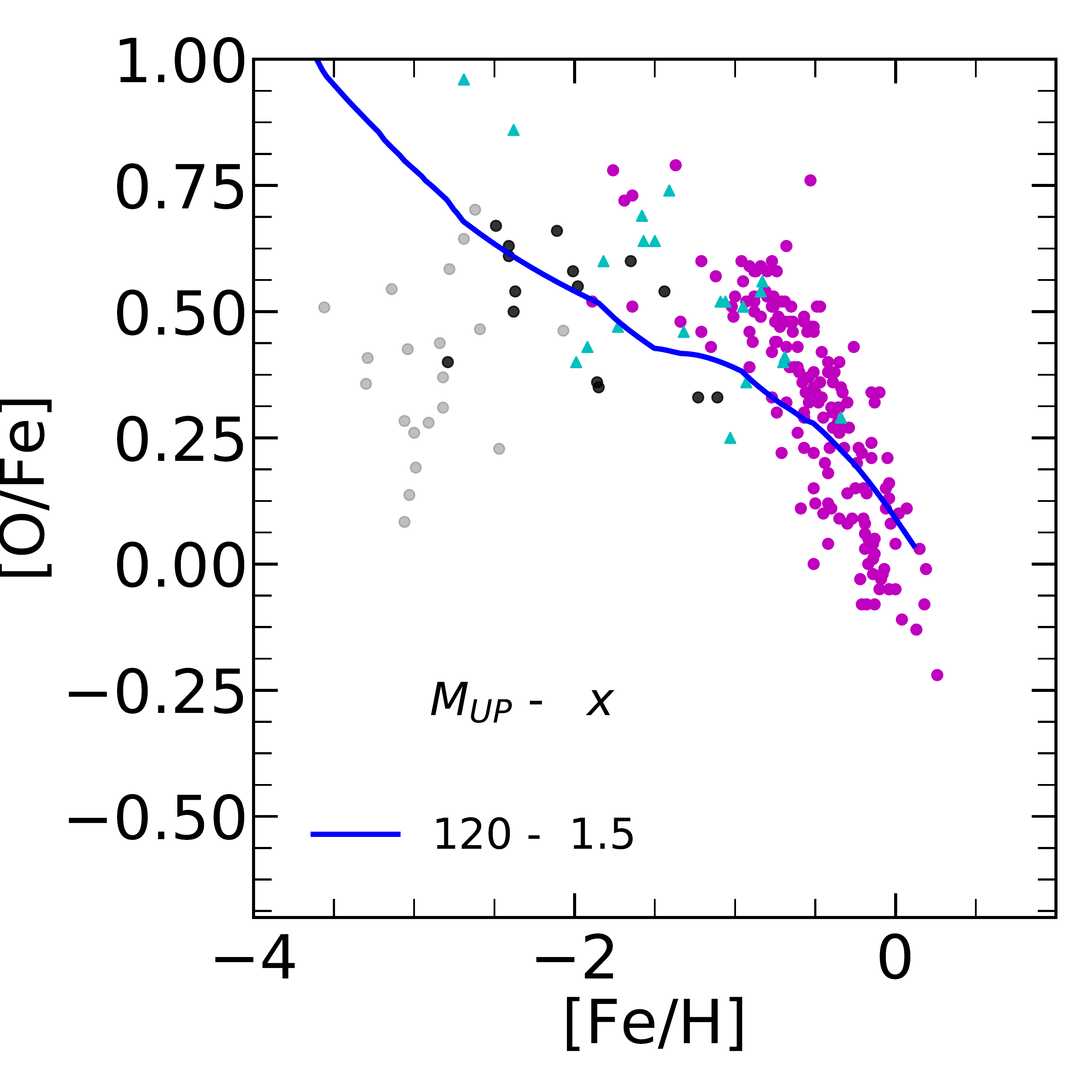}}
\caption{Comparison of the observed \ofe\ vs. \feh\ ratios of the thick disc
(magenta dots) with those predicted by model TD1 discussed in the text (see Table\ref{model_comp}).}
\label{oxy-thick}
\end{figure}

Contrary to the thin disc case, where we can constrain the chemical evolution models on a variety of observational data such as the current SFR, gas fraction etc., for the thick disc  the most robust constraint is the MDF. 
The model that best reproduces the observed thick disc MDF, keeping  the IMF parameters identical to our model MTW for the thin disc, \MUP \ = 120 and $x$ = 1.5 and the same chemical evolution parameters  $k$= 1 and A$_{SNIa}$=0.04,  is obtained using $\tau_{inf}$=0.5 Gyr and $\nu$=1.4Gyr$^{-1}$. The resulting theoretical thick disc MDF  (blue histogram) is compared with  the observed one (yellow histogram) in Fig.~\ref{hist_thick_MTW}. The parameters of this model, named TD1 are listed in Table \ref{model_comp}. In order to fit the high \feh \ tail of the observed MDF we need to include a  galactic wind at an epoch of  t$_{GW}$=2.5~Gyr, to expel the  residual gas (about 10\% of the total mass)  and stop the SF. While this assumption is introduced to cope with the simplicity of our chemical evolution code, we note that the adopted t$_{GW}$ corresponds to the epoch of the end of the second infall episode in the \citet{Micali_etal2013} model.
We note that models with shorter $\tau_{inf}$ show a clear star number excess below \feh \ $\sim$~-1, with respect to the observed one while, models with a larger  $\tau_{inf}$ show an excess in the high metallicity tail of the MDF.
The \ofe \ vs. \feh \ abundance patterns obtained with model TD1, are shown in Figure \ref{oxy-thick}. This model is able to fit the observed region occupied by thick disc stars but the slope is different from the observed one.
In the \ofe \ vs. \feh \ diagram, it underestimates the high \ofe \ values at low \feh \ and, it overestimates the low \ofe \ values at high \feh . This model is, however, able to reproduce the region occupied by the most metal-poor stars, likely belonging to the halo population, that will be  discussed later.


\subsubsection{Models with stellar rotation}
\begin{figure} 
\centering
\resizebox{0.9\hsize}{!}{\includegraphics[angle=0]{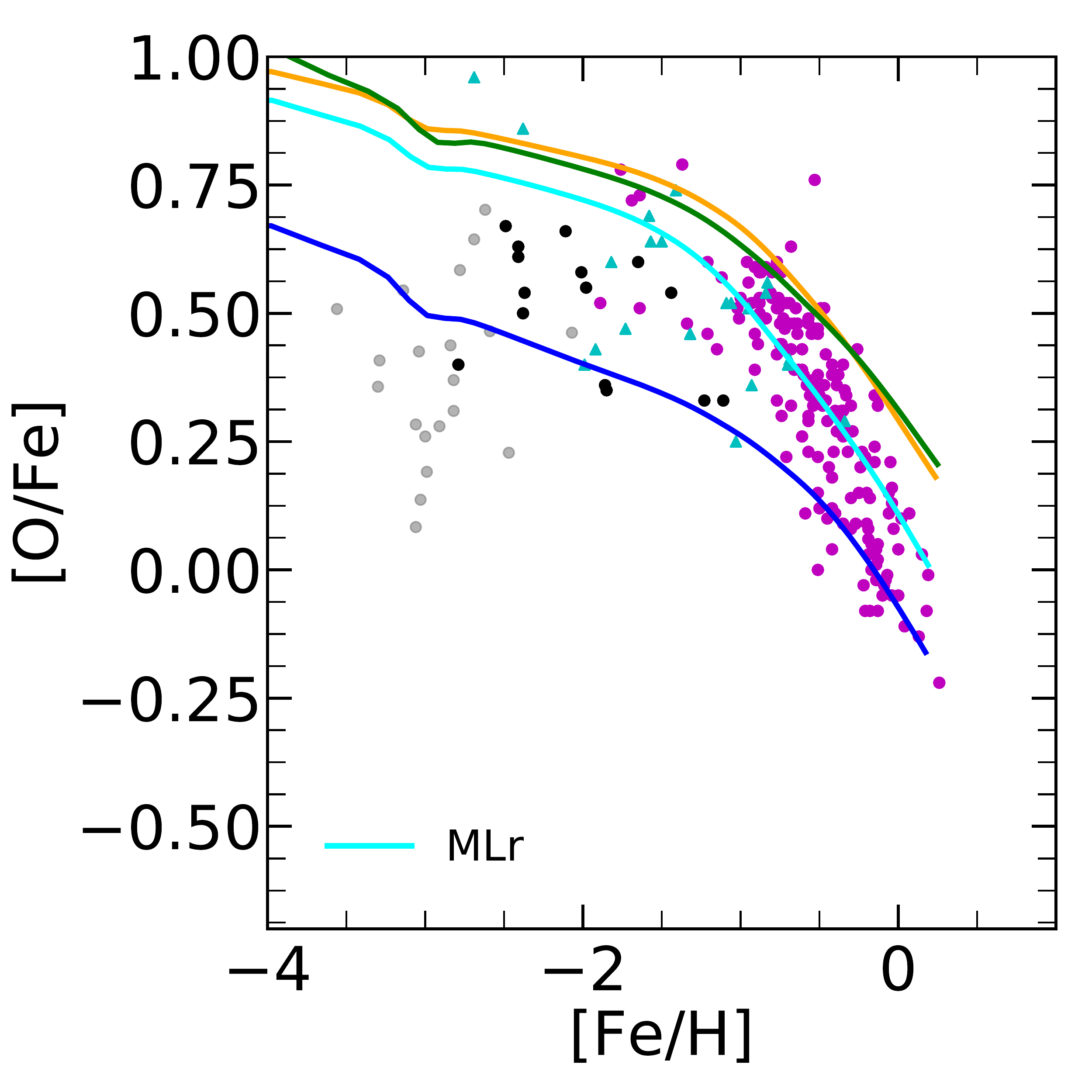}}
\caption{\ofe\ vs. \feh\ abundances predicted by massive star models without and with rotation.  The blue, orange and green lines refer to \cite{Limongi_etal18} yields for initial rotational velocities $V_{\rm rot}$=0, $V_{\rm rot}$=150 and $V_{\rm rot}$=300 km/s, respectively. The cyan line represents the MLr model with rotation-averaged yields. \label{oxy-rot} }
\end{figure}

{An improvement of the fit of the model TD1 in the \ofe \ vs. \feh \ diagram could be obtained adopting yields that account for rotational mixing in massive stars, as recently suggested by \cite{Donatella_etal19, Romano2020}.}
In support of this indication, we recall that, only  yields including stellar rotation, e.g. model MLr,  are actually able to produce slopes similar to the ones observed for the thick-disc stars (except for Mg; see Fig.~\ref{met-TW}). 
For this purpose we calculated new models based on the chemical evolution parameters of model TD1, but using  the  \cite{Limongi_etal18} yields for different initial rotational velocities,  $V_{\rm rot}=0$, $V_{\rm rot}=150$, $V_{\rm rot} =300\,\mathrm{km\, s^{-1}}$ and the averaged rotation yields, already discussed in model MLr. 
The \ofe \ vs. \feh \ diagram obtained with the new models is shown in Fig.~\ref{oxy-rot}.
Using yields from stars with higher rotational velocity we obtain higher [O/Fe] ratios. This simply reflects the fact that massive stars produce larger amount of oxygen at increasing $V_{\rm rot}$ (see Figs.~\ref{ejecta-comp0001}-\ref{ejecta-comp006}), while iron, being mainly contributed by SNIa, barely changes. 
Figure~\ref{oxy-rot} shows that enhancing the fraction of stars with high rotational velocity may help explaining the observed [O/Fe], especially at  metallicities $\feh\lesssim -1$. Looking at Fig.~\ref{oxy-rot}, we may see that the non rotating model computed with \cite{Limongi_etal18} yields fits the lower envelope of the thin disc stars, of some thick disc outliers at 
$\feh \simeq -2$ and of the few halo stars with $\feh < -2$.
By increasing the rotational velocity the models shifts toward higher \feh\ values. In particular using the model computed adopting the averaged yields obtained from the three different rotational velocity values (cyan line, \citet{Prantzos_18}), our model fits the border between thin and thick disc stars. Then, by further increasing the rotational velocity, all the thick disc stars can be fitted. These models are also able to reproduce the observed \ofe\ values of the halo stars.

\begin{table}
\caption{IMF parameters of the chemical evolution models for the thick disc.$n_*$ denotes the corresponding number of stars born in the selected mass range per 10$^6$\Msun\ of gas converted into stars.}
\label{model_comp}
\centering
\begin{tabular}{|c|c|c|c|c|c|} 
\hline\hline
\multicolumn{1}{|c|}{{model}} &
\multicolumn{1}{|c|}{{$\MUP\ [\Msun]$}} &
\multicolumn{1}{|c|}{{$x$}} &
\multicolumn{1}{|c|}{$\frac{n_*}{1E6\Msun}$}&
\multicolumn{1}{|c|}{$\frac{n_*}{1E6\Msun}$}\\
\multicolumn{1}{|c|}{} &
\multicolumn{1}{|c|}{} &
\multicolumn{1}{|c|}{} &
\multicolumn{1}{|c|}{(10-120\Msun)}&
\multicolumn{1}{|c|}{($120-\MUP$)}\\
\hline

TD1 & 120  & 1.5  &  5448 & 0\\
\hline
TD2 & 120  &  1.7 &  3465 & 0  \\
\hline
TD3 & 200  & 1.7 &  3449 & 30\\
\hline
TD4 & 200  & 1.4 &  6626 & 108 \\
\hline
TD5 & 350  & 1.2 &  9226 & 356 \\
\hline
\multicolumn{5}{|c|}{
all models: $\nu$=1.4Gyr$^{-1}$; $k=1$;  $\tau_{inf}$=0.5 Gyr; A$_{SNIa}$=0.04 }\\
\hline

\end{tabular} 
\end{table}
\begin{figure} 
\centering
\vskip-0.3truecm
\resizebox{0.99\hsize}{!}{
\includegraphics[angle=0]{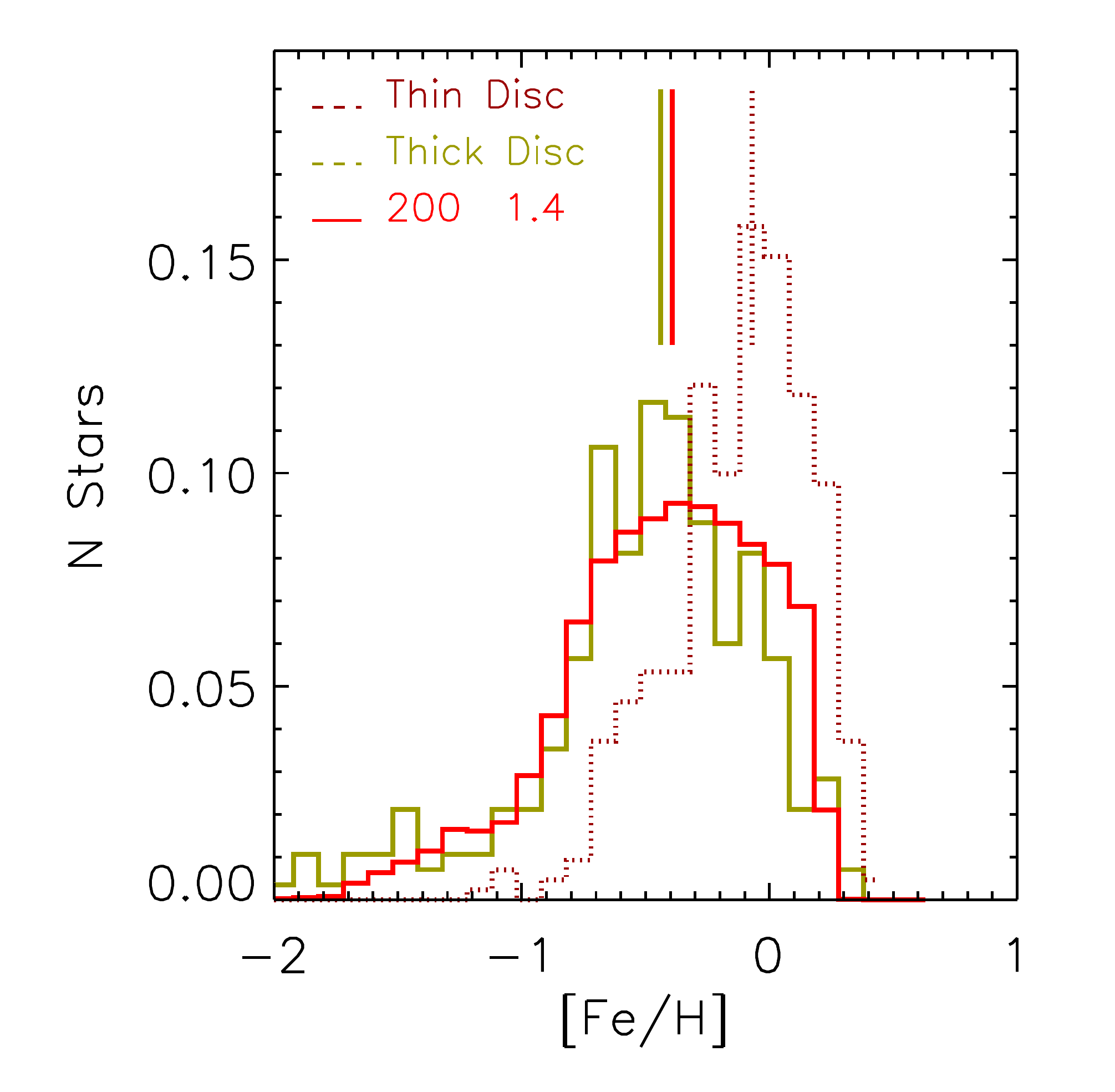}}
\vskip-0.3truecm
\caption{Same as Figure \ref{hist_thick_MTW} but for the thick disc model TD4.}
\label{hist_thick_TD4}
\end{figure}

\begin{figure*} 
\centering
\resizebox{0.9\hsize}{!}{\includegraphics[angle=0]{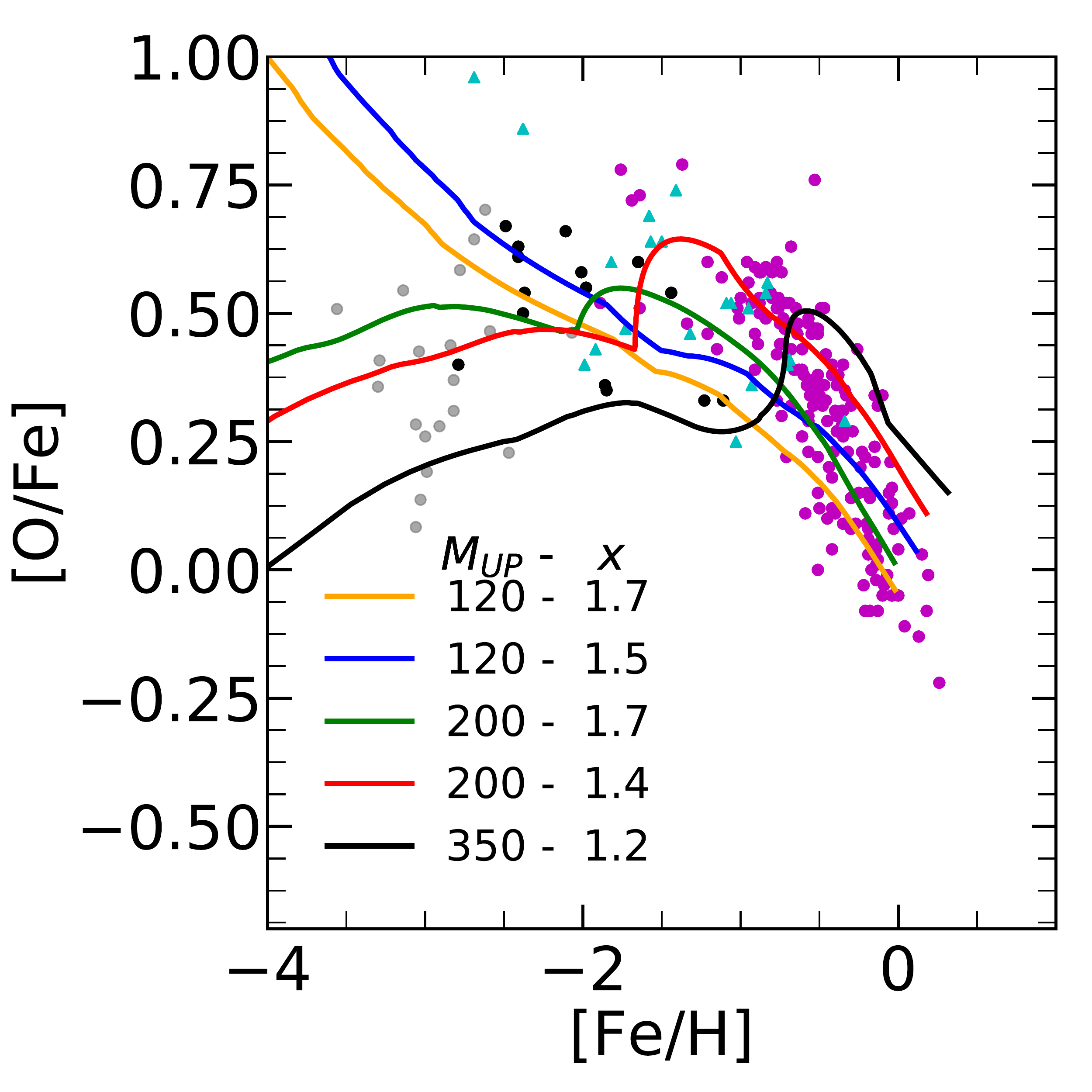}\includegraphics[angle=0]{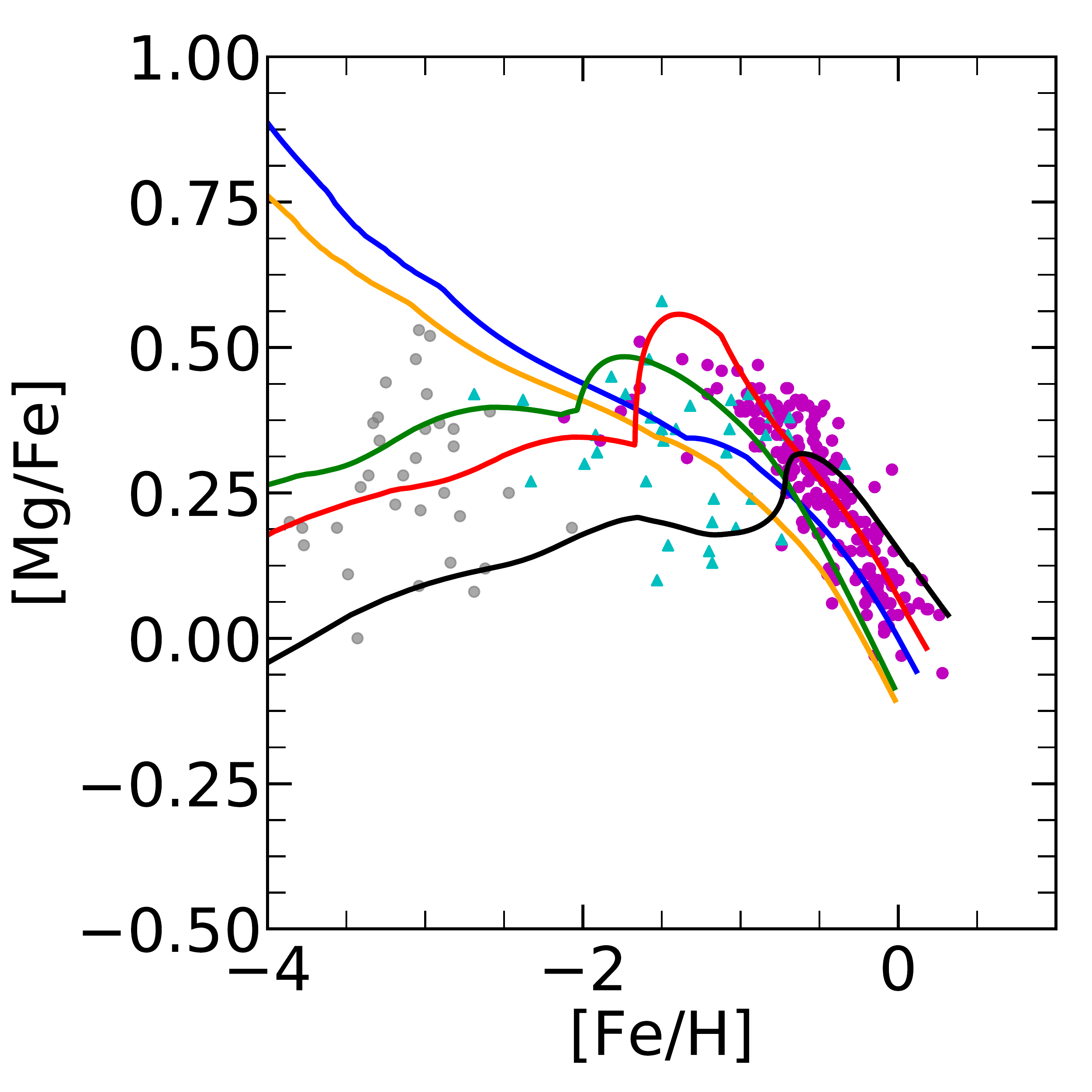}}
\caption{Comparison of the observed \ofe\ vs. \feh\  (left panel) and \mgfe\ vs. \feh\ (right panel) ratios of the thick disc stars, 
 with the predictions of models with different IMF parameters as listed in Table\ref{model_comp}.
We see that increasing $\MUP$,  so as to include  the contributions of VMO (winds, PPISN and PISN), thick-disc stars can be reproduced pretty well.
Moreover, changing  both the IMF slope and $\MUP$, different populations can be recovered.}
\label{oxy-imf}
\end{figure*}

\subsubsection{Effects due to IMF variations and VMO}
Another possible explanation for the different [O/Fe] evolution of thin- and thick-disc stars could be linked to different IMFs of the two populations.
As already discussed in Sect.~\ref{ssec_pisn}, the IGIMF predicts that, at low metallicity and high star formation rates, the IMF may become steeper at low masses and flatter at high masses than the canonical one \citep{marks_imf_2012MNRAS,jerabkova2018}. This suggests that, in the early phases of the thick disc evolution, the IMF could be well populated up to  \MUP .  
To investigate this aspect, we make use of our new MTW set of yields, which includes also the chemical contributions of VMO (winds, PPISN/PISN explosions and DBH).
This gives us the possibility to investigate the effects of changing both the IMF exponent $x$ (for $\Mi > 1\,\Msun$), and the upper mass limit \MUP, that can be pushed up to $350\,\Msun$. 

Starting from the chemical evolution parameters calibrated to reproduce the MDF shown in Figure \ref{hist_thick_MTW}, we  vary the IMF parameters ($x$ and \MUP), shifting the upper mass limit into the range ($120 \le \MUP/\Msun \leq 350$). 
The adopted values, listed in Table~\ref{model_comp}, are meant to test slopes typically found in the literature ($1.3 \leq x \leq 1.7$ for $\Mi > 1\,\Msun$; see Sect.~\ref{ssec_chemevol}), and to obtain models that are able to bracket the observed thick disc data shown in Figure \ref{oxy-thick}.
The theoretical thick-disc MDFs somewhat change with the IMF variation (Fig.~\ref{hist_thick_TD4}), however, the impact is not dramatic within the explored ranges of the IMF parameters, and the corresponding models actually bracket the median \feh\ of the observed distribution. 

The  abundance patterns of thick disc stars predicted by these new models 
are compared with observations in  the  [O/Fe] and [Mg/Fe] vs. \feh \ diagrams shown in Figure~\ref{oxy-imf}. 
We note that, in the \ofe \ vs. \feh \ diagram, model TD2, with x=1.7, runs almost parallel, below model TD1, eventually being able to reproduce the branch with low \ofe \ ratios at \feh \ =0. 
Models TD1 and TD2 are also able to reproduce 
the lower envelope of observed halo stars with \feh \ $\geq$ -3.
However, while below this metallicity, the data show a plateau or even a decreasing branch, the \ofe \ values of the above two models continue to increase. We recall that the majority of these halo stars are sub-giant or giant stars from \cite{Cayrel2004} and their \ofe \ values  have been  corrected for 3D effects, as suggested in \cite{Nissen2002}. 

The behaviour of the models change significantly if they include the ejecta from VMO. In Table \ref{model_comp}, we report also
the expected number of stars per 10$^6$ \Msun\  of gas converted into stars, that are born with $10 \ \Msun \ \le \Mi  \ \leq \ 120$ \Msun\  and with $120 \ \Msun \ \leq \Mi   \leq$  \MUP , resulting from the corresponding IMF parameters. It is easy to see from Table \ref{model_comp}, that the fractional contribution of VMOs to the total number of massive stars  progressively increase as we move from model TD3 to model TD4 and TD5.
We first note that even a small fraction of stars born as VMO is enough to change the initial evolution of the \ofe \ ratios (model TD3). This is due to the fact that low metallicity VMOs are able to produce enough Fe to decrease the \ofe \ ratio. 
Then, as the fractional number of VMOs increases, the models are able to reach lower and lower \ofe \ ratios.
Models TD3, TD4 and TD5 reproduce well the distribution  of halo stars with \feh \ $\leq$ -2. 

When the metallicity rises above \Zi \ $\sim$ 0.0001 there is a jump in the $^{16}$O production with respect to that of Fe. 
Indeed, while $^{16}$O is copiously produced in almost all PISN models, for Fe this is true only for the more massive ones, with E$_{\rm{j}}$(Fe) $\sim$ 20 \Msun \ for \Mhe \ $\sim$ 115  \Msun \ and E$_{\rm{j}}$(Fe) $\sim$ 0.5 \Msun \ for \Mhe \ $\sim$ 85  \Msun \citep{Heger_Woosley2002y}. This alternative behaviour of $^{16}$O and Fe production by PISNe at decreasing \Mhe ,  gives rise to a rapid jump in the \ofe \ ratio at \feh \ $\sim$ -2.  After the jump, the \ofe \ ratio reaches a peak well within the observed thick disc data, and then decreases following the observed slope of the data. At \feh \ $\geq$ -2,  models TD3 and TD4 bracket the majority of the thick disc, data leaving outside only a few objects with high \ofe \ ratios. Model TD5 can be considered as an extreme case, that is able to reproduce the lower and upper envelopes of the \ofe \ ratios, at the lower and higher metallicities, respectively.

A similar evolution can be seen also in the case of the  \mgfe \ vs. \feh \ ratios, as shown in the right panel of Figure \ref{oxy-imf}.  We note that models without VMOs, TD1 and TD2, are not able to reproduce the \mgfe \ ratios observed at very low metallicity \feh \ $\leq$ -2. On the contrary, models TD3, TD4 and TD5 run across the observed metal-poor halo stars, well reproducing the growing \mgfe \ ratio at increasing \feh . Thus, in the very early stages of the thick disc chemical evolution, the models with VMOs are able to reproduce the observed $\alpha$-poor (Mg) stars, for instance, those with $\mgfe\sim 0.0$ at $\feh\lesssim -1$.
For \feh \ $>$ -2, models TD3, TD4 and TD5 jump inside the thick disc \mgfe \ observations, providing an overall good fit to the evolution of halo and thick disc stars. 

In summary, while no single TD model is able to reproduce the \ofe \ and \mgfe \ observations of the halo and thick disc stars at once,
a combination of TD models with varying IMF, in line with the predictions of the IGIMF already discussed, is able to reproduce the location and the dispersion observed among halo stars and the thick disc chemical evolution pattern. 

Finally, we note that, the inclusion of PISNe may also help in solving the issue of Mg underprediction present in most of the published yields tables \citep{Timmes1995,Portinari1998,Prantzos_18}.

In conclusion, our analysis shows that IMF variations, in terms of slope but, more importantly, of \MUP, may significantly affect the predictions of chemical evolution models. In turn, this may impact on our understanding of the observed MW populations.

\section{Summary and concluding remarks}
\label{conclusions}
In this work we investigate the impact of the ejecta of massive and very massive stars on the predictions of chemical evolution models. We constructed different sets of chemical yields that include both  winds and explosion contributions.
To this aim, we collected various explosion yields  available in the literature and combine them with the wind ejecta computed with our code \texttt{PARSEC} or other authors.
A novelty of this work is that we investigate the effect of increasing the IMF upper mass limit up to $\MUP = 350\,\Msun$, which is well beyond the standard value of $\MUP \sim  100\,\Msun$, typically used in most studies.
This allows us to explore the impact of VMO, which are expected to eject significant amounts of newly produced elements  through both powerful stellar winds and PPISN/PISN explosions.
For completeness, we also include the chemical yields of AGB stars computed with our \texttt{COLIBRI} \citep{Marigo2013} and from other studies. We note that the impact of AGB stars on the chemical species considered in this work (Fe, O, Mg, Si, Ca) is minor. An analysis aimed to highlight the role of the chemical yields from low- and intermediate-mass stars is postponed to a dedicated future work.

The different sets of chemical yields are then incorporated in our chemical evolution code \texttt{CHE-EVO} \citep{Silva1998} to analyse the chemical evolution of thin- and thick-disc stars in the solar vicinity, for which we can rely on accurate abundance measurements and kinematical classification \citep{Bensbyetal14}.

For each set of chemical yields, we ran grids of chemical evolution models to single out the parameters that best fit the main observational constraints of the MW thin disc, namely:
the present-day star formation rate and gas fraction,  the rates of CCSN and SNIa, the metallicity of the Sun at its birth epoch $\simeq 4.6$ Gyr ago, and the observed MDF.
For all sets of chemical yields, we were able to find suitable combinations of input parameters that match all the constraints reasonably well (see Table \ref{yields_comp}).
In the best-fit models the Schmidt star formation law,
has a typical efficiency in the range  $0.4<\nu<0.8$  and an exponent $k=1$. The gas infall time-scale generally varies in the range   $3 < \tau/\mathrm{Gyr} < 6$.

Once the chemical evolution model is calibrated for each set of yields, we move to test them against the observations.
We focus on the trends of the abundance ratios for a few \textalpha-elements ([O/Fe], [Mg/Fe], [Si/Fe], [Ca/Fe]) as a function of the metallicity, traced by [Fe/H].

A general agreement is found for [O/Fe] and [Si/Fe], while the predictions for [Mg/Fe] and [Ca/Fe] run below the observed data.
The best results are obtained when including the yields from rotating massive stars \citep{Limongi_etal18} and those based on our \texttt{PARSEC} models for the hydrostatic phases.
With the \cite{Ritter_etal18} yields, all abundance ratios are significantly below the observed values, despite the fact that the corresponding calibrated chemical evolution model reproduces the thin-disc constraints. After carefully examining the problem, we conclude that a possible cause is the large iron production of these sets of yields.

Most best-fit models are able to recover reasonably well the slope of the bulk of the thin-disc stars. With our MTW  set of yields, the calibrated IMF exponent is $x = 1.5$, which is intermediate between the results of
\citet[][$x = 1.7$]{1993MNRAS.262..545K}  and the top-heavier IMFs of \citet[][$x = 1.3$]{Kroupa01} and \citet[][$x = 1.3$]{Chabrier2003}.

In this study  we did not find evidence that the IMF exponent is degenerate with other chemical evolution parameters, e.g. $\nu$ or $k$, but we cannot exclude that more complex chemical evolution models may provide a different indication.
Recently \cite{Valentini2019}  pointed out that hydrodynamical models support either of the Kroupa IMF slopes, depending on the set of observational data adopted for comparison.
In this respect, we note that a steeper slope of the IMF is more suitable to reproduce thin-disc stars, as suggested by various other investigations \citep{Matteucci1989, Matte2001, Grisoni2018, Matte2019, Matte2020}. 

Thick-disc stars exhibit a \ofe\ vs. \feh\ pattern different from that of the  thin-disc stars, characterised by a steeper slope and a larger degree of \textalpha-enhancement, with higher \ofe\ ratios.
This feature  is much less evident in the \mgfe\ diagram.
In this respect, we note that there are  stars, likely belonging to the thin disc according to  the kinematical classification, that has higher [O/Fe] values and fall in the region populated by the thick-disc stars. Our thin-disc models do not reproduce such stars. 
Possible explanations of this {\sl anomalous} behaviour 
are that, either they have formed in different environmental conditions compatible with a top-heavier IMF  (e.g. in the IGIMF scenario for a low metallicity and/or high SFR) or, rather, the kinematical parameters of these stars varied in such a way that they are now classified as thin-disc members. \ 

{None of the chemical evolution models  calibrated on the thin disc is able to reproduce the \ofe\ trend of the thick disc.  
To reproduce the higher values of the $\alpha$- enhancement seen in thick disc stars one need to change the
chemical evolution parameters in particular  
reduce the infall time-scale to  $\tau_{inf}$=0.5 Gyr
and increase the star formation efficiency to $\nu$=1.4 Gyr$^{−1}$, in agreement with the literature \citep{Grisoni2019,Grisoni2020a,Grisoni2020b}. The MDF of thick disc stars is well reproduced with these parameters but, in order to better fit its fast decreasing tail at the higher metallicity  with our simple chemical evolution model, we need to impose a galactic wind at an age of 2.5 Gyr. Tough the MDF is well reproduced, the tun on the \ofe \ vs. \feh \ diagram is not satisfactory. The model enters the region populated by thick disc stars but with a slope that is flatter than that of the trend of the data. In particular, the model cannot reproduce the high $\alpha$- enhanced thick disc stars at low metallicities. The model cannot reproduce also the pattern of halo stars, which have been included here to check its early chemical evolution.

Since it has been recently suggested that a higher $\alpha$-enhancement could be obtained by considering yields from fast rotating stars \citep{Limongi_etal18,Donatella_etal19}, we have calculated new models changing only the yields and using those that include effects of rotation \citep{Limongi_etal18, Prantzos_18}.
Our analysis shows that we can well reproduce the 
\ofe \ vs. \feh \ diagram of thick disc by increasing only the fraction of rotating stars. The LC18 models without rotation run
along the lower envelope of thick disc while those with intermediate and high rotational velocity reproduce its upper envelope as shown in Figure~\ref{oxy-rot}. This in fact may give rise to a degeneracy between chemical evolution parameters ($\tau_{inf}$ and $\nu$) and  stellar rotational velocities. The low metallicity halo stars in this diagram cannot be reproduced by our model.}

{An alternative way to reproduce the observations of  the thick disc  is 
offered by the recent growing evidence for 
a significant dependence of the IMF on the environment \citep{marks_imf_2012MNRAS,jerabkova2018}. In the IGIMF theory  
the IMF may become top heavier at decreasing metallicity and increasing star formation rate.  If this is true, the different MDFs of the thin and thick discs could be not only the result but also one of the reasons of the different chemical evolution patterns of the two MW components.  In particular  in the early phases of the thick disc evolution, the IMF could be well populated up to an \MUP \ that is able to switch on the chemical contribution from PISNe. 
Motivated by the above reasons we have used our MTW yields to analyzed the effects of varying both the slope and $\MUP$.  Using the same chemical evolution parameters of model TD1, we computed other models to check the effects of such IMF variations. Among the calculated new models we have selected four representative cases that are able to match the properties of thick disc stars. One with a steeper IMF (x=1.7) and canonical \MUP \ =120 which defines the lower boundary of the thick disc in the \ofe \ vs. \feh \ diagram, other two with  \MUP \ =200 and x=1.7 and 1.4, respectively and a final one with \MUP \ =350 and x=1.2.

The latter three models, characterized by a \MUP \ that falls within the VMOs, are able to bracket the thick disc observations in both the \ofe \ vs. \feh and \mgfe \ vs. \feh \ diagrams. 
At the same time, while with canonical \MUP \ we are not able to reproduce the properties of metal-poor halo stars in neither of the above diagrams, this is possible with the models that include the contribution of PISNe. Halo stars, in the \ofe \  and \mgfe \ vs. \feh \ diagrams, show a constant or even decreasing trend of the $\alpha$-enhancement at decreasing \feh. This trend can be reproduced if the models include enough Fe contribution coming from PISNe at lower metallicities (\feh \ $\leq$-2). While there is a dispute concerning the reality of this decreasing branch in the \ofe \ diagram \citep{Nissen2002, Cayrel2004, Micali_etal2013}  this constant or even decreasing trend is seen also in the 
\mgfe \ diagram. In any case recent analysis of O and Fe measurements in disc and halo stars by \cite{Amarsi2019} indicate that the \ofe \ ratio reaches a constant plateau 
of \ofe \ $\sim$0.6  below \feh \ $\sim$-1 and it is thus not compatible with the constant growth, at decreasing \feh , indicated by our models with canonical \MUP . It is worth to note here that the models that include the effects of rotation are facing the same problem. Instead, an increasing fraction of VMOs, eventually in line with the prediction of the IGIMF theory, will produce models that progressively show lower [$\alpha$/Fe] ratios, thus naturally explaining the halo stars dispersion and, at the same time, matching the high [$\alpha$/Fe] ratios of thick disc stars.}


In conclusion, our study provides a clear indication that PISN may have played a significant role in shaping the chemical evolution path of halo and  thick-disc stars. Especially in view of the possible flattening of the IMF at low metallicity and high SFR,
their contribution should not be neglected in chemical evolution models.
In the light of these results, we may reasonably expect that similar data, which actually exist for nearby extremely metal-poor star bursting galaxies \citep{Kojima2020}, might host the chemical signature of very massive stars, and that would witness  their key role in the early phases of galaxy evolution.

The complete TW sets of chemical ejecta for massive and very massive stars, described in Appendix~\ref{tables_ejecta}, are made publicly available through the URL: xxxxx.

\color{black}
\begin{acknowledgements}
We thank F. Matteucci, D. Romano and the anonymous referee for useful suggestions that helped to improve the manuscript.
PM and AS acknowledge
the support from the ERC Consolidator Grant funding
scheme (project STARKEY, grant agreement n. 615604).
AB and LS acknowledge 
PRIN MIUR 2017 - 20173ML3WW\_001. MS acknowledges funding from the European Union’s Horizon 2020 research and innovation programme under the Marie-Sk\l{}odowska-Curie grant agreement No. 794393.
This research made use of Astropy, a community-developed core
Python package for Astronomy (Astropy Collaboration et al. 2013,
2018) and matplotlib, a Python library for publication quality graphics
(Hunter 2007). This work has been partially supported by PRIN MIUR 2017 prot. 20173ML3WW 002, `Opening the ALMA window on the cosmic evolution of gas, stars and supermassive black holes'. AL acknowledges the MIUR grant `Finanziamento annuale individuale attivita base di ricerca' and the EU H2020-MSCA-ITN-2019 Project 860744 `BiD4BEST: Big Data applications for Black hole Evolution STudies'. VG acknowledges financial support at SISSA from the European Social Fund operational Programme 2014/2020 of
the autonomous region Friuli Venezia Giulia.
\end{acknowledgements}



\bibliography{pp_p1}
\appendix
\section{Tables of chemical ejecta}
\label{tables_ejecta}




Here we present the ejecta tables (set TW) for massive and very massive stars used in this work that will be available on-line, for 4 values of the initial metallicity ($\Zi=0.0001,0.001,0.006,0.02$) and 30 values of the initial mass ($8 \leq \Mi/\Msun \leq 350$).
Each table corresponds to one selected value of \Zi.
The row labelled $X_{j,0}$ gives the initial abundances (in mass fraction) corresponding to a scaled-solar composition  for elements heavier than He \citep{Caffau_etal11}.
The initial abundances of H and He as a function of \Zi\ are derived from the enrichment law calibrated with \texttt{PARSEC} models \citep{Bressan2012}.
The complete ejecta tables include the following chemical species: 
$^1$H, $^3$He, $^{4}$He, $^{7}$Li, $^{7}$Be, $^{12}$C, $^{13}$C, $^{14}$N, $^{15}$N, $^{16}$O, $^{17}$O, $^{18}$O, $^{19}$F, $^{20}$Ne, $^{21}$Ne, $^{22}$Ne, $^{23}$Na, $^{24}$Mg, $^{25}$Mg, $^{26}$Mg, $^{26}$Al, $^{27}$Al, $^{28}$Si, $^{29}$Si, P,   S,  Cl,  Ar,   K,  Ca,  Sc,  Ti,   V,  Cr,  Mn,  Fe,  Co,  Ni,  Cu,  Zn.
For elements without the indication of the atomic mass, we give the sum of the ejecta of their stable isotopes.

We remind that mass loss in the \texttt{PARSEC} code is applied only to stars with $\Mi \geq 14\,\Msun$. It follows that the wind ejecta for stars with $\Mi < 14\,\Msun$ are set to zero.
It is worth specifying that the ejecta of the VMO that avoid the explosion and directly collapse to black holes (DBH, for $\Mhe > 130\, \Msun)$ are included in the wind tables.
The explosion ejecta tables contain the chemical contribution of all layers that extend from \Mcut\ to \Mfin. This applies to both successful CCSN, PPISN and PISN. We recall that in the case of a failed core-collapse supernova we have $\Mcut=\Mrem=\Mfin$ and the explosion ejecta are zero for all species. As to PISN  we set $\Mcut=0$, as the associated thermonuclear explosion leaves no remnant.

Other relevant stellar parameters are tabulated (in units of \Msun), namely: the pre-SN mass ($\mathrm{M_{fin}}$), the mass of He core (\Mhe), the mass of the C-O core (\Mco) and remnant mass ($M_{\rm rem}$). 
We also provide the pre-SN phase (see also Sect.~\ref{sec_parsec}):
\begin{itemize} 
\item RSG: red supergiant
\item BSG: blue supergiant
\item WC: Wolf-Rayet stars enriched in carbon
\item WN: Wolf-Rayet stars enriched in nitrogen
\item WO: Wolf-Rayet stars enriched in oxygen
\item LBV: luminous blue variables
\end{itemize}
and the final fate: 
\begin{itemize} 
\item ECSN: electron capture SN
\item sCCSN: successful core collapse SN
\item fCCSN: failed core collapse SN
\item PPISN: pulsation-pair instability SN
\item PISN: pair instability SN
\item DBH: stars that direct collapse into black hole without explosion
\end{itemize}

As an example, a reduced version of an ejecta table,  containing a smaller number of elements, is presented in Table~\ref{ejetot}.

\begin{table*}[h]
\centering
\caption{An example of a table containing the total ejecta, $E_j=E_j^{\rm w} + E_j^{\rm sn}$ (in \Msun, see Sect.~\ref{EJECTA}), of massive and very massive stars used  in this work (extracted from the set MTW), for $\Zi = 0.02$ and a few selected chemical species. The complete tables, available online, include more nuclides, from H to Zn.}
\label{ejetot}
\begin{adjustbox}{angle=90, max width=0.5\textwidth}
\begin{tabular}{|r|r|r|r|r|l|l|l|l|l|l|l|l|l|l|l|l|l|l|l|l|l|l|l|l|l|l|l|l|l|l|}
\hline
  \multicolumn{1}{|c|}{\Mi}  &
  \multicolumn{1}{|c|}{\Mfin}  &
  \multicolumn{1}{|c|}{\Mhe}  &
  \multicolumn{1}{|c|}{\Mco}  &
  \multicolumn{1}{|c|}{\Mrem}  &
  \multicolumn{1}{|c|}{pre-SN}  &
  \multicolumn{1}{|c|}{SN-type}  &
  \multicolumn{1}{|c|}{$^1$H}  &
  \multicolumn{1}{|c|}{$^4$He}  &
  \multicolumn{1}{|c|}{$^{12}$C}  &
  \multicolumn{1}{|c|}{$^{14}$N}  &
  \multicolumn{1}{|c|}{$^{16}$O}  &
  \multicolumn{1}{|c|}{$^{19}$F}  &
  \multicolumn{1}{|c|}{$^{20}$Ne}  &
  \multicolumn{1}{|c|}{$^{23}$Na}  &
  \multicolumn{1}{|c|}{$^{24}$Mg}  &
  \multicolumn{1}{|c|}{.....}  \\
  \hline

    \multicolumn{1}{|c|}{$X_{j,0}$}  &  \multicolumn{1}{|c|}{--}   &  \multicolumn{1}{|c|}{--}  & \multicolumn{1}{|c|}{--} & \multicolumn{1}{|c|}{--} & \multicolumn{1}{|c|}{--}  &  \multicolumn{1}{|c|}{--} &  6.95E-001 &    2.83E-001 &      3.59E-003 &    9.70E-004  &     8.80E-003  &     6.61E-007 &   1.88E-003  &     4.71E-005 &   6.99E-004  & .....  \\
  \hline
8.00  & 8.00 &1.38& 1.38 & 1.366 &RSG & ~ECSN & 4.213E+00 & 2.280E+00 & 1.409E-02 & 2.544E-02 & 4.879E-02 & 3.540E-06 &  5.803E-05 & 6.126E-04 & 4.307E-03 & .....\\
9.00  & 9.00 &1.38& 1.38 & 1.366 &RSG & ~ECSN & 4.984E+00 & 2.445E+00 & 1.657E-02 & 2.765E-02 & 5.750E-02 & 4.179E-06 &  1.378E-02 & 6.694E-04 & 4.957E-03 & .....\\
    
    11.00   &     11.00   &    3.79     &   2.11     &   1.51       &     RSG     &   sCCSN   &  4.85E+000 &    3.78E+000 &      1.25E-001 &    3.94E-002  &     3.34E-001  &     2.88E-006 &   4.72E-002  &     1.15E-003 &   3.47E-002  & .....	\\
    14.00   &     13.03   &    4.56     &   2.90     &   1.67       &     RSG     &   sCCSN   &  6.04E+000 &    4.66E+000 &      2.31E-001 &    4.57E-002  &     7.32E-001  &     4.58E-006 &   1.69E-001  &     4.45E-003 &   6.03E-002  & .....	\\
    16.00   &     13.94   &    5.44     &   3.75     &   1.66       &     RSG     &   sCCSN   &  6.53E+000 &    5.29E+000 &      2.87E-001 &    4.95E-002  &     1.17E+000  &     6.39E-006 &   4.28E-001  &     1.04E-002 &   1.12E-001  & .....	\\
    18.00   &     14.96   &    6.31     &   4.38     &   1.69       &     RSG     &   sCCSN   &  7.23E+000 &    5.91E+000 &      3.59E-001 &    5.40E-002  &     1.53E+000  &     8.55E-006 &   5.61E-001  &     1.47E-002 &   1.32E-001  & .....	\\
    20.00   &     17.74   &    7.12     &   5.02     &   1.76       &     RSG     &   sCCSN   &  7.96E+000 &    6.44E+000 &      4.68E-001 &    5.67E-002  &     1.93E+000  &     1.12E-005 &   6.28E-001  &     1.89E-002 &   1.30E-001  & .....	\\
    24.00   &     19.41   &    9.15     &   6.69     &   1.82       &     RSG     &   sCCSN   &  9.07E+000 &    7.54E+000 &      6.20E-001 &    6.33E-002  &     3.04E+000  &     1.80E-005 &   9.62E-001  &     3.08E-002 &   1.68E-001  & .....	\\
    28.00   &     20.76   &   11.01     &   8.32     &   1.85       &     RSG     &   sCCSN   &  1.03E+001 &    8.66E+000 &      7.00E-001 &    7.52E-002  &     4.12E+000  &     2.36E-005 &   1.29E+000  &     4.09E-002 &   2.16E-001  & .....	\\
    30.00   &     16.71   &   12.22     &   8.44     &  16.71       &     RSG     &   fCCSN   &  8.35E+000 &    4.65E+000 &      3.34E-002 &    5.03E-002  &     9.27E-002  &     6.71E-006 &   2.47E-002  &     1.16E-003 &   9.17E-003  & .....	\\
    35.00   &     13.31   &   13.31     &   9.77     &  13.31       &      WO     &   fCCSN   &  1.14E+001 &    9.62E+000 &      1.33E-001 &    1.23E-001  &     1.61E-001  &     8.19E-006 &   4.01E-002  &     2.84E-003 &   1.49E-002  & .....	\\
    40.00   &     12.75   &   12.39     &   9.28     &  12.75       &      WC     &   fCCSN   &  1.28E+001 &    1.31E+001 &      6.74E-001 &    1.62E-001  &     1.86E-001  &     9.12E-006 &   5.02E-002  &     4.19E-003 &   1.88E-002  & .....	\\
    45.00   &     13.89   &   13.50     &  10.10     &  13.89       &      WC     &   fCCSN   &  1.38E+001 &    1.56E+001 &      9.21E-001 &    1.96E-001  &     2.05E-001  &     9.24E-006 &   5.72E-002  &     5.24E-003 &   2.15E-002  & .....	\\
    50.00   &     15.03   &   14.61     &  10.93     &  15.03       &      WC     &   fCCSN   &  1.48E+001 &    1.81E+001 &      1.16E+000 &    2.31E-001  &     2.23E-001  &     9.37E-006 &   6.42E-002  &     6.30E-003 &   2.41E-002  & .....	\\
    55.00   &     16.55   &   16.08     &  12.03     &  16.55       &      WC     &   fCCSN   &  1.58E+001 &    2.02E+001 &      1.42E+000 &    2.60E-001  &     2.59E-001  &     9.71E-006 &   7.06E-002  &     7.19E-003 &   2.66E-002  & .....	\\
    60.00   &     18.10   &   17.60     &  13.13     &  18.10       &      WC     &   fCCSN   &  1.68E+001 &    2.23E+001 &      1.69E+000 &    2.89E-001  &     3.00E-001  &     1.00E-005 &   7.68E-002  &     8.09E-003 &   2.89E-002  & .....	\\	
    65.00   &     19.93   &   19.37     &  14.40     &  19.93       &      WC     &   fCCSN   &  1.78E+001 &    2.40E+001 &      1.98E+000 &    3.17E-001  &     3.54E-001  &     1.02E-005 &   8.26E-002  &     8.92E-003 &   3.11E-002  & .....   \\
    70.00   &     22.29   &   21.67     &  16.09     &  22.29       &      WC     &   fCCSN   &  1.89E+001 &    2.52E+001 &      2.27E+000 &    3.43E-001  &     4.35E-001  &     1.03E-005 &   8.74E-002  &     9.58E-003 &   3.30E-002  & .....	\\
    75.00   &     23.43   &   22.77     &  16.91     &  23.43       &      WC     &   fCCSN   &  1.99E+001 &    2.76E+001 &      2.57E+000 &    3.70E-001  &     4.90E-001  &     1.10E-005 &   9.44E-002  &     1.05E-002 &   3.56E-002  & .....	\\
    80.00   &     24.56   &   23.87     &  17.73     &  24.56       &      WC     &   fCCSN   &  2.10E+001 &    2.99E+001 &      2.87E+000 &    3.97E-001  &     5.45E-001  &     1.16E-005 &   1.01E-001  &     1.15E-002 &   3.83E-002  & .....	\\
    90.00   &     26.90   &   26.15     &  19.39     &  26.90       &      WC     &   fCCSN   &  2.30E+001 &    3.46E+001 &      3.53E+000 &    4.65E-001  &     7.07E-001  &     1.14E-005 &   1.15E-001  &     1.38E-002 &   4.36E-002  & .....	\\
    95.00   &     27.58   &   26.81     &  19.90     &  27.58       &      WC     &   fCCSN   &  2.40E+001 &    3.75E+001 &      3.75E+000 &    5.01E-001  &     7.54E-001  &     1.17E-005 &   1.23E-001  &     1.50E-002 &   4.66E-002  & .....	\\
   100.00   &     27.75   &   26.97     &  19.93     &  27.75       &      WC     &   fCCSN   &  2.51E+001 &    4.11E+001 &      3.84E+000 &    5.48E-001  &     7.76E-001  &     1.18E-005 &   1.31E-001  &     1.63E-002 &   5.00E-002  & .....	\\
   120.00   &     22.32   &   21.69     &  16.11     &  22.32       &      WC     &   fCCSN   &  3.01E+001 &    6.27E+001 &      2.54E+000 &    9.20E-001  &     4.61E-001  &     1.07E-005 &   1.77E-001  &     2.38E-002 &   6.76E-002  & .....	\\
   150.00   &     16.57   &   16.10     &  12.03     &  16.57       &      WC     &   fCCSN   &  4.09E+001 &    8.81E+001 &      1.52E+000 &    1.41E+000  &     3.10E-001  &     9.78E-006 &   2.42E-001  &     3.37E-002 &   9.24E-002  & .....	\\
   200.00   &     16.11   &   15.66     &  11.70     &  16.11       &      WC     &   fCCSN   &  5.74E+001 &    1.21E+002 &      1.44E+000 &    2.04E+000  &     3.21E-001  &     9.18E-006 &   3.33E-001  &     4.77E-002 &   1.27E-001  & .....	\\
   250.00   &     16.26   &   15.80     &  11.83     &  16.26       &      WC     &   fCCSN   &  7.40E+001 &    1.53E+002 &      1.47E+000 &    2.67E+000  &     3.41E-001  &     8.43E-006 &   4.23E-001  &     6.16E-002 &   1.61E-001  & .....	\\
   300.00   &     17.13   &   16.96     &  12.46     &  17.13       &      WC     &   fCCSN   &  9.20E+001 &    1.83E+002 &      1.57E+000 &    3.22E+000  &     4.21E-001  &     1.15E-005 &   5.12E-001  &     7.41E-002 &   1.95E-001  & .....	\\
   350.00   &     16.84  &    16.37    &   12.23     &  16.84       &      WC     &   fCCSN   &  1.07E+002 &    2.17E+002 &      1.59E+000 &    3.91E+000  &     3.83E-001  &     7.28E-006 &   6.02E-001  &     8.94E-002 &   2.30E-001  & .....	\\

\hline
\end{tabular}
\end{adjustbox}
\end{table*}

\section{Comparison of the ejecta from various authors}
\label{app_compeje}
\begin{figure*} 
\centering
\resizebox{\hsize}{!}{\includegraphics[]{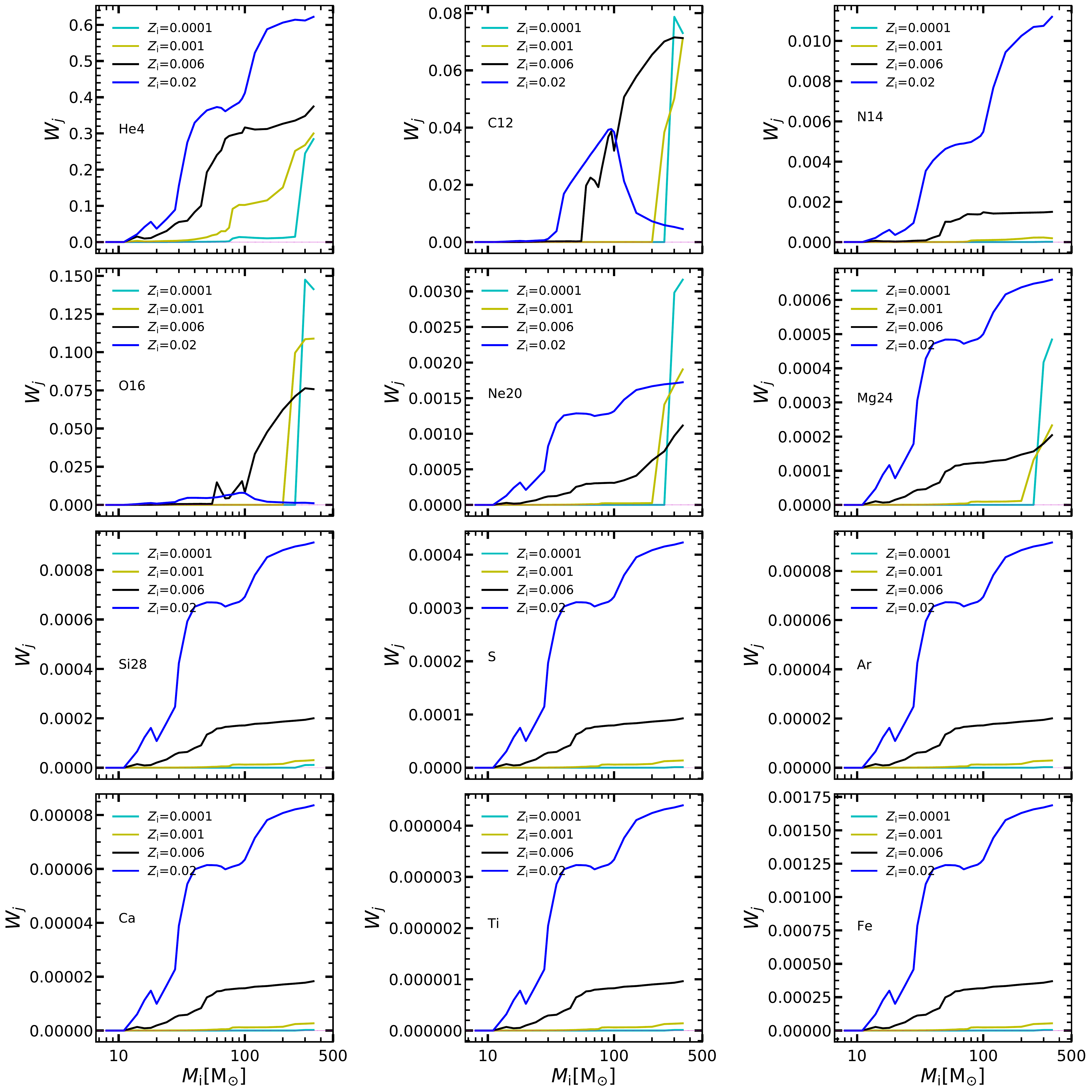}}
\caption{Fractional wind ejecta, $W_j$, derived from \texttt{PARSEC} stellar models, as a function of \Mi\ and \Zi.
The chemical species shown are $^4$He, $^{12}$C, $^{14}$N, $^{16}$O, $^{20}$Ne, $^{24}$Mg, $^{28}$Si, S, Ar, Ca, Ti, Fe.}
\label{ej-wind1}
\end{figure*}

To compare the  wind ejecta over a very large range of initial masses, it is convenient to define the quantity
\begin{equation*}W_j(\Mi)= E^{\rm w}_j(\Mi) /\Mi\, ,
\end{equation*} 
which corresponds to the fraction of \Mi\ lost through stellar winds in the form of the species $j$.
Figure~\ref{ej-wind1} shows $W_j(\Mi)$ derived from \texttt{PARSEC} models at varying \Mi\ and \Zi, for a few relevant species: $^4$He, $^{12}$C, $^{14}$N, $^{16}$O, $^{20}$Ne, $^{24}$Mg, $^{28}$Si, S, Ca, Ar, Ti, Fe.

\begin{figure*} 
\centering
\resizebox{\hsize}{!}{\includegraphics[]{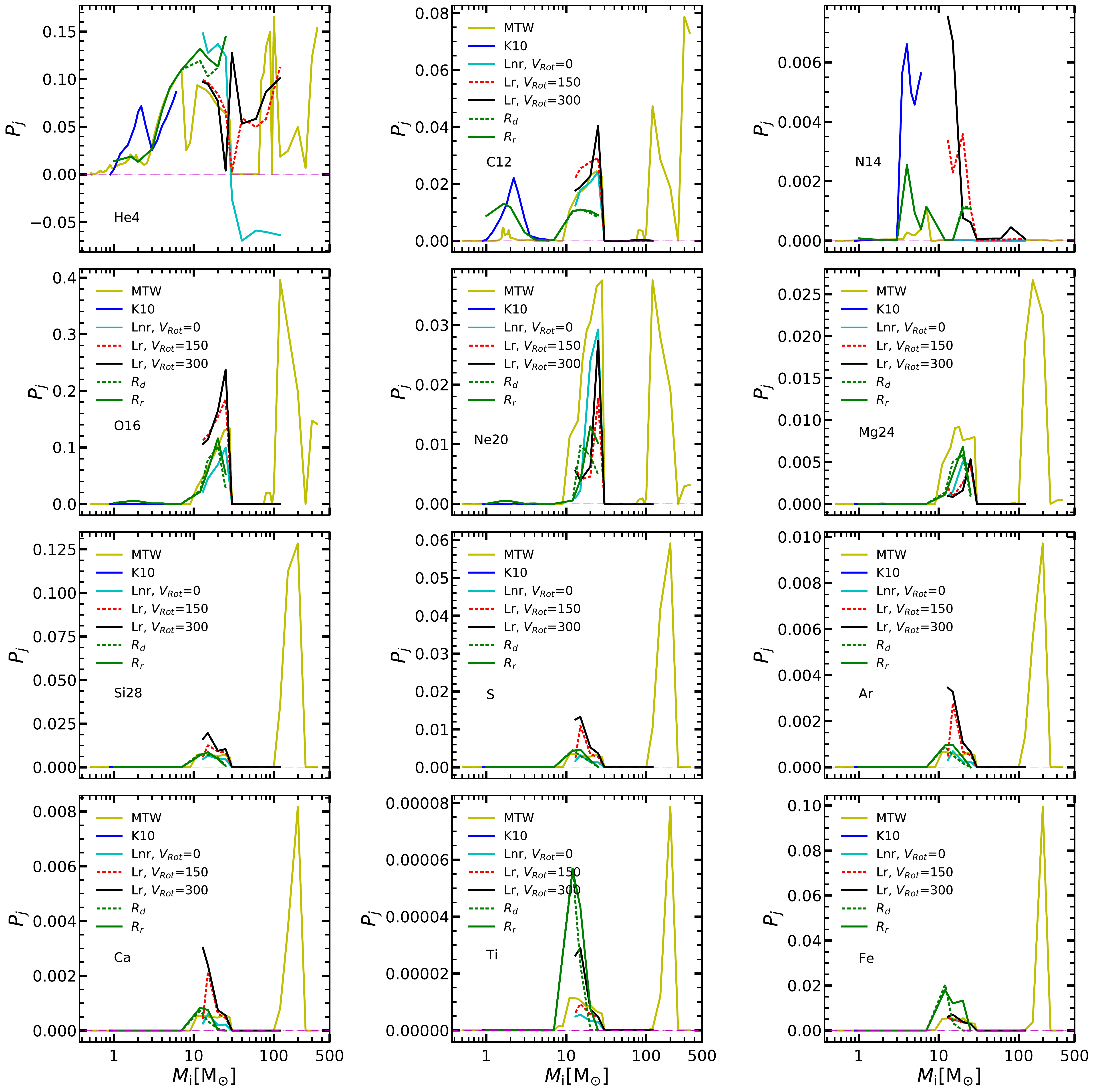}}
\caption{Fractional total ejecta (winds and explosion) of new production, $P_j$, for $\Zi = 0.0001$ as a function of \Mi.
Our MTW ejecta are compared with those of other literature works. 
The chemical species shown are $^4$He, $^{12}$C, $^{14}$N, $^{16}$O, $^{20}$Ne, $^{24}$Mg, $^{28}$Si, S, Ar, Ca, Ti, Fe.}
\label{ejecta-comp0001}
\end{figure*}

\begin{figure*} 
\centering
\resizebox{\hsize}{!}{\includegraphics[]{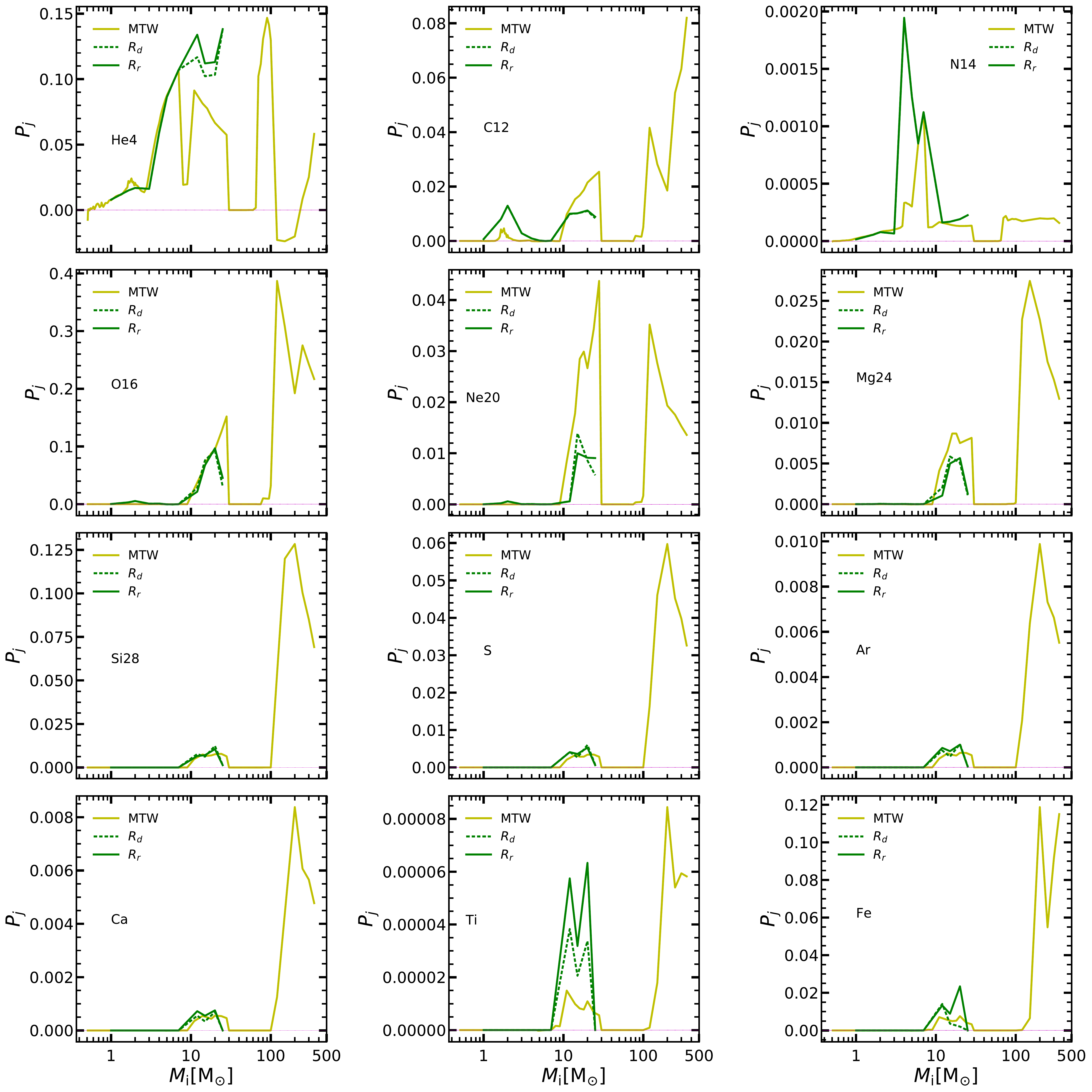}}
\caption{The same as in Fig.~\ref{ejecta-comp0001} but for $\Zi = 0.001$.}
\label{ejecta-comp001}
\end{figure*}

\begin{figure*} 
\centering
\resizebox{\hsize}{!}{\includegraphics[]{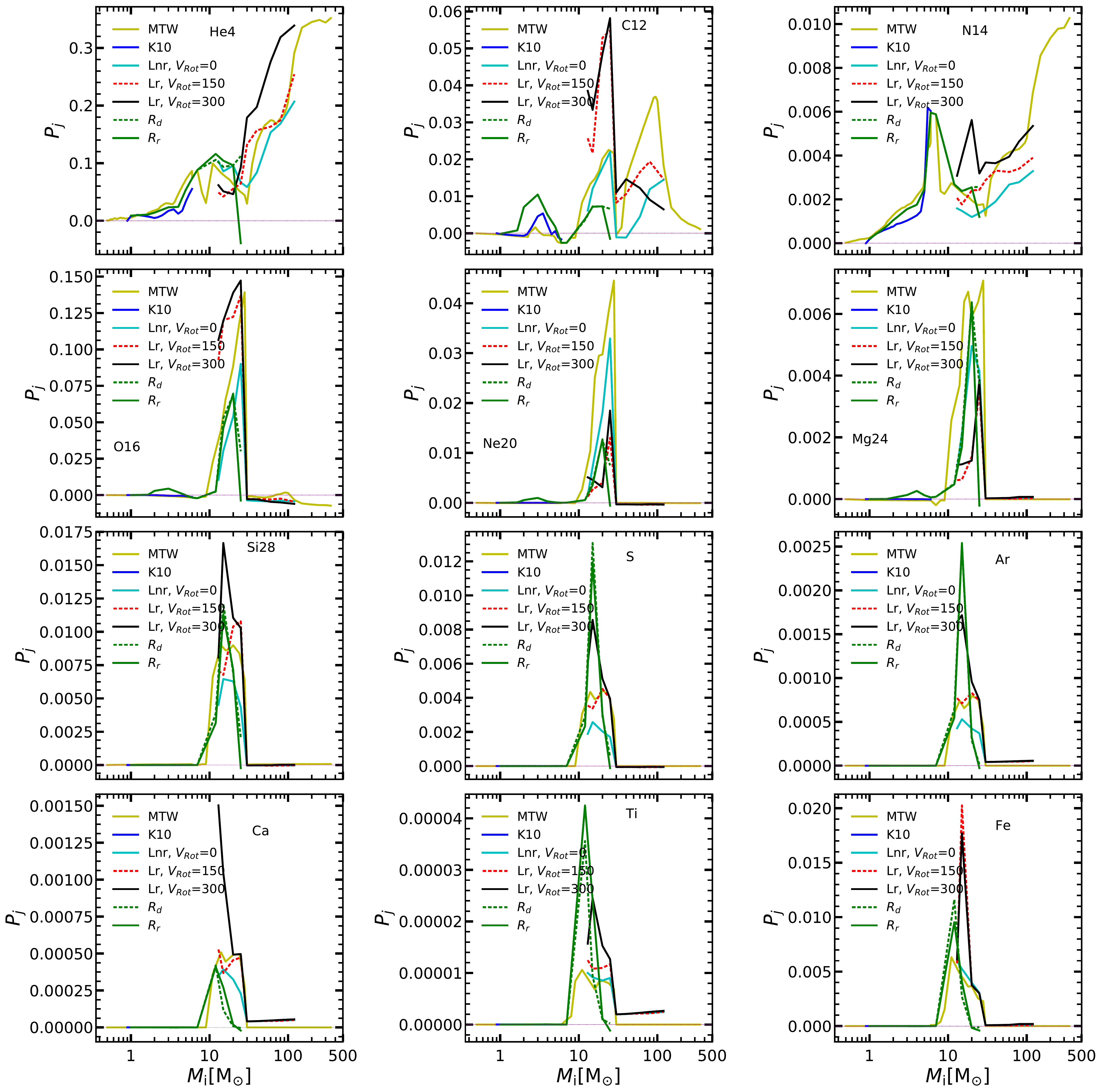}}
\caption{The same as in Fig.~\ref{ejecta-comp0001} but for $\Zi= 0.02$.}
\label{ejecta-comp02}
\end{figure*}

Similarly, the newly synthesised total ejecta (winds plus explosion) can be suitably analysed with the aid of the quantity 
\begin{equation*}
P_j(\Mi)= \left[(E_j^{\rm w} + E_j^{\rm sn}) - (\Mi-\Mrem) X_{j,0} \right]/\Mi\, ,
\end{equation*}
which represents the fraction of the initial stellar mass that is ejected in the form of the species $j$.
From the definition it follows that $P_j(\Mi)> 0$ corresponds to a net production of the species $j$, while $P_j(\Mi) < 0$ means that the species $j$ is effectively destroyed compared to its initial content.
Figures~\ref{ejecta-comp0001} - \ref{ejecta-comp02}  show 
$P_j(\Mi)$ as a function of \Mi\ and \Zi\ for the same chemical species as Fig.~\ref{ej-wind1}, and compare
the predictions obtained with the various sets of yields discussed in this work.

\begin{figure*} 
\centering
\resizebox{\hsize}{!}{\includegraphics[]{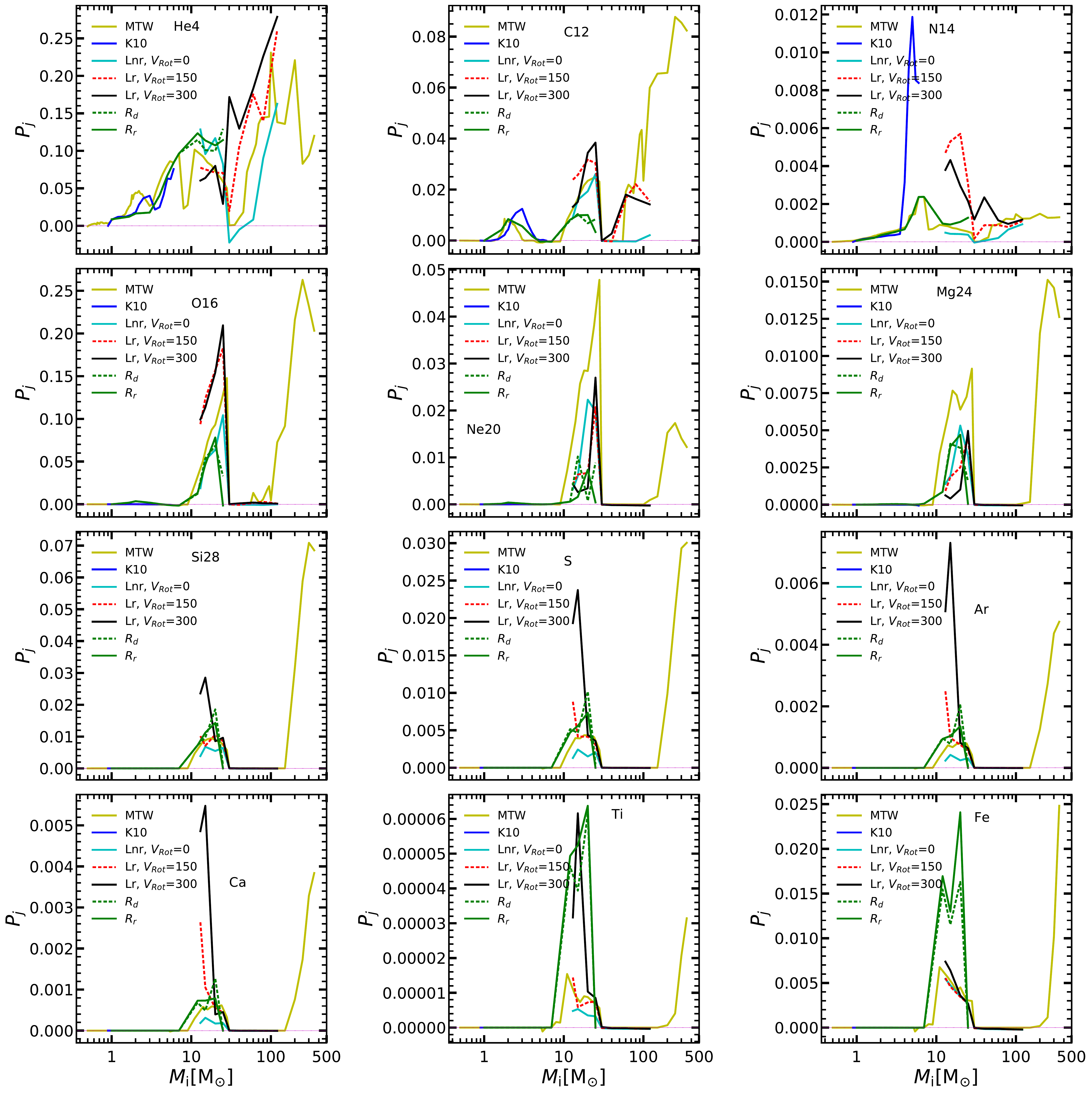}}
\caption{The same as in Fig.~\ref{ejecta-comp0001} but for $\Zi= 0.006$. Note that,  LC18 ejecta are taken from their closest lower metallcitiy, $\Zi=0.004$. }
\label{ejecta-comp006}
\end{figure*}

\begin{figure*} 
\centering
\resizebox{\hsize}{!}{\includegraphics[]{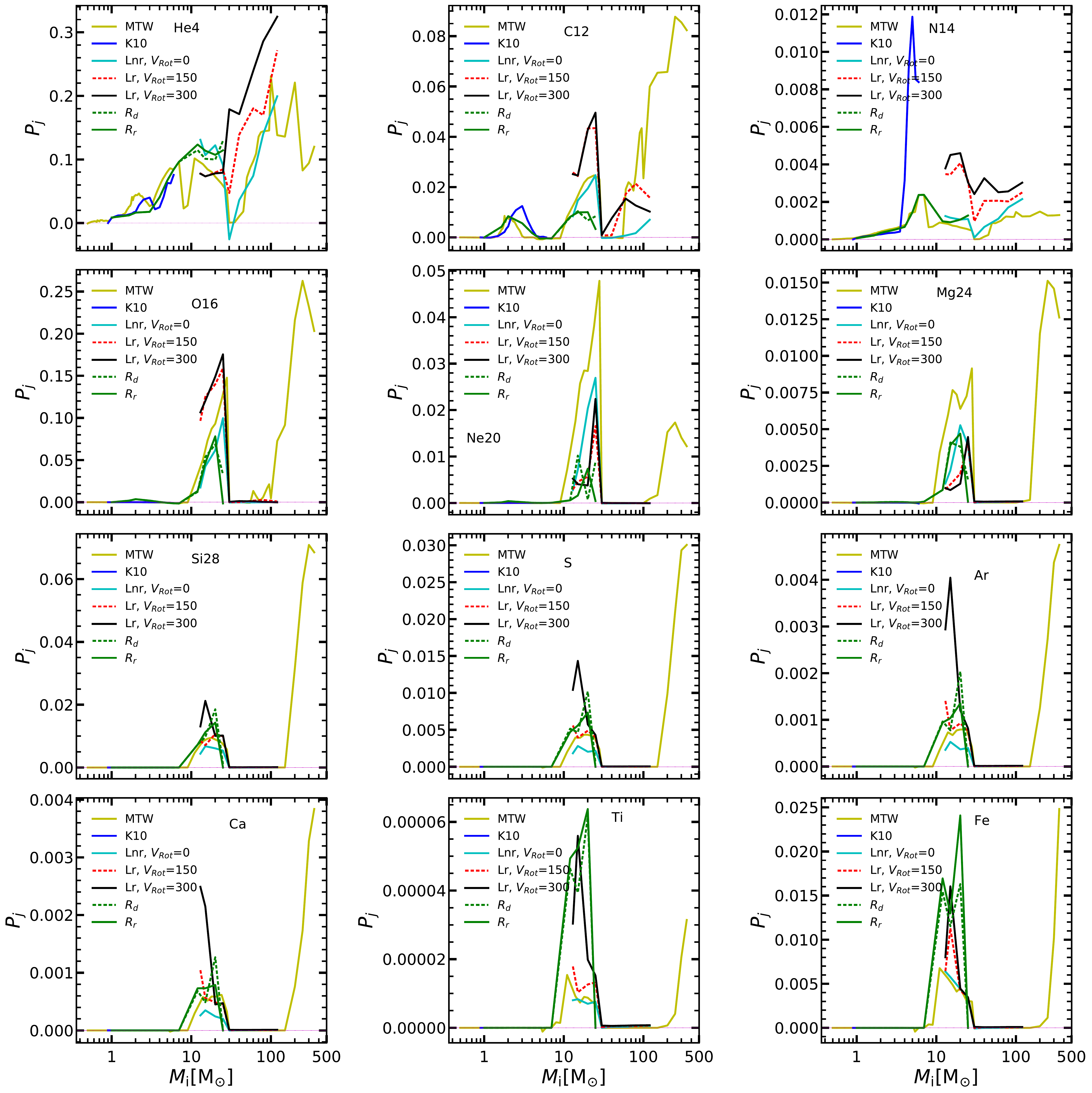}}
\caption{The same as in Fig.~\ref{ejecta-comp0001} but for $\Zi= 0.006$. Note that,  LC18 ejecta are taken from their closest larger metallcitiy, $\Zi=0.008$.}
\label{ejecta-comp006}
\end{figure*}

\makeatletter
\def\thebiblio#1{%
 \list{}{\usecounter{dummy}%
         \labelwidth\z@
         \leftmargin 1.5em
         \itemsep \z@
         \itemindent-\leftmargin}
 \reset@font\small
 \parindent\z@
 \parskip\z@ plus .1pt\relax
 \def\newblock{\hskip .11em plus .33em minus .07em}
 \sloppy\clubpenalty4000\widowpenalty4000
 \sfcode`\.=1000\relax
}
\let\endthebiblio=\endlist
\makeatother




\label{lastpage}

\end{document}